\documentclass[notitlepage,11pt]{article}
\usepackage{amsmath}
\usepackage{amssymb}
\usepackage{mathrsfs}
\usepackage{setspace, subfig}
\usepackage{achicago}
\usepackage{graphicx}
\usepackage{subfig}
\usepackage{caption}
\usepackage{color}
\usepackage{hyperref}
\usepackage{booktabs}
\usepackage{array}
\usepackage{subfloat}
\usepackage{tabularx}
\usepackage{rotating}
\usepackage{cite}
\usepackage[normalem]{ulem}
\useunder{\uline}{\ul}{}

\renewcommand{\cite}{\citeasnoun}
\def\bsk{\bigskip}                      
\def\msk{\medskip}
\def\ssk{\smallskip}

\begin{document}
\begin{titlepage}

\title{\textbf{Predicting Consumer Default:  \\ A Deep Learning Approach}\thanks{ We are grateful to Dokyun Lee, Sera Linardi, Yildiray Yildirim, Albert Zelevev  and  seminar participants at the Financial Conduct Authority, the University of Pittsburgh, the European Central Bank, Baruch College and Goethe University for useful comments and suggestions. This research was supported by the National Science Foundation under Grant No. SES 1824321. This research was also supported in part by the University of Pittsburgh Center for Research Computing through the resources provided. Correspondence to: stefania.albanesi@gmail.com.} }

\author{Stefania Albanesi, University of Pittsburgh, NBER and CEPR 
\and
Domonkos F. Vamossy, University of Pittsburgh
}

\maketitle
\thispagestyle{empty}

\begin{abstract}
We develop a  model to predict consumer default based on deep learning. We show that the model consistently outperforms standard credit scoring models, even though it uses the same data. Our model is interpretable and is able to provide a score to a larger class of borrowers relative to standard credit scoring models while accurately tracking variations in systemic risk. We argue that these properties can provide valuable insights for the design of policies targeted at reducing consumer default and alleviating its burden on borrowers and lenders, as well as macroprudential regulation.
\msk
\bsk


\bsk
\msk

\end{abstract}

\end{titlepage}
\newpage

\begin{onehalfspacing}
 
\section{Introduction}\label{sec:intro}
\begin{figure}[!h]
\begin{center}
\subfloat[Total consumer credit balances]{\includegraphics[scale=0.375]{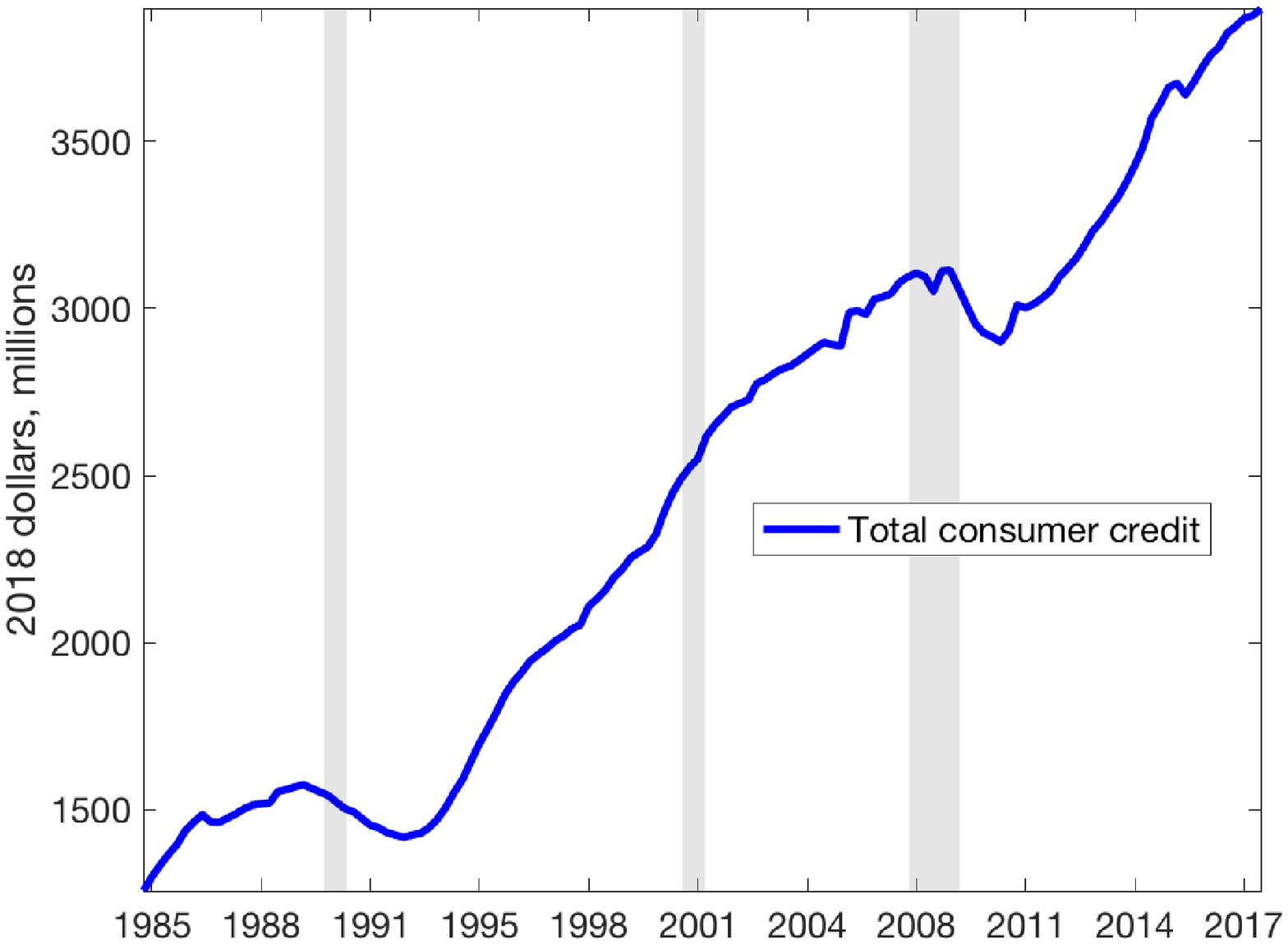}}
\subfloat[Delinquency rate on consumer loans]{\includegraphics[scale=0.375]{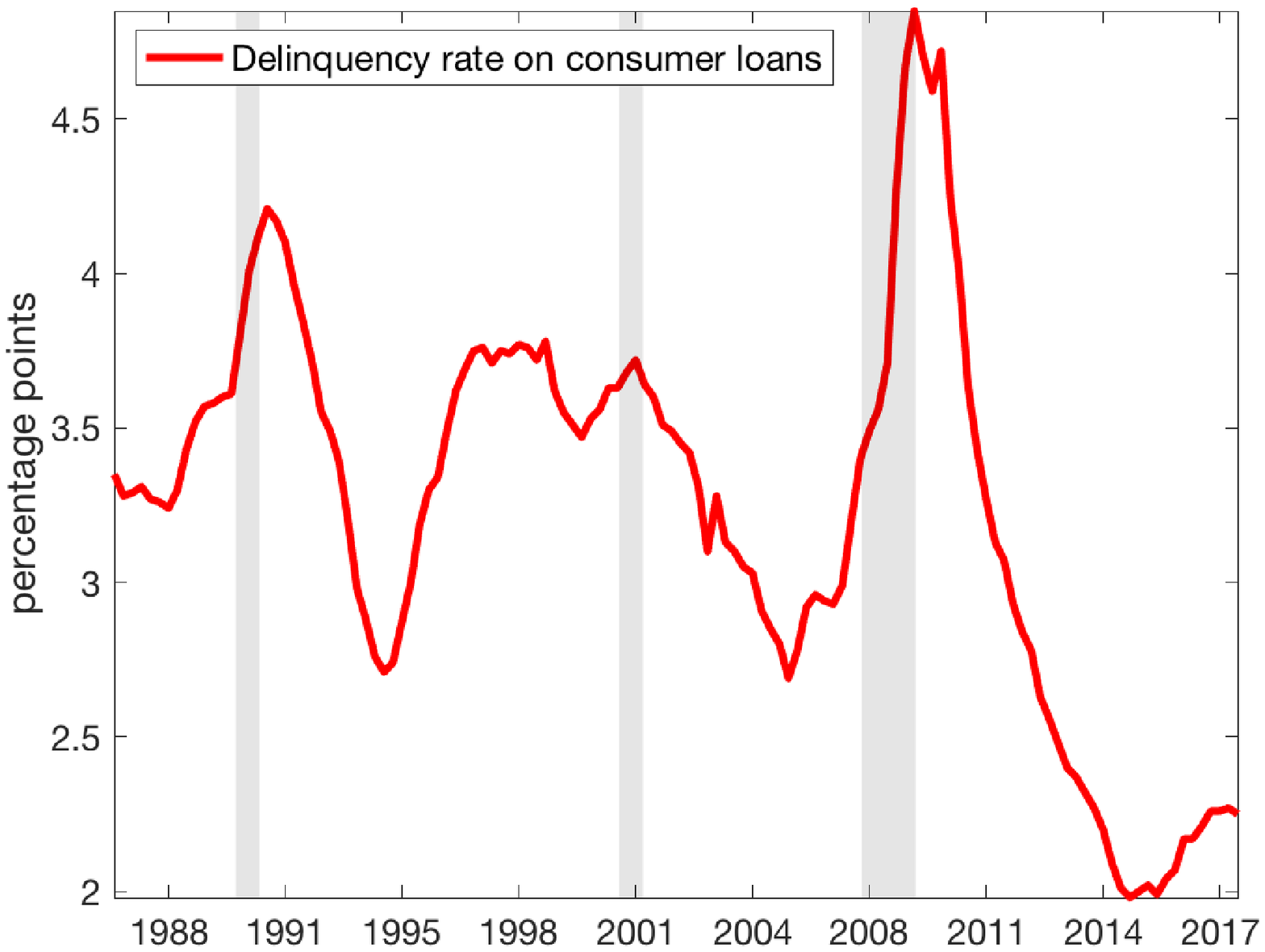}}
\caption{Source: Author's calculations based on Federal Reserve Board data.}\label{fig:intro}
\end{center}
\end{figure}

The dramatic growth in household borrowing since the early 1980s has increased the macroeconomic impact of consumer default. Figure \ref{fig:intro} displays total consumer credit balances in millions of 2018 USD and the delinquency rate on consumer loans starting in 1985. The delinquency rate mostly fluctuates between 3 and 4\%, except at the height of the Great Recession when it reached a peak of over 5\%, and in its aftermath when it dropped to a low of 2\%. With the rise in consumer debt, variations in the delinquency rate have an ever larger impact on household and financial firm balances sheets. 
Understanding the determinants of consumer default and  predicting its variation over time and across types of consumers can not only improve the allocation of credit, but also lead to important insights for the design of policies aimed at preventing consumer default or alleviating its effects on borrowers and lenders. They are also critical for macroprudential policies, as they can assist with the assessment  of the impact of consumer credit on the fragility of the financial system.

This paper proposes a novel approach to predicting consumer default based on deep learning. We rely on deep learning as this methodology is specifically designed for prediction in environments with high dimensional  data and complicated non-linear patterns of interaction among factors affecting the outcome of interest, for which standard regression approaches perform poorly. 
Our methodology  uses the same information as standard credit scoring models, which are one of the most important factors in the allocation of consumer credit. 
We show that our model improves the accuracy of default predictions while increasing transparency and accountability. It is also able to track variations in systemic risk, and is able to identify the most important factors driving defaults and how they change over time.  Finally, we show that adopting our model can accrue substantial savings to borrowers and lenders.

Credit scores constitute one of the most important factors in the allocation of consumer credit in the United States. They are proprietary measures designed to rank borrowers based on their probability of future default. Specifically, they target the probability of a 90 days past due delinquency in the next 24 months.\footnote{The most commonly known is the FICO score, developed by the FICO corporation and launched in 1989. The  three  credit reporting companies or CRCs, Equifax, Experian and TransUnion have also partnered to produce VantageScore, an alternative score, which was launched in 2006. Credit scoring models are updated regularly. More information on credit scores is reported in Section \ref{sec:credit_score_comp} and Appendix \ref{app:score}.} 
Despite their ubiquitous use in the financial industry, there is very little information on credit scores, and emerging evidence suggests that  as currently formulated credit scores have severe limitations. For example, \citeN{New_Narrative_NBER} show that during the 2007-2009 housing crisis there was a marked rise in mortgage delinquencies and foreclosures among high credit score borrowers, suggesting that credit scoring models at the time did not accurately reflect the probability of default for these borrowers. Additionally, it is well known that credit scores and indiscriminately low for young borrowers,  and a substantial fraction of borrowers are unscored, which prevents them from accessing conventional forms of consumer credit.

The Fair Credit Reporting Act, a legislation passed in 1970, and the Equal Opportunity in Credit Access Act of 1984 regulate credit scores and in particular determine which information can be included and must be excluded in credit scoring models. Such models can incorporate information in a borrower's credit report, except age and location. These restrictions are intended to prevent discrimination by age and factors related to location, such as race.\footnote{Credit scoring models are also restricted by law from using information on race, color, gender, religion, marital status, salary, occupation, title, employer, employment history, nationality.} The law also mandates that entities that provide credit scores make public the four most important factors affecting scores.  In marketing information, these are reported to be payment history, which is stated to explain about 35\% of variation in credit scores, followed by amounts owed, length of credit history, new credit and credit mix, explaining 30\%, 15\%, 10\% and 10\% of the variation in credit scores respectively. Other than this, there is very little public information on credit scoring models, though several services are now available that allow consumers to simulate how various scenarios, such as paying off balances or taking out new loans, will affect their scores.

The purpose of our analysis is to propose a model to predict consumer default that uses the same data as conventional credit scoring models, improves on their performance, benefiting both lenders and borrowers, and provides more transparency and accountability. To do so, we resort to deep learning, a type of machine learning ideally suited to high dimensional data, such as that available in consumer credit reports.\footnote{For excellent reviews of how machine learning can be applied in economics, see \citeN{mullainathan2017machine} and \citeN{athey2019machine}.} Our model uses inputs as features, such as debt balances and number of trades, delinquency information, and attributes related to the length of a borrower's credit history, to produce an individualized estimate that can be interpreted as a probability of default. We target the same default outcome as conventional credit scoring models, namely a 90+ days delinquency in the subsequent 8 quarters. For most of the analysis, we train the model on data for one quarter and test it on data 8 quarters ahead, in keeping with the default outcome we are considering, so that our predictions are truly out of sample. We present a variety of performance metrics suggesting that our model has very strong predictive ability. Accuracy, that is percent of observations correctly classified, is above 86\% for all periods in our sample, and the AUC-Score, a commonly used metric in machine learning, is always above 92\%.

To better assess the validity of our approach, we compare our deep learning model to logistic regression and a number of other machine learning models. Deep learning models feature multiple hidden layers, designed to capture multi-dimensional feature interactions. By contrast, logistic regression can be interpreted as a neural network without any hidden layers. Our results suggest that deep learning is necessary to capture the complexity associated with default behavior, since all deep models perform substantially better than logistic regression. The importance of feature interaction reflects the complexity associated with default behavior.
Additionally, our optimized model combines a deep neural network and gradient boosting and  outperforms  other machine learning models, such as random forests and decision trees, as well as deep neural networks and gradient boosting in isolation. However, all approaches show much stronger performance than logistic regression, suggesting that the main advantage is the adoption of a deep framework.

We also compare the performance of our model to a conventional credit score. By construction, credit scores only provide an ordinal ranking of consumers based on their default risk, and are not associated to a specific default probability. Yet, it is still possible to compare performance by assessing whether borrowers fall in different points of the distribution with the credit score compared to our model predictions. We find that our model performs significantly better than conventional credit scores. The rank correlation between realized default rates and the credit score is about 98\%, where it is close to 1 for our model. Additionally, the Gini coefficient for the credit score, a measure of the ability to differentiate borrowers based on their credit score is approximately 81\% and drops during the 2007-2009 crisis, while the Gini coefficient for our model is approximately 86\% and stable over time. Perhaps most importantly, the credit score generates large disparities between the implied predicted probability of default and the realized default rate for large groups of customers, particularly at the low end of the credit score distribution. As an illustration, among Subprime borrowers, 17\% display default behavior which is consistent with Near Prime borrowers  and 15\% display default behavior consistent with Deep Subprime. The default rates for Deep Subprime, Subprime and Near Prime borrowers are respectively 95\%, 79\% and 44\%, so this misclassification is large, and it would imply large losses for lenders and borrowers in terms of missed revenues or higher interest rates. By contrast, the discrepancy between predicted and realized default rates for our model is never more than 4 percentage points for categories with at least a percent share of default risk.

Another advantage of our approach when compared to conventional credit scoring models is that we can generate a predicted probability of default for a much larger class of borrowers. Borrowers may be unscored because they do not have sufficient information in their credit report or because the information is stale, and approximately 8\% of borrowers fall into this category.\footnote{See \citeN{CFPB_2016_unscored}. For more information, see Section \ref{sec:unscored}.} The absence of a credit score implies that these borrowers do not qualify for most types of credit and is very consequential. Our model can generate a predicted probability of default for all borrowers with a non-empty credit record. We achieve this in part by not including lags in our specification, which implies that only current information in a borrower's credit report is used. This is not costly from a performance standpoint as many attributes used as inputs in the model are temporal in nature and capture lagged behavior, such as "worst status on all trades in the last 6 months."  

We also examine the ability of our model to capture the evolution of aggregate default risk. Since our data set is nationally representative and we can score all borrowers with a non-empty credit record, the average predicted probability of default in the population based on our model corresponds to an estimate of aggregate default risk. We find that our model tracks the behavior of aggregate default rates remarkably well. It is able to capture the sharp rise in aggregate default rates in the run up and during the 2007-2009 crisis and also captures the inversion point and the subsequent drastic reduction in this variable. With the growth in consumer credit, household balance sheets have become very important for macroeconomic performance.  Having an accurate assessment of the financial fragility of the household sector, as captured by the predicted probability of default on consumer credit has become crucially important and can aid in macro prudential regulation, as well as for designing fiscal and monetary  policy responses to adverse aggregate economic shocks. This is another advantage of our model compared to credit scores, since the latter only provides an ordinal ranking of consumers with respect to their probability of default. Our model can provide such a ranking but in addition also provides an individual prediction of the default rate which can be aggregated into a systemic measure of default risk for the household sector.

As a final application, we compute the value to borrowers and lenders of using our model. For consumers, the comparison is made relative to the credit score. Specifically, we compute the credit card interest rate savings of being classified according to our model relative to the credit score. Being placed in a higher default risk category substantially increases the interest rates charged on credit cards at origination and increasingly so as more time lapses since origination, whereas being placed in a lower risk category reduces interest rate costs. We choose credit cards as they are a very popular form of unsecured debt, with 73\% of consumers holding at least one credit  or bank card.   In percentage of credit cards balances, average net interest rate expense savings are approximately 5\% for low credit score borrowers. These values constitute lower bounds as they do not include the higher fees and more stringent restrictions associated with credit cards targeted to low credit score borrowers and the increased borrowing limits available to higher credit score borrowers. For lenders, we calculated the value added by using our model in comparison to not having a prediction of default risk or having a prediction based on logistic regression. We use logistic regression for this exercise as it is understood to be the main methodology for conventional credit scoring models. Over a loan with a three year amortization period, we find that the gains relative to no forecast are in the order of 75\% with a 15\% interest rate, while the gains for relative to a model based on logistic regression are approximately 5\%. These results suggest that both borrowers and lenders would experience substantial gains from switching to our model.

Our analysis contributes to the literature on consumer default in a variety of ways. We are the first to develop a prediction model of consumer default using credit bureau data that complies with all of the restrictions mandated by U.S. legislation in this area, and we do so using a large and temporally extended panel of data. This enables us to evaluate model performance in a setting that is closer to the one prevailing  in the industry and to train and test our model in a variety of different macroeconomic conditions. Previous contributions either focus on particular types of default or use transaction data that is not admissible in conventional credit scoring models. The closest contributions to our work are \citeN{khandani}, \citeN{butaru}  and  \citeN{sirignano}.
\citeN{khandani} apply a decision tree approach to forecast credit card delinquencies with data for 2005-2009. They estimate cost savings of cutting credit lines based on their forecasts and calculate implied time series patterns of estimated delinquency rates.
\citeN{butaru} apply machine learning techniques to combined consumer trade line, credit bureau, and macroeconomic variables for 2009-2013 to predict delinquency. They find substantial heterogeneity in risk factors, sensitivities, and predictability of delinquency across lenders, implying that no single model applies to all  institutions in their data.  
\citeN{sirignano} examine over 120 million mortgages between 1995 to 2014 to develop prediction models of multiple states, such as probabilities of prepayment, foreclosure and various types of delinquency. They use loan level and zip code level aggregate information, and provide a review of the literature using machine learning and deep learning in  financial economics. 
\citeN{kvamme2018} predict mortgage default using use convolutional neural networks and emphasize the advantages of deep learning, but they do not evaluate their models out of sample the way we do.
Finally, \citeN{lessmann} reviews the recent literature on credit scoring, which is based on substantially smaller datasets than the one we have access to, and recommends random forests as a possible benchmark. However, we find that our hybrid  model as well as our model components, a deep neural network and gradient boosted trees, improves substantially over random forests, possibly owing to recent methodological advances in deep learning, including the use of dropout, the introduction of new activation functions and the ability to train larger models.

Our model is interpretable, which implies that we are able to assess the most important factors associated with default behavior and how they vary over time. This information is important for lenders, and can be used to comply with legislation that requires lenders and credit score providers to notify borrowers of the most important factors affecting their credit score. Additionally, it can be used to formulate economic models of consumer default. The literature on consumer default\footnote{ Some notable contributions include \citeN{chatterjee2007quantitative}, \citeN{livshits2007consumer}, and \citeN{athreya2012quantitative}.} suggests that the determinants of default are related to preferences, such as impatience which increases the propensity to borrow, or adverse expenditure of income shocks. Based on these theories, it is then possible to construct theoretical models of credit scoring, of which \citeN{chatterjee2016theory} is a leading example. We find that the number of trades and the balance on outstanding loans are the most important factors associated with an increase in the probability of default, in addition to outstanding delinquencies and length of the credit history. This information can be used to improve models of consumer default risk and enhance their ability to be used for policy analysis and design.

We also identify and quantify a variety of limitations of conventional credit scoring models, particularly  their tendency to misclassify borrowers by default risk, especially for relatively risky borrowers. This implies that our default predictions could help improve the allocation of credit in a way that benefits both lenders, in the form of lower losses, and borrowers, in the form of lower interest rates. Our results also speak to the perils associated with using conventional credit scores outside on the consumer credit sphere. As it is well known, credit scores are used to screen job applicants, in insurance applications, and a variety of additional settings. Economic theory would suggest that this is helpful, as long as credit score provide information which is correlated with characteristics that are of interest for the party using the score (\citeN{Corbae_Glover_2018}). However, as we show, conventional credit scores misclassify borrowers by a very large degree based on their default risk, which implies that they may not be accurate and may not include appropriate information or use adequate methodologies. The broadening use of credit scores would amplify the impact of these limitations.


The paper is structured as follows. Section \ref{sec:data} describes our data. Section \ref{sec:patterns} discusses the patterns of consumer default that motivate our adoption of deep learning. Section \ref{sec:model} describes our prediction problem and our model. Section \ref{sec:results} provides a comprehensive performance assessment of our model, compares it to other approaches, and uses a variety of interpretability techniques to understand which factors are strongly associated with  default behavior. Section \ref{sec:applications} compares our model to conventional credit scores, illustrates its performance in predicting and quantifying aggregate default risk and calculates the value added of adopting our model over alternatives for lenders and borrowers. 

\section{Data}\label{sec:data}
We use anonymized credit file data from the Experian credit bureau. The data is quarterly, it starts  in 2004Q1 and ends in 2015Q4.  The data comprises over 200 variables for an anonymized panel of 1 million households. The panel is nationally representative, constructed from a random draw for the universe of borrowers with an Experian credit report. The attributes available comprise information on credit cards, bank cards, other revolving credit, auto loans, installment loans, business loans, first and second mortgages, home equity lines of credit, student loans and collections. There is information on the number of trades for each type of loan, the outstanding balance and available credit, the monthly payment, and whether any of the accounts are delinquent, specifically 30, 60, 90, 180 days past due, derogatory or charged off. All balances are adjusted for joint accounts to avoid double counting. Additionally, we have the number of hard inquiries by type of product, and public record items, such as bankruptcy by chapter, foreclosure and liens and court judgments. For each quarter in the sample, we also have each borrowers's credit score. The data also includes an estimate of individual and household labor income based on IRS data. Because this is data drawn from credit reports, we do not know gender, marital status or any other demographic characteristic, though we do know a borrower's address at the zip code level. We also do not have any information on asset holdings. 

Table \ref{tab:desc_stats} reports basic demographic information on our sample, including age, household income, credit score and incidence of default, which here is defined as the fraction of households who report a 90 or more days past due delinquency on any trade. This will be our baseline definition of default, as this is the outcome targeted by credit scoring models. Approximately 34\% of consumers display such a delinquency.

\begin{table}[htbp]
\caption{Descriptive Statistics}\label{tab:desc_stats}
\normalsize 
\begin{center}
\begin{tabular}{lllllllll}
Feature & Mean & Std. Dev. & Min &  25\% & 50\% & 75\% & Max   \\ \hline 
&&&&&&& \\
Age       & 45.8   & 16.3  & 18 & 32.2     & 45.1     & 57.8      & 83 \\
Household Income       & 77.1   & 55.0  & 15 & 42     & 64      & 90      & 325 \\
Credit Score & 678.4 & 111.0 & 300 & 588 & 692 & 780 & 839 \\
Default within 8Q                         & 0.339 & 0.473 & 0 & 0 & 0 & 1 & 1 \\  \hline 
\end{tabular}
\end{center}

\msk

\footnotesize{Credit score corresponds to Vantage Score 3. Household income is in USD thousands, trimmed at the 99th percentile.  
Source: Authors' calculations based on Experian Data.}

\end{table}

\section{Patterns in Consumer Default}\label{sec:patterns}
We now illustrate the complexity of the relation between the various factors that are considered important drivers of consumer default. Our point of departure are standard credit scoring models. While these models are proprietary, the Fair Credit Reporting Act of 1970 and the Equal Opportunity in Credit Access Act of 1984 mandate that the 4 most important factors determining the credit scores be disclosed, together with their importance in determining variation in credit scores. These include credit utilization and number of hard inquiries, which are supposed to capture a consumer's demand for credit, the variety of debt products, which capture the consumer's experience in managing credit, and the number and severity of delinquencies. Each of these factors is stated to account for 10-35\% of the variation in credit scores. The length of the credit history is also seen as a proxy on a consumer's experience in managing credit, and this is reported as accounting for 15\% of the variation in credit scores.\footnote{For an overview of the information available to borrowers about the determinants for their credit score, see \url{https://www.myfico.com/resources/credit-education/whats-in-your-credit-score}.} The models used to determine credit scores as a function of these attributes are not disclosed, but they are widely believed to be based on linear and logistic regression as well as score cards. Additionally, available credit scoring algorithms typically do not score all borrowers.  

Subsequently, we illustrate the properties of consumer default that suggest deep learning might be a good candidate for developing a prediction model. Specifically, we show that default is a relatively rare but very persistent outcome, there are substantial non-linearities in the relation between default and plausible covariates, as well as high order interactions between covariates and default outcomes.

\subsection{Default Transitions}
The default outcome we consider is a 90+ days delinquency, which occurs if the borrower has missed scheduled payments on any product for 90 days or more.\footnote{For instance, if no payment has been made by the last day of the month within the past three months and the payment was due on the first day of the month three months ago. For credit cards, this occurs if the borrower does not make at least their minimum payment.} This is the default outcome targeted by the most widely used credit scoring models, which rank consumers based on their probability of becoming 90+ days delinquent in the subsequent 8 quarters. We refer to borrowers who are either current or up to 60 days delinquent on their payments as \emph{current}.

The  transition matrix from current to 90+ days past due in the subsequent 8 quarters is given in Table \ref{tab:transition}. Clearly, the two states are both highly persistent, with a 77\% of current customers remaining current in the next 8 quarters, and 93\% of customers in default remaining in that state over the same time period. The probability of transition from current to default is 23\%, while the probability of curing a delinquency with a transition from default to current is only 7\%. These results suggest that default is a particularly persistent state, and predicting a transition into default is very valuable form the lender's perspective, since they are unlikely to be able to recuperate their losses. But it is also quite difficult, as the current state is also very persistent.

\begin{table}[htbp]
\caption{Default Transitions}\label{tab:transition}
\normalsize
\begin{center}
\begin{tabular}{lll}
Current/Next 8Q & No default & Default \\ \hline
&& \\ 
No default & 0.776 & 0.224 \\
Default & 0.073 & 0.927 \\ \hline 
\end{tabular}
\msk
\end{center}

\footnotesize{Quarterly frequency of transition from current to default. Current corresponds to 0-89 day past due on any trade, Default corresponds to 90+ day past due on any  trade in the subsequent 8 quarters. Source: Authors' calculations based on Experian Data.}

\end{table}

\subsection{Non-linearities}
Our model includes a relatively large list of features, which is presented in Table \ref{tab:features}. The summary statistics for these features are reported in Table \ref{tab:sum_stats} in the Appendix. As is demonstrated in the table, there is a wide dispersion in the distribution of these variables. For example, the average balance on credit card trades is approximately \$4,500, but the standard deviation, at \$9,800, is more than twice as large. Similarly, average total debt balances are approximately \$77,000, while the standard deviation is \$170,000 and the 75th percentile \$95,000, suggesting a high upper tail dispersion of this variable. The other features display similar patterns.  

\begin{figure}[htbp]
\caption{Nonlinear Relation Between Default and Covariates}\label{fig:non_linear}
\begin{center}

\subfloat[]{\includegraphics[scale=0.475]{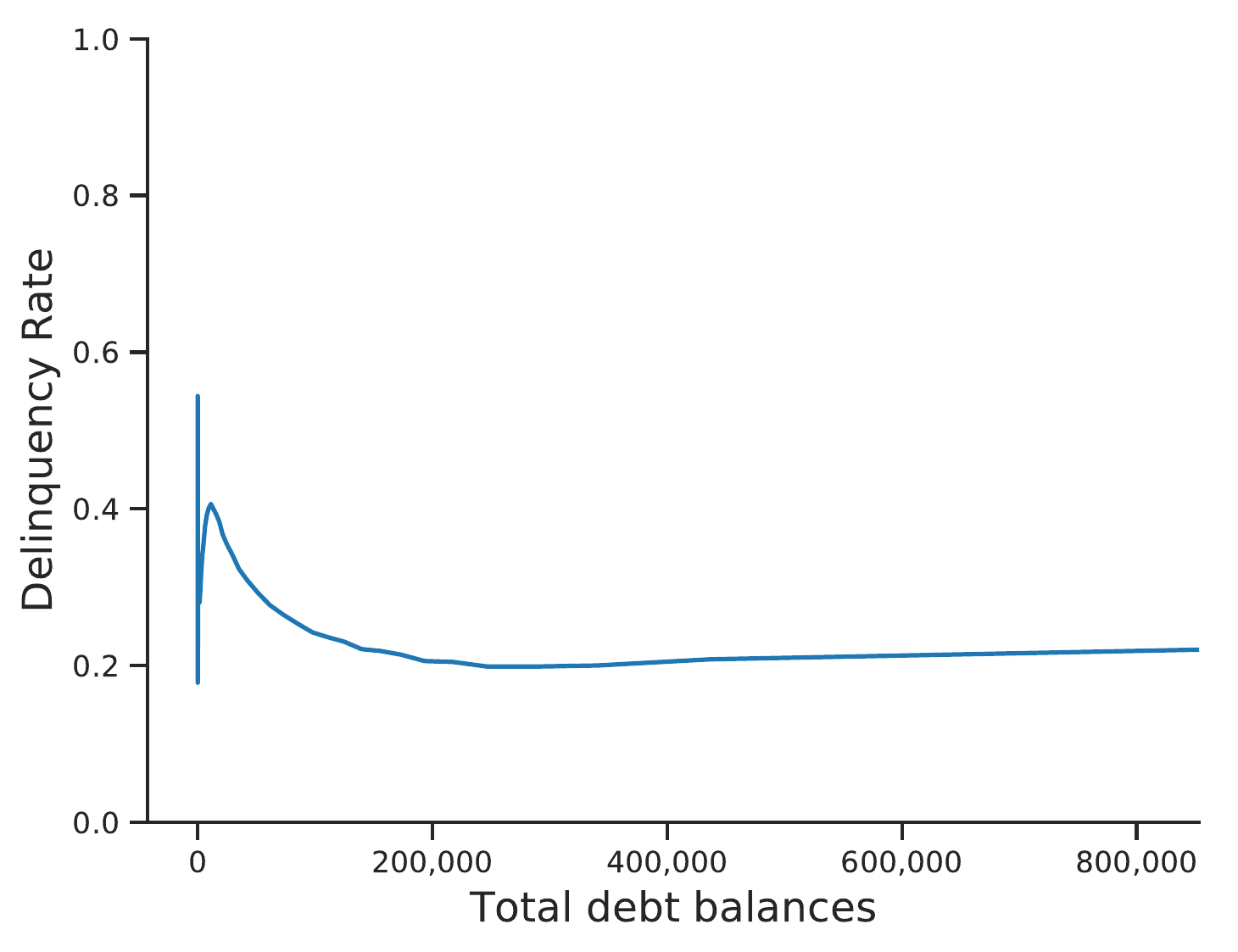}}
\subfloat[]{\includegraphics[scale=0.475]{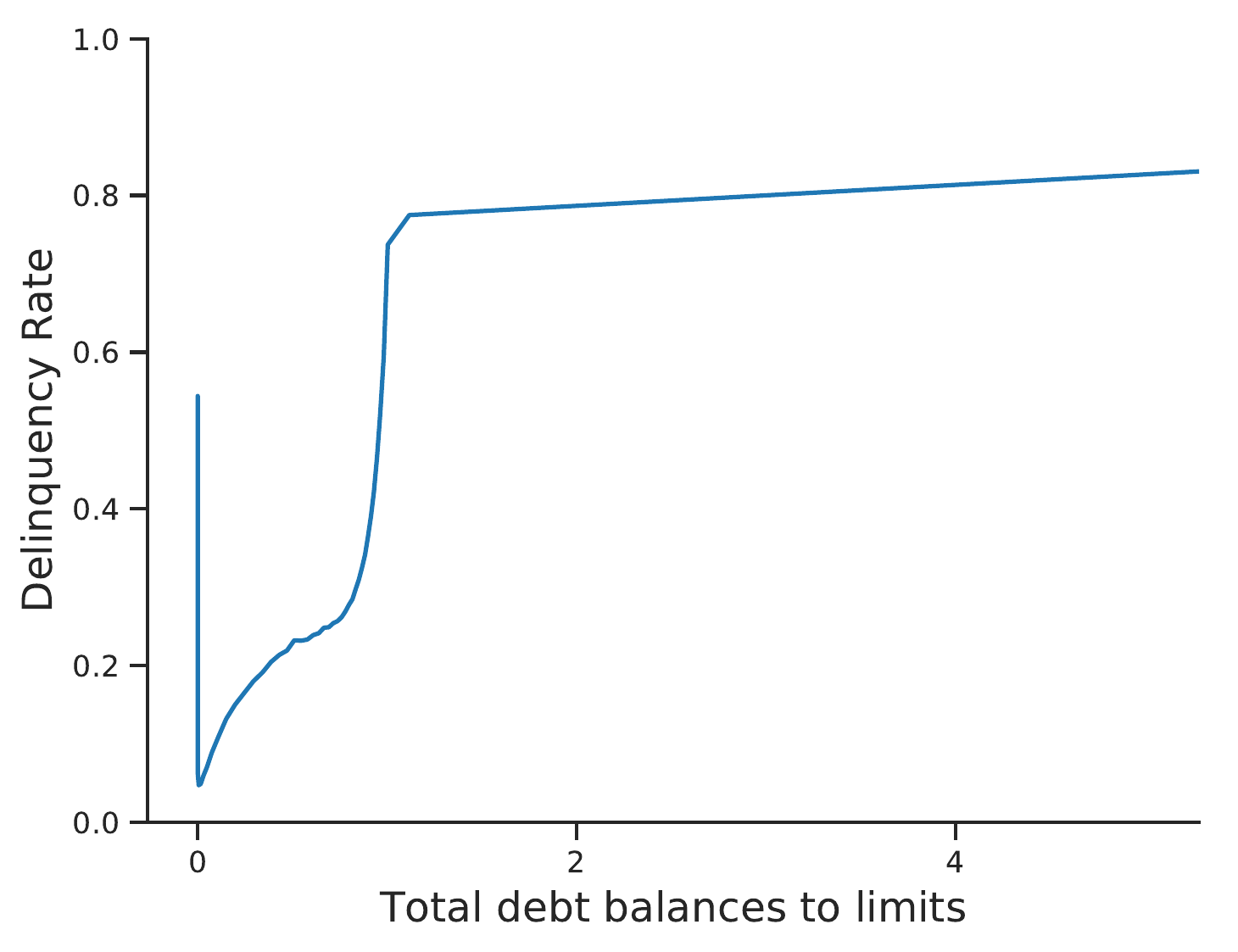}}

\subfloat[]{\includegraphics[scale=0.475]{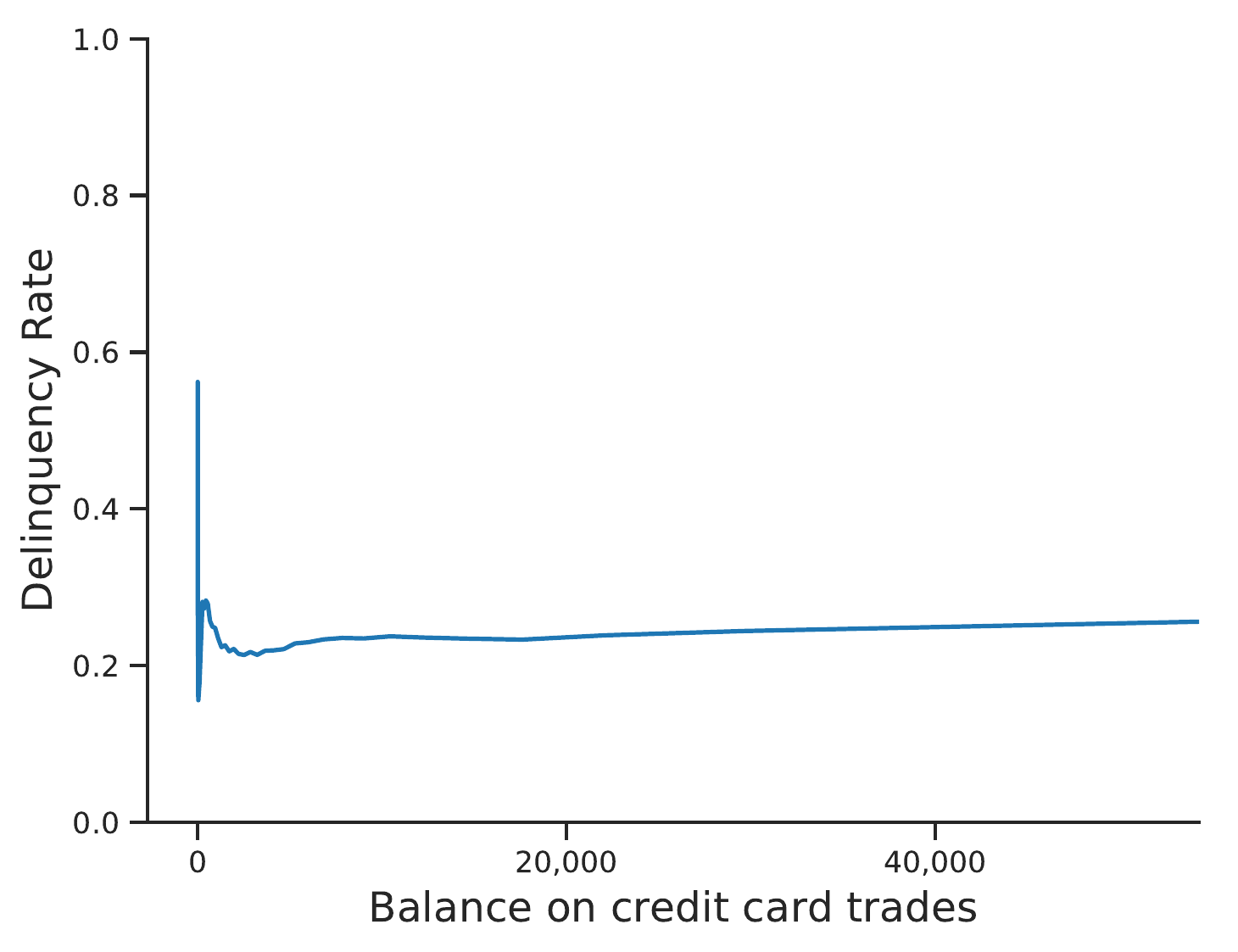}}
\subfloat[]{\includegraphics[scale=0.475]{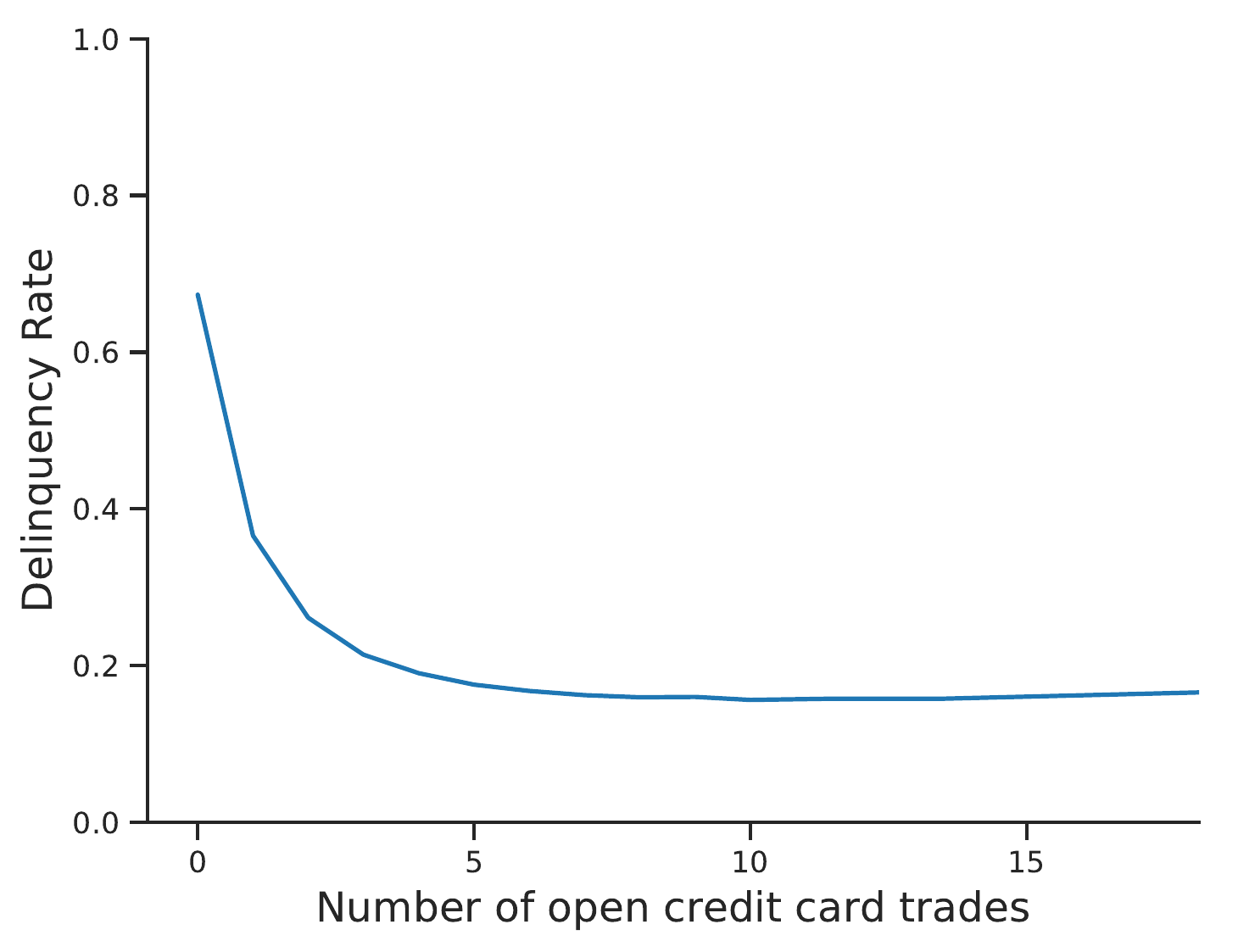}}

\subfloat[]{\includegraphics[scale=0.475]{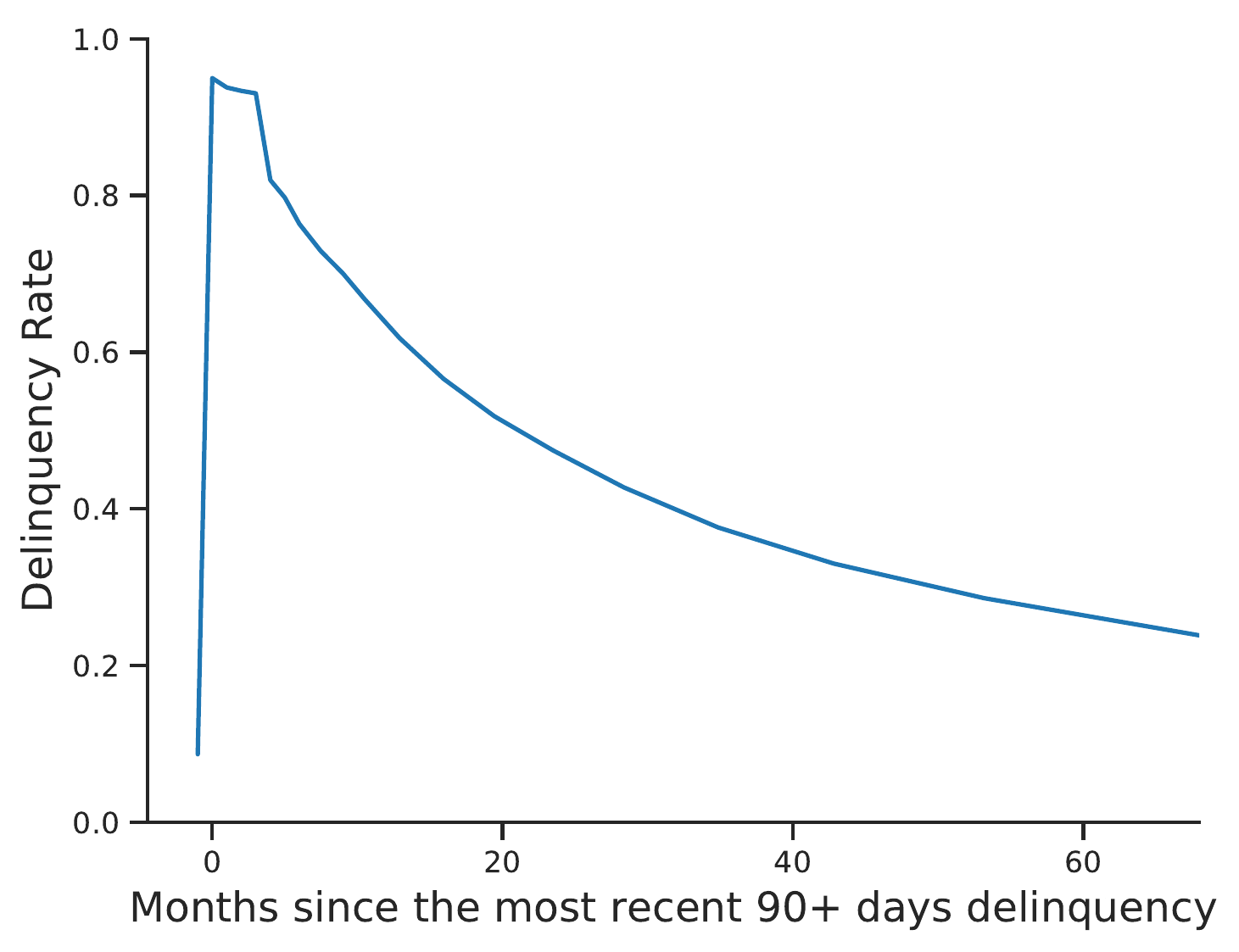}}
\subfloat[]{\includegraphics[scale=0.475]{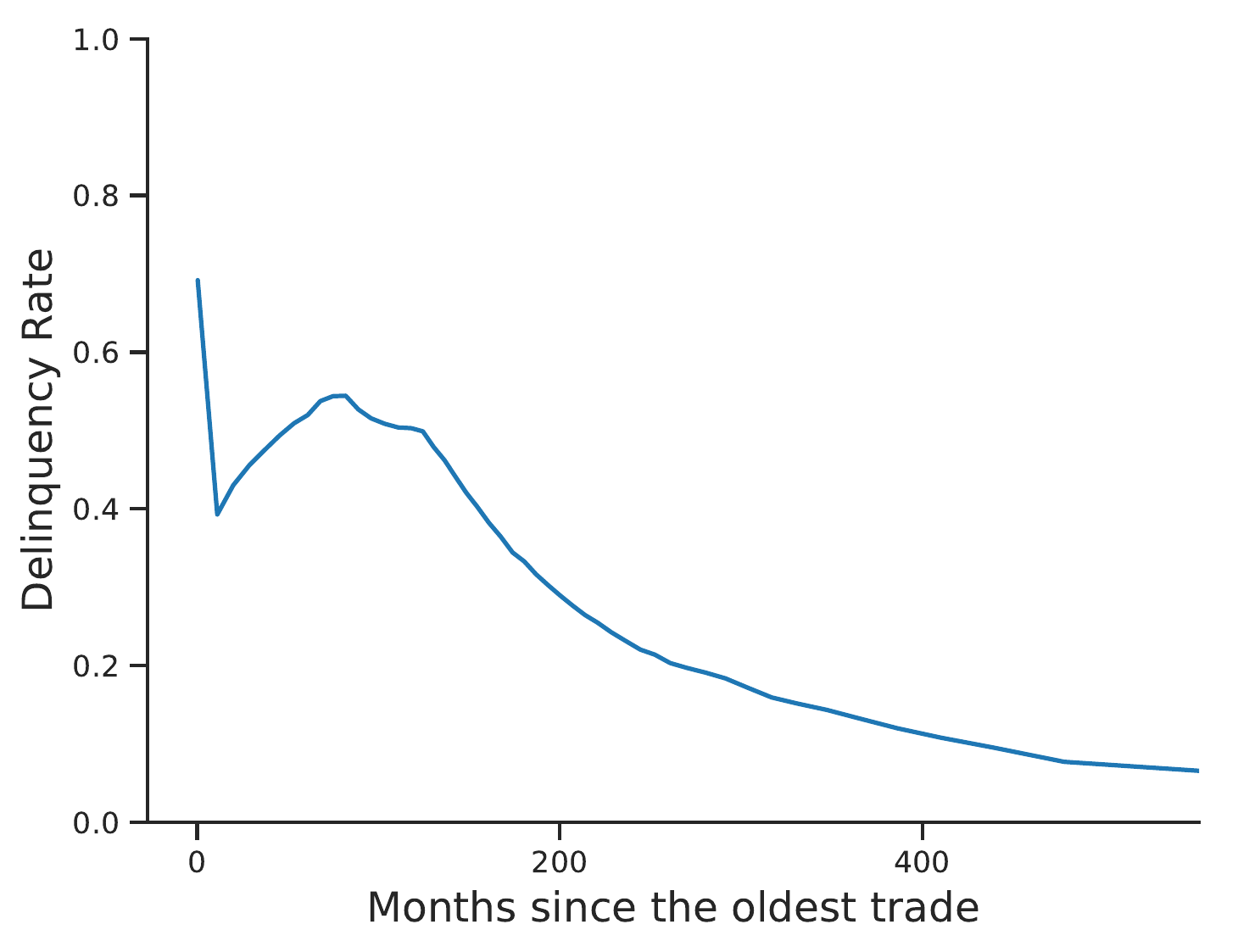}}
\end{center}

\footnotesize{Delinquency rate is the fraction with 90+ days past due trades in subsequent 8 quarters. In panel (e) and (f), -1 implies no past delinquency. Source: Authors' calculations based on Experian Data.}

\end{figure}

The features are used to predict the probability of default. We now illustrate the highly non-linear relation between the features and  the incidence of default. Figure \ref{fig:non_linear} shows how the default rate, defined as the fraction of borrowers with a 90+ day past due delinquency in the subsequent 8 quarters, varies with total debt balances, credit utilization,  the credit limit on credit cards,  the number of open credit card trades, the number of months since the most recent 90+ day past due delinquency and the months since the oldest trade was opened. The figures show that while the relation between the features and the incidence of default is mostly monotone, it is highly nonlinear, with vary little variation in the incidence of default for most intermediate values of the variable and much higher or lower values at the extremes of the range of each covariate. The variables in the figure are just illustrative, a similar pattern holds for most plausible features.

\subsection{High Order Interactions}
\begin{figure}[htpb]
\caption{Multidimensional Relation Between Default and Covariates}\label{fig:multi_dim}
\begin{center}

\subfloat[]{\includegraphics[scale=0.5]{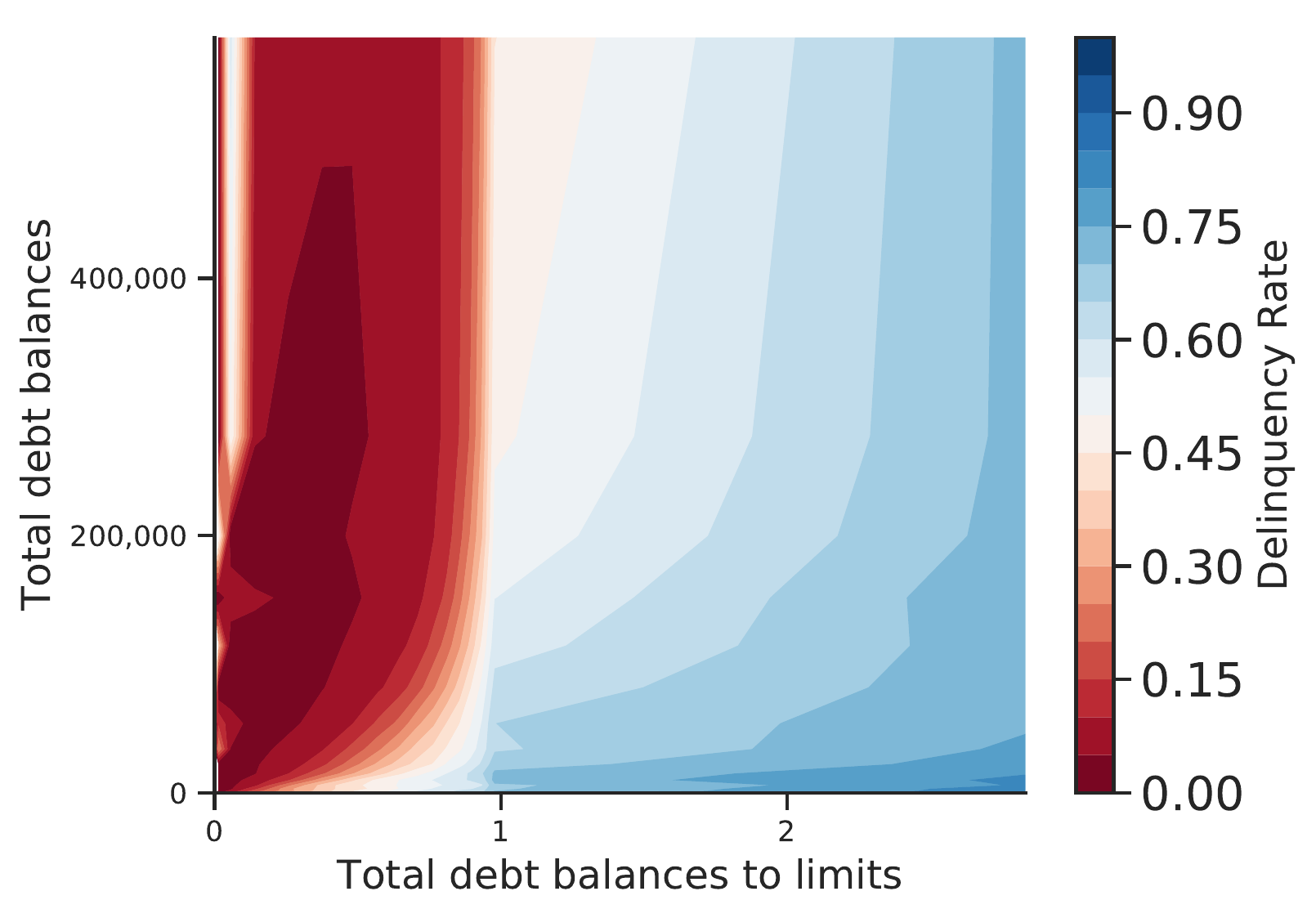}}
\subfloat[]{\includegraphics[scale=0.5]{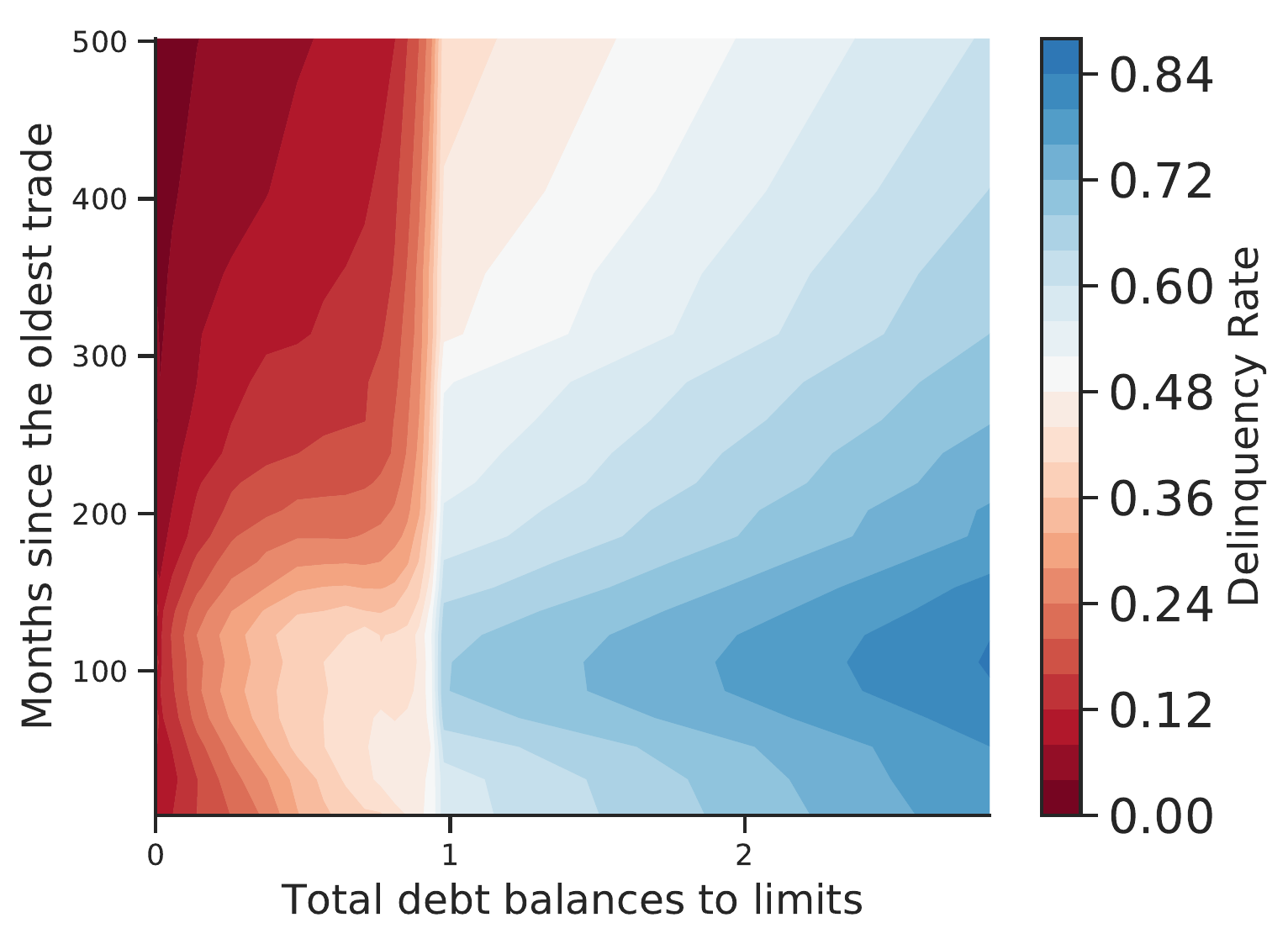}}

\subfloat[]{\includegraphics[scale=0.5]{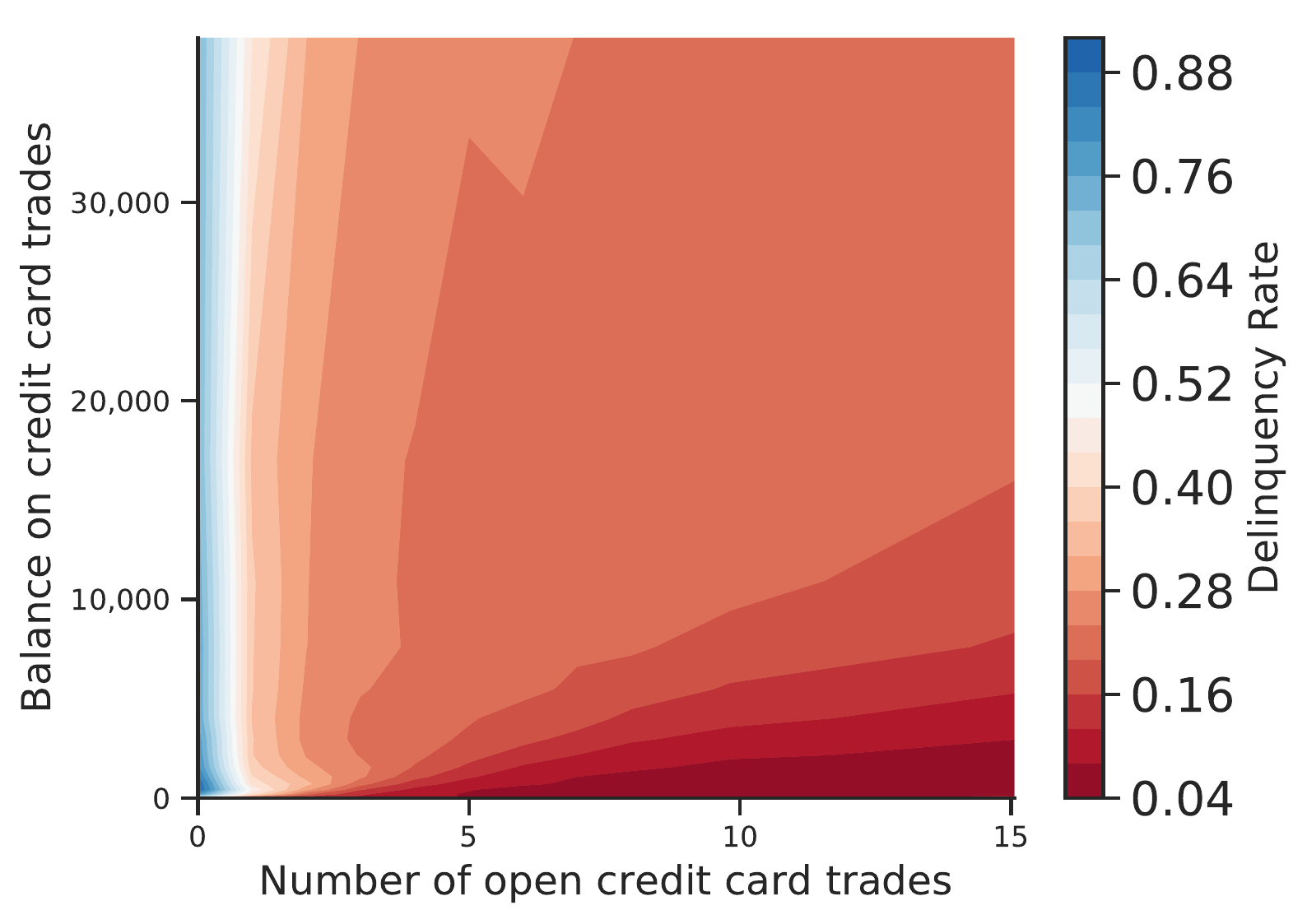}}
\subfloat[]{\includegraphics[scale=0.5]{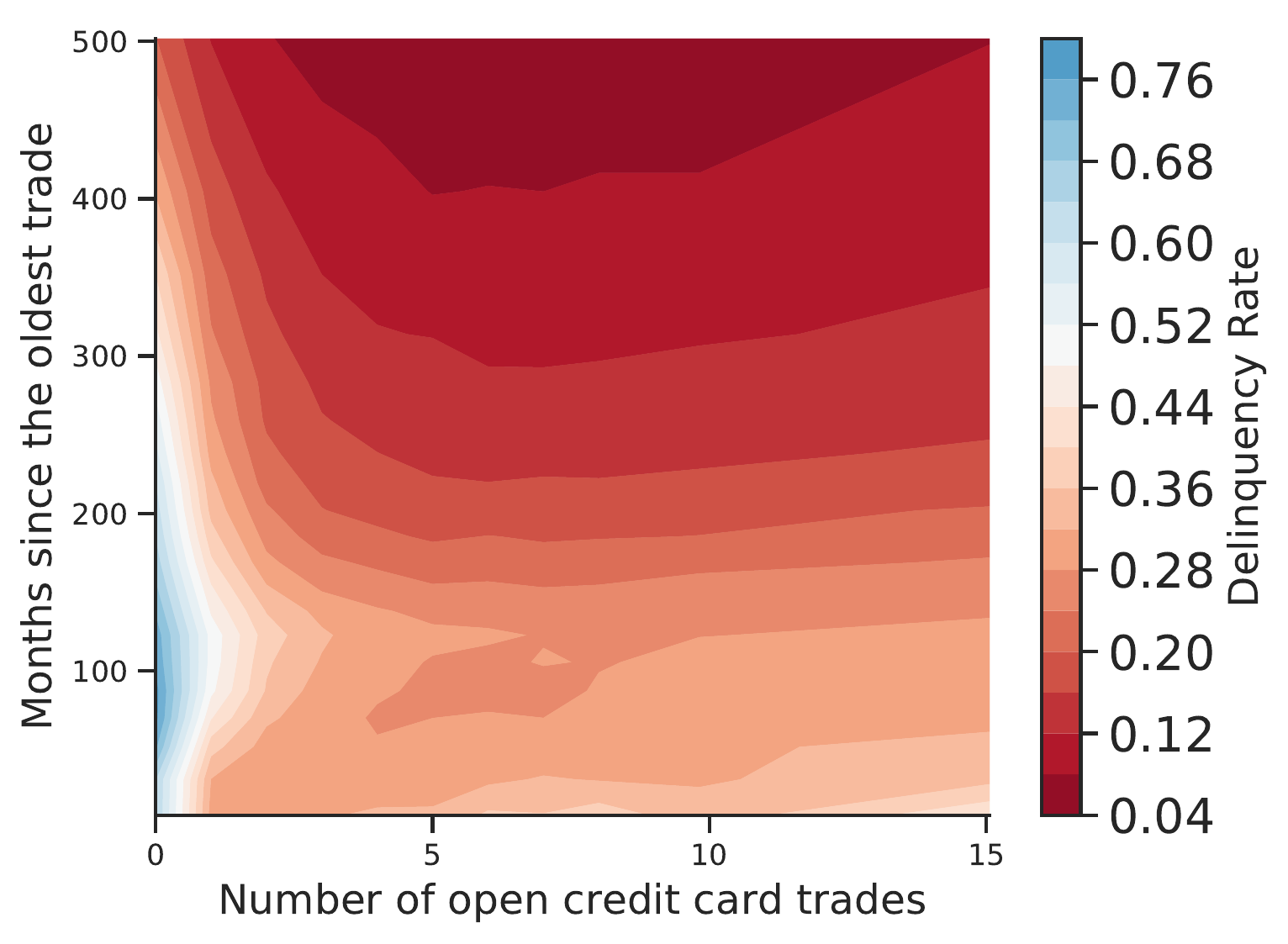}}

\end{center}

\msk
\footnotesize{Relationship between 90+ days past due delinquency rate and pairs of covariates. 
Source: Authors' calculations based on Experian Data.}

\end{figure}

Multidimensional interactions are another feature of the relation between default and plausible covariates, that is default behavior is simultaneously related with multiple variables. To see this, Figure \ref{fig:multi_dim} presents contour plots of the relation between the incidence of default and couples of covariates. The covariates reported here are chosen since they are important driving factors in  default decisions, based on our model, as discussed in  Section \ref{sec:interpretation}. Panels (a) and (b) explore the joint variation in the incidence of default with total debt balances, credit utilization (total debt balances to limits), and credit history. Blue values correspond to high delinquency rates while red values to low delinquency rates. As can be seen from both panels, higher credit utilization corresponds to higher delinquency rate, but  for given credit utilization, an increase in total debt balances first decreases then increases the delinquency rate, where the switch in sign depends on the utilization rate. For given utilization rates, a longer credit history first increases then decreases the delinquency rate, provided the utilization rate is smaller than 1.\footnote{Utilization rates above 1 can arise for a delinquent borrower if fees and other penalty add to their balances for given credit limits.} Panels (c) and (d) explore the relation between default and credit card borrowing. Default rates decline with the number of credit cards, though for a given number of credit card trades, they mostly increase with credit card balances. This relation, however varies  with the level of both variables. An increase in the length of credit history is typically associated with lower default rates, however, if the number of open credit cards is low, this relation is non-monotone. The variables reported in the figures are illustrative of a general pattern in the joint relation between couples of covariates and default rates.      

This pattern of multidimensional non-linear interactions across covariates is fairly difficult to model using standard econometric approaches. For this reason, we propose a deep learning approach to be explained below.

\section{Model}\label{sec:model}
Predicting consumer default maps well into a supervised learning framework, which is one of the most widely used techniques in the machine learning literature. In supervised learning, a learner takes in pairs of input/output data. The input data, which is typically a vector, represent pre-identified attributes, also known as features, that are used to determine the output value. Depending on the learning algorithm, the input data can contain continuous and/or discrete values with or without missing data. The supervised learning problem is referred to as a "regression problem" when the output is continuous, and as a "classification problem" when the output is discrete. Once the learner is presented with input/output data, its task is to find a function that maps the input vectors to the output values. A brute force way of solving this task is to memorize all previous values of input/output pairs. Though this perfectly maps the input data to the output values in the training data set, it is unlikely to succeed in forecasting the output values if (1) the input values are different from the ones in the training data set or (2) when the training data set contains noise. Consequently, the goal of supervised learning is to find a function that generalizes beyond the training set, so that it correctly forecasts out-of-sample outcomes. Adopting this machine-learning methodology, we build a  model that predicts defaults for individual consumers. We define default as a 90+ days delinquency on any debt in the subsequent 8 quarters, which is the outcome targeted by conventional credit scoring models. Our model outputs a continuous variable between 0 and 1 that can be interpreted under certain conditions as an estimate of the probability of default for a particular borrower at a given point in time, given input variables from their credit reports.      

We start by formalizing our prediction problem. We adopt a discrete-time formulation for periods 0,1,...,T, each corresponding to a quarter. We let the variable $D_t^i$ prescribe the state at time $t$ for individual $i$ with $D \subset \mathbb{N}$ denoting the set of states. We define $D_1^i=1$ if a consumer is 90+ days past due on any trade and $D_1^i=0$ otherwise. Consumers will transition between these two states over their lifetime.      

Our target outcome is 90+ days past due in the subsequent 8 quarters, defined as:
\begin{equation}
Y_t^i=\left\{ \begin{array}{cl}
0 & \textrm{if }\sum_{n=t}^{t+7} D_n^i = 0\\
1 & \textrm{otherwise}\\
\end{array}\right.
\end{equation}
We allow the dynamics of the state process to be influenced by a vector of explanatory variables $X_{t-1}^i$ $\in \mathbb{R}^{d_X}$, which includes the  state $D_{t-1}^i$. In our empirical implementation, $X_{t-1}^i$ represents the features in Table \ref{tab:features}.  
We fix a probability space $(\Omega,\mathcal{F},\mathbb{P})$ and an information filtration $(\mathcal{F}_t)_{(t=0,1,...,T)}$. Then, we specify a probability transition function $h_\theta : \mathbb{R}^{d_X} \rightarrow [0,1]$ satisfying
\begin{equation}\label{eq:trans}
    \mathbb{P}[Y_t^i = y| \mathcal{F}_{t-1}] = h_\theta(X_{t-1}^i), y \in D 
\end{equation}
where $\theta$ is a parameter to be estimated. Equation \ref{eq:trans} gives the marginal conditional probability for the transition of individual $i$'s debt from its state $D_{t-1}^i$ at time $t-1$ to state $y$ at time $t$ given the explanatory variables $X_{t-1}^i$.\footnote{The state $ y$ encompasses realizations of the state between time $t$ and $t+7$.} Let $g$ denote the standard \emph{softmax} function: 
\begin{equation}
g(z) =  \Bigg(\displaystyle \dfrac{1}{1+e^{-z}}\Bigg), z \in \mathbb{R}^K,
\end{equation}
where $K = |D|$. The vector output of the function $g$ is a probability distribution on $D$.      

The marginal probability defined in equation \ref{eq:trans}  is the theoretical  counterpart of the empirical transition matrix reported in Table \ref{tab:transition}. We propose to model the transition function $h_\theta$ with a hybrid deep neural network/gradient boosting  model, which combines the predictions of a deep neural network and an extreme gradient boosting model. We explain each of the component models and their properties and the rationale for combining them below.

\subsection{Deep Neural Network}
One component of our model is based on deep learning, in the class used by \citeN{sirignano}. We restrict attention to feed-forward neural networks, composed of an input layer, which corresponds to the data, one or more interacting hidden layers that non-linearly transform the data, and an output layer that aggregates the hidden layers into a prediction. Layers of the networks consist of neurons with each layer connected by synapses that transmit signals among neurons of subsequent layers. A neural network is in essence a sequence of nonlinear relationships. Each layer in the network takes the output from the previous layer and applies a linear transformation followed by an element-wise non-linear transformation.

\begin{figure}[htbp]
\begin{center}
\includegraphics[scale=0.33]{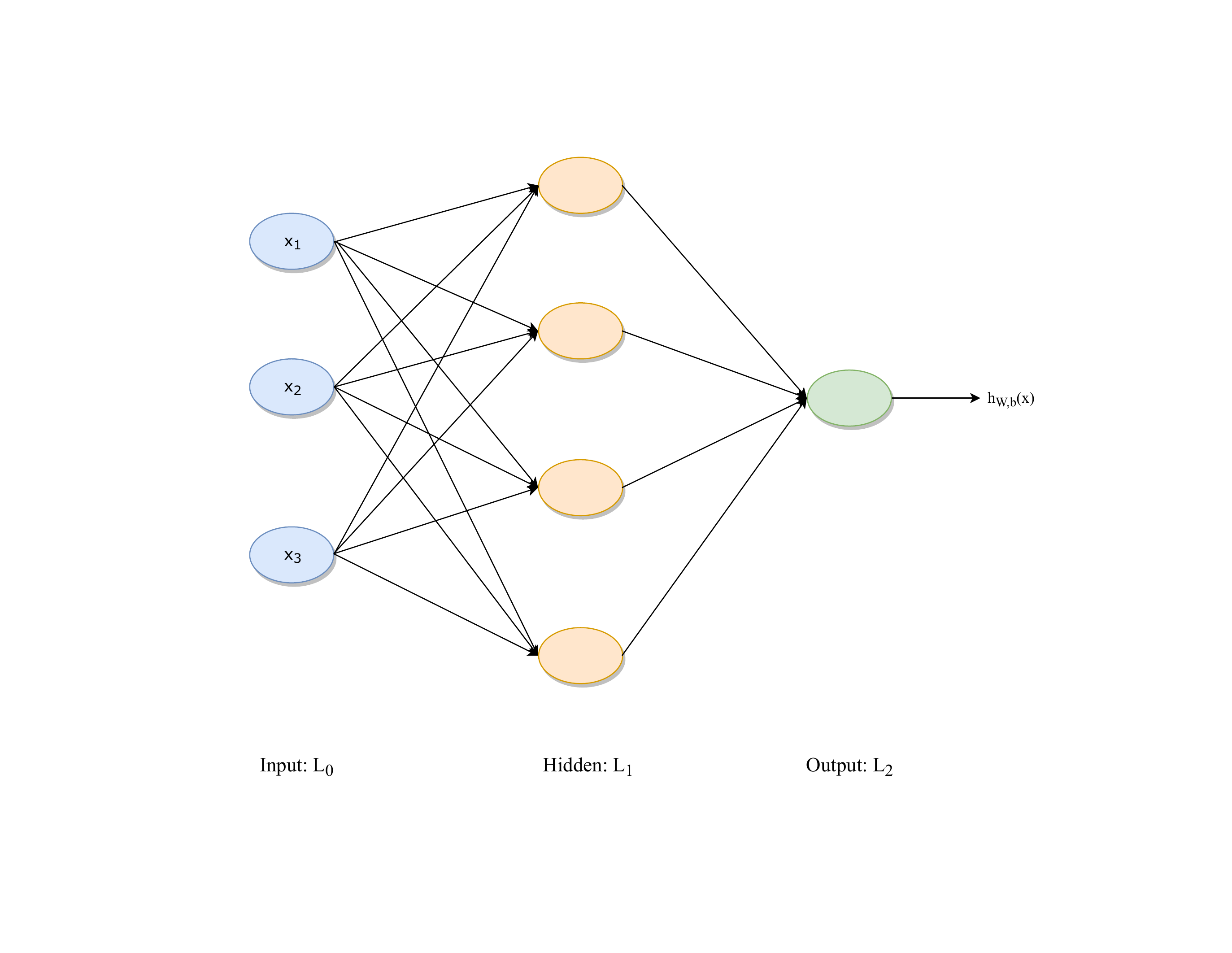}
\caption{Two Layer Neural Network Example}\label{fig:nn}
\end{center}
\end{figure}

Figure \ref{fig:nn} illustrates an example of a two layer neural network.   
This neural network has 3 input units (denoted $x_1, x_2, x_3$), 4 hidden units, and 1 output unit. Let $n_l$ denote the number of layers in this network ($n_l=2$). We label layer $l$ as $L_l$, where layer $L_0$ is the input layer, and layer $L_{L=2}$ is the output layer. The layers between the input ($l = 0$) and the output layer ($l = L$) are called hidden layers. Given this notation, there are $L - 1$ hidden layers, 1 in this specific example.  A neural network without any hidden layers ($L = 1$) is a logistic regression model. 

There are two ways to increase the complexity a neural network: (1)  increase the number of hidden layers and (2) increase the number of units in a given layer. Lower tier layers in the neural network learn simpler patterns, from which higher tier layers learn to produce more complex patterns. Given a sufficient number of  neurons, neural networks can approximate continuous functions on compact sets arbitrarily well (see \citeN{hornik1989} and \citeN{hornik1991}). This includes approximating interactions (i.e., the product and division of features). There are two main advantages of adding more layers over increasing the number of units to existing layers; (1) later layers build on early layers to learn features of greater complexity and (2) deep neural networks-- those  with three or more hidden layers--  need exponentially fewer neurons than shallow networks (\citeN{bengiolecun} and \citeN{montufar}).      

In the neural network represented in Figure \ref{fig:nn}, the parameters  to be estimated are $(W,b) = (W^{(0)}, b^{(0)}, W^{(1)}, b^{(1)})$, where $W_{ij}^{(l)}$ denotes the weight associated with the connection between unit $j$ in layer $l$ and unit $i$ in layer $l+1$, and $b_i^{(l)}$ is the bias associated with unit $i$ in layer $l+1$. Thus, in this example $W^{(0)} \in \mathbb{R}^{3 \times 4}, b^{(0)} \in \mathbb{R}^{4 \times 1} \textrm{ and } W^{(1)} \in \mathbb{R}^{1 \times 4}, b^{(1)} \in \mathbb{R}$. This implies that there are a total of 21 = (3+1)*4+5 parameters (four parameters to reach each neuron and five weights to aggregate the neurons into a single output). In general, the number of weight parameters in each hidden layer $l$ is $N^{(l)}(1+N^{(l-1)})$, plus $1+N^{(L-1)}$ for the output layer, where $N^{(l)}$ denotes the number of neurons in each layer $l = 1$,\ldots, $L$.       

Let $a_i^{(l)}$ denote the activation (e.g., output value) of unit $i$ in layer $l$. Fix $W$ and $b$, our neural network defines a hypothesis $h_{W,b}(x)$ that outputs a real number between 0 and 1.\footnote{This is a property of the sigmoid activation function.} Let $f(\cdot)$ denote the activation function that applies to vectors in an element-wise fashion. The computation this neural network represents, often referred to as forward propagation, can be written as:
$$ z^{(1)} = W^{(0), T}x + b^{(0)}$$
$$ a^{(1)} = f(z^{(1)}) $$
$$ z^{(2)} = W^{(1), T}a^{(1)} + b^{(1)}$$
$$ h_{W,b}(x) = a^{(2)} = f(z^{(2)}) $$

There are many choices to make when structuring a neural network, including the number of hidden layers, the number of neurons in each layer, and the activation functions. We built a number of network architectures having up to fifteen hidden layers.\footnote{The number of layers and the number of neurons in each layer, along with other hyperparameters of the model, are chosen by Tree-structured Parzen Estimator (TPE) approach. See Appendix \ref{app:estimation}  for more details.} All architectures are fully connected so each unit receives an input from all units in the previous layer.       

Neural networks tend to be low-bias, high-variance models, which imparts them a tendency to over-fit the data. We apply dropout to each of the layers to avoid over-fitting (see \citeN{srivastava}). During training, neurons are randomly dropped (along with their connections) from the neural network with probability $p$ (referred to as the dropout rate), which prevents complex co-adaptations on training data.      

We apply the same activation function (rectified linear unit or RELU) at all nodes, which is obtained via hyperparameter optimization,\footnote{There are many potential choices for the nonlinear activation function, including the sigmoid, relu, and tanh.} and defined as:
\begin{equation}
\textrm{RELU}(x)= \left\{ \begin{array}{cl}
x & \textrm{if }x \ge 0 \\
0 & \textrm{otherwise}\\
\end{array}\right.
\end{equation}

Let $N^{(l)}$ denote the number of neurons in each layer $l = 1$,\ldots, $L$. Define the output of neuron $k$ in layer $l$ as $z_k^{(l)}$. Then, define the vector of outputs (including the bias term $z_0^{(l)}$) for this layer as $z^{(l)} = (z_0^{(l)},z_1^{(l)},\ldots,z^{(l)}_{N^{(l)}})'$. For the input layer, define $z^{(0)} = (x_0^{(l)},x_1^{(l)},\ldots,x^{(l)}_{N^{(l)}})'$. Formally, the recursive output of the $l-th$ layer of the neural network is:
\begin{equation}
    z^{(l)} = \textrm{RELU} ( W^{(l-1), T}z^{(l-1)} +  b^{(l-1)} ),
\end{equation}
with final output:
\begin{equation}
    h_\theta(x) = g( W^{(L-1), T}z^{(L-1)} +  b^{(L-1)} ).
\end{equation}
The parameter specifying the neural network is:
\begin{equation}
    \theta = (W_0,b_0,\ldots,W_{L-1},b_{L-1})
\end{equation}

\subsection{Decision Tree Models}\label{sec:tree}
The second component of our model is Extreme Gradient Boosting, which builds on decision tree models. Tree-based models split the data several times based on certain cutoff values in the explanatory variables.\footnote{Splitting means that different subsets of the dataset are created, where each observation belongs to one subset. For a review on decision trees, see \citeN{khandani}.} A number of such models have become quite prevalent in the literature, most notably random forests (see \citeN{breiman} and \citeN{butaru}) and Classification and Regression Trees, known as CART. We briefly review CART and then explain gradient boosting.
 \subsubsection{CART}
There are a number of different decision tree-based algorithms. As an illustration of the approach, we describe Classification and Regression Trees or CART.
CART models an outcome $y_i$ for an instance $i$ as follows:
\begin{equation}
    \hat{y_i} = \hat{f}(x_i) = \sum_{m=1}^{M}c_m I\{x_i \in R_m\},
\end{equation}
where each  observation $x_i$ belongs to exactly one subset $R_m$. The identity function $I$ returns 1 if $x_i$ is in $R_m$ and 0 otherwise. If $x_i$ falls into $R_l$, the predicted outcome is $\hat{y} = c_l$, where $c_l$ is the mean of all training observations in $R_l$. 

The estimation procedure takes a feature and computes the cut-off point that minimizes the Gini index of the class distribution of $\mathbf{y}$, which makes the two resulting subsets as different as possible. Once this is done for each feature, the algorithm uses the best feature to split the data into two subsets. The algorithm is then repeated until a stopping criterium is reached.

Tree-based models have a number of advantages that make them popular in applications. They are invariant to monotonic feature transformations and can handle categorical and continuous data in the same model. Like deep neural networks, they are well suited to capturing  interactions between variables in the data. Specifically, a tree of depth $L$ can capture $(L-1)$ interactions. The interpretation is straightforward, and provides immediate counterfactuals: "If feature $x_j$ had been bigger / smaller than the split point, the prediction would have been $\bar{y}_0$ instead of $\bar{y}_1$."
However, these models also have a number of limitations.  They are poor at handling linear relationships, since tree algorithms rely on splitting the data using step functions, an intrinsically non-linear transformation. Trees also tend to be unstable, so that small changes in the training dataset might generate a different tree. They are also prone to overfitting to the training data. For more information on tree-based models see \citeN{molnar}.

\subsubsection{eXtreme Gradient Boosting (XGBoost)}
Gradient Boosted Trees (GBT) are an ensemble learning method that corrects for  tree-based models' tendency to overfit to training data by recursively combining the forecasts of many over-simplified trees. Though shallow trees are "weak learners" on their own with little predictive power, the theory behind boosting proposes that a collection of weak learners, as an ensemble, creates a single strong learner with improved stability over a single complex tree.

At each step m, $1 \le m \le M$, of gradient boosting, an estimator, $h_m$, is computed on the residuals from the previous models predictions. A critical part of gradient boosting method is regularization by shrinkage as proposed by \citeN{friedman}. This consists in modifying the update rule as follows:
\begin{equation}
    F_m(x) = F_{m-1}(x) + \nu \gamma_m h_m(x),
\end{equation}
where $h_m(x)$ represents a weak learner of fixed depth, $\gamma_m$ is the step length and $\nu$ is the learning rate or shrinkage factor. \newline 

XGBoost is a fast implementation of Gradient Boosting, which has the advantages of fast speed and high accuracy. For classification, XGBoost combines the principles of decision trees and logistic regression, so that the output of our XGBoost model is a number between 0 and 1. For the remainder of the paper we refer to XGBoost as GBT.\footnote{For more on XGBoost, see \citeN{chen2018hybrid} and \citeN{ren2017novel}.}  

\subsection{Hybrid DNN-GBT Model}\label{subsec:hybrid}
We examined two techniques to create a hybrid DNN-GBT ensemble model. Ensemble models combine multiple learning algorithms to generate superior predictive performance than could be obtained from any of the constituent learning algorithms alone. The first method combines the two models by replacing the final layer of the neural network with a gradient boosted trees model. Examples of this approach are \citeN{chen2018hybrid} and \citeN{ren2017novel}. The second, uses both models separately and then averages out the final predicted probabilities of the two models. We found the latter to perform better on our dataset. This method is similar to \citeN{kvamme2018}, who combined a convolutional neural network with a random forest by averaging. Thus, our methodology relies on combining the output of the deep neural network with the output of a gradient boosted trees model. This is achieved in two steps:
\begin{enumerate}
    \item For each observation, run DNN and GBT separately and obtain predicted probabilities for each of the models;
    \item Take the arithmetic mean of the predicted probabilities.\footnote{We have investigated alternative weighting schemes, and the results are reported in Table \ref{tab:weighting}.}
\end{enumerate}

\section{Implementation}\label{sec:results}
Table \ref{tab:features} lists the features from the credit report data we use as inputs in the model. These covariates are chosen based on economic theory (see for example \citeN{chatterjee2007quantitative}) as well as based on information from currently used credit scoring models. They include information on balances and credit limits for different types of consumer debt, severity and number of delinquencies, credit utilization by type of product, public record items such as bankruptcy filings by chapter and foreclosure, collection items, and length of the credit history. In order to be consistent with the restrictions of the Fair Credit Reporting Act on 1970 and the Equal Opportunity in Credit Access Act of 1984 we do not include information on age or zip code, and we do not include any information on income, to be consistent with current credit scoring models. Table \ref{tab:features} lists the full set of features used in our machine learning models. 

It is important to note that we do not use any lagged features. This is because many of the features have a temporal dimension, for example "worst present status on any traded in the last 6 months." Importantly, excluding lags enables us to provide a default prediction to any borrower with a non-empty credit record, which implies that we can score virtually all consumers.

\subsection{Classifier Performance}\label{sec:performance}
In this section, we describe the performance of our hybrid model under various training and testing windows. First, we evaluate our model on the pooled sample (2004Q1-2013Q4), where we apply a random 60\%-20\%-20\% split to our training, validation, and testing sets. Then, to account for look-ahead bias, we train and test our models based on 8 quarter windows that were observable at the time of forecast. In particular, we require our training and testing sets to be separated by 8 quarters to avoid overlap. For instance, the second out-of-sample model was calibrated using input data from 2004Q2, from which the parameter estimates were applied to the input data in 2006Q2 to generate forecasts of delinquencies over the 8 quarter window from 2006Q3-2008Q2. This gives us a total of 32+1 calibration and testing periods reported in Table \ref{tab3}.  The percentage of 90+ days past due  accounts within 8 quarters varies from 32.5\% to 35.9\%.     

The hybrid model outputs a continuous variable that, under certain circumstances, can be interpreted as an estimate of the probability of an account becoming 90+ days delinquent during the subsequent 8 quarters. One measure of the model's success is its ability to differentiate between accounts that did become delinquent and those that did not; if these two groups have the same forecasts, the model provides no value. Table \ref{tab:full_sample} presents the average forecast for accounts that did and did not fall into the 90+ days delinquency category over the 32+1 evaluation periods. For instance, during the testing period for 2010Q4, the model's average prediction among the 35.44\% of accounts that became 90+ days delinquent was 73.17\%, while the average prediction among the 64.56\% of accounts that did not was 16.03\%. We should highlight that these are truly out-of-sample predictions, since the model is calibrated using input data from 2008Q4. This shows the forecasting power of our model in distinguishing between accounts that will and will not become delinquent within 8 quarters. Furthermore, this forecasting power seems to be stable over the 32+1 calibration and evaluation periods, partly driven by the frequent re-calibration of the model that captures some of the changing dynamics of consumer behavior.     

\begin{table}[!h]
\caption{1 Quarter Ahead Predictions, Full Sample-- Hybrid DNN-GBT }\label{tab:full_sample}
\begin{center}
\footnotesize
\label{tab3}
\begin{tabular}{cccccc}
Training Window & Testing Window & Data  & Predicted & Delinquents & Non-Delinquents \\  \hline 
 &  &  &  &  &    \\
2004Q1-2013Q4 & 2004Q1-2013Q4 & 0.3396 & 0.3354 & 0.7516 & 0.1213 \\
 &  &  &  &  &  \\
2004Q1 & 2006Q1 & 0.3248 & 0.2919 & 0.6508 & 0.1192 \\
2004Q2 & 2006Q2 & 0.3274 & 0.3042 & 0.6732 & 0.1246 \\
2004Q3 & 2006Q3 & 0.3306 & 0.3102 & 0.6838 & 0.1256 \\
2004Q4 & 2006Q4 & 0.3347 & 0.3128 & 0.6843 & 0.1260 \\
2005Q1 & 2007Q1 & 0.3410 & 0.3160 & 0.6851 & 0.1251 \\
2005Q2 & 2007Q2 & 0.3444 & 0.3196 & 0.6861 & 0.1271 \\
2005Q3 & 2007Q3 & 0.3469 & 0.3201 & 0.6847 & 0.1265 \\
2005Q4 & 2007Q4 & 0.3505 & 0.3307 & 0.6972 & 0.1329 \\
2006Q1 & 2008Q1 & 0.3535 & 0.3370 & 0.7090 & 0.1335 \\
2006Q2 & 2008Q2 & 0.3545 & 0.3340 & 0.6982 & 0.1341 \\
2006Q3 & 2008Q3 & 0.3558 & 0.3338 & 0.7019 & 0.1305 \\
2006Q4 & 2008Q4 & 0.3587 & 0.3429 & 0.7121 & 0.1364 \\
2007Q1 & 2009Q1 & 0.3588 & 0.3483 & 0.7223 & 0.1391 \\
2007Q2 & 2009Q2 & 0.3580 & 0.3507 & 0.7259 & 0.1415 \\
2007Q3 & 2009Q3 & 0.3573 & 0.3525 & 0.7279 & 0.1437 \\
2007Q4 & 2009Q4 & 0.3589 & 0.3540 & 0.7277 & 0.1448 \\
2008Q1 & 2010Q1 & 0.3589 & 0.3612 & 0.7359 & 0.1514 \\
2008Q2 & 2010Q2 & 0.3568 & 0.3630 & 0.7366 & 0.1558 \\
2008Q3 & 2010Q3 & 0.3559 & 0.3635 & 0.7365 & 0.1574 \\
2008Q4 & 2010Q4 & 0.3544 & 0.3628 & 0.7317 & 0.1603 \\
2009Q1 & 2011Q1 & 0.3541 & 0.3577 & 0.7282 & 0.1545 \\
2009Q2 & 2011Q2 & 0.3511 & 0.3591 & 0.7265 & 0.1603 \\
2009Q3 & 2011Q3 & 0.3500 & 0.3555 & 0.7248 & 0.1566 \\
2009Q4 & 2011Q4 & 0.3484 & 0.3538 & 0.7242 & 0.1558 \\
2010Q1 & 2012Q1 & 0.3467 & 0.3559 & 0.7331 & 0.1557 \\
2010Q2 & 2012Q2 & 0.3434 & 0.3498 & 0.7264 & 0.1528 \\
2010Q3 & 2012Q3 & 0.3396 & 0.3498 & 0.7295 & 0.1546 \\
2010Q4 & 2012Q4 & 0.3358 & 0.3488 & 0.7326 & 0.1547 \\
2011Q1 & 2013Q1 & 0.3341 & 0.3481 & 0.7350 & 0.1540 \\
2011Q2 & 2013Q2 & 0.3317 & 0.3440 & 0.7305 & 0.1522 \\
2011Q3 & 2013Q3 & 0.3298 & 0.3440 & 0.7342 & 0.1520 \\
2011Q4 & 2013Q4 & 0.3275 & 0.3400 & 0.7299 & 0.1501 \\ \hline 
\end{tabular}
\end{center}

\msk
\footnotesize{Performance metrics for our model of default risk over 32+1 testing windows. For each testing window, the model is calibrated on data over the period specified in the training window, and predictions are based on the data available as of the data in the training window. For example, the fourth row reports the performance of the model calibrated using input data available in 2004Q3, and applied to 2006Q3 data to generate forecasts of delinquencies for within 8 quarter delinquencies. Average model forecasts over all customers, and customers that (ex-post) did and did not become 90+ days delinquent over the testing window are also reported. Source: Authors' calculations based on Experian Data.}
\end{table}

We also look at accounts that are current as of the forecast date but become 90+ days delinquent within the subsequent 8 quarters. In particular, we contrast the model's average prediction among individuals who were current on their accounts but became 90+ days delinquent with the average prediction among customers who were current and did not become delinquent. Given the difficulty of predicting default among individuals that currently show no sign of delinquency, we anticipate the model's performance to be less impressive than the values reported in Table \ref{tab:full_sample}. Nonetheless, the values reported in Table \ref{tab:current} indicate that the model is able to distinguish between these two populations. For instance, using input data from 2008Q4, the average model prediction for individuals who were current on their debts and became 90+ days delinquent is 45.13\%, contrasted with 12.48\% for those who did not. As in Table \ref{tab:full_sample}, the model's ability to distinguish between these two classes is consistent across the 32+1 evaluation periods listed in Table \ref{tab:current}.   

\begin{table}[!h]
\caption{1 Quarter Ahead Predictions, Current-- Hybrid DNN-GBT }\label{tab:current}
\begin{center}
\footnotesize
\label{tab4}
\begin{tabular}{cccccc}
Training Window & Testing Window & Data & Predicted & Delinquent & Non-delinquent \\ \hline 
 &  &  &  &  &    \\
2004Q1-2013Q4 & 2004Q1-2013Q4 & 0.1676 & 0.1616 & 0.5253 & 0.088 \\
 &  &  &  &  &  \\
2004Q1 & 2006Q1 & 0.1844 & 0.1540 & 0.4232 & 0.0931 \\
2004Q2 & 2006Q2 & 0.1702 & 0.1467 & 0.3970 & 0.0954 \\
2004Q3 & 2006Q3 & 0.1695 & 0.1467 & 0.3974 & 0.0956 \\
2004Q4 & 2006Q4 & 0.1727 & 0.1473 & 0.3995 & 0.0947 \\
2005Q1 & 2007Q1 & 0.1805 & 0.1515 & 0.4012 & 0.0964 \\
2005Q2 & 2007Q2 & 0.1813 & 0.1521 & 0.3948 & 0.0983 \\
2005Q3 & 2007Q3 & 0.1831 & 0.1502 & 0.3873 & 0.0971 \\
2005Q4 & 2007Q4 & 0.1847 & 0.1567 & 0.4031 & 0.1008 \\
2006Q1 & 2008Q1 & 0.1890 & 0.1628 & 0.4177 & 0.1033 \\
2006Q2 & 2008Q2 & 0.1896 & 0.1619 & 0.4077 & 0.1044 \\
2006Q3 & 2008Q3 & 0.1872 & 0.1558 & 0.3979 & 0.1000 \\
2006Q4 & 2008Q4 & 0.1817 & 0.1588 & 0.4043 & 0.1043 \\
2007Q1 & 2009Q1 & 0.1781 & 0.1618 & 0.4167 & 0.1066 \\
2007Q2 & 2009Q2 & 0.1752 & 0.1638 & 0.4223 & 0.1089 \\
2007Q3 & 2009Q3 & 0.1713 & 0.1660 & 0.4290 & 0.1116 \\
2007Q4 & 2009Q4 & 0.1661 & 0.1627 & 0.4170 & 0.1120 \\
2008Q1 & 2010Q1 & 0.1683 & 0.1717 & 0.4396 & 0.1175 \\
2008Q2 & 2010Q2 & 0.1668 & 0.1772 & 0.4519 & 0.1221 \\
2008Q3 & 2010Q3 & 0.1661 & 0.1793 & 0.4580 & 0.1238 \\
2008Q4 & 2010Q4 & 0.1644 & 0.1785 & 0.4513 & 0.1248 \\
2009Q1 & 2011Q1 & 0.1674 & 0.1764 & 0.4529 & 0.1208 \\
2009Q2 & 2011Q2 & 0.1668 & 0.1805 & 0.4593 & 0.1247 \\
2009Q3 & 2011Q3 & 0.1669 & 0.1769 & 0.4555 & 0.1211 \\
2009Q4 & 2011Q4 & 0.1597 & 0.1716 & 0.4431 & 0.1200 \\
2010Q1 & 2012Q1 & 0.1604 & 0.1725 & 0.4500 & 0.1195 \\
2010Q2 & 2012Q2 & 0.1622 & 0.1694 & 0.4478 & 0.1155 \\
2010Q3 & 2012Q3 & 0.1598 & 0.1678 & 0.4450 & 0.1151 \\
2010Q4 & 2012Q4 & 0.1575 & 0.1667 & 0.4459 & 0.1145 \\
2011Q1 & 2013Q1 & 0.1606 & 0.1708 & 0.4601 & 0.1154 \\
2011Q2 & 2013Q2 & 0.1603 & 0.1695 & 0.4579 & 0.1144 \\
2011Q3 & 2013Q3 & 0.1578 & 0.1660 & 0.4525 & 0.1123 \\
2011Q4 & 2013Q4 & 0.1548 & 0.1622 & 0.4442 & 0.1106 \\ \hline 
\end{tabular}
\end{center}

\msk
\footnotesize{Performance metrics for our model of default risk over 32+1 testing windows for customers who are current as of the forecast date but become 90+ days delinquent in the following 8 quarters. For each testing window, the model is calibrated on data over the period specified in the training window columns, and predictions are based on the data available as of the data in the training window. For example, the fourth row reports the performance of the model calibrated using input data available in 2004Q3, and applied to 2006Q3 data to generate forecasts of delinquencies for within 8 quarter delinquencies. Average model forecasts over all current customers, and all current customers that did and did not become 90+ days delinquent over the testing window are also reported. Source: Authors' calculations based on Experian Data.}
\end{table}

Under certain conditions, the forecasts generated by our model can be converted to binary decisions by comparing the forecast to a specified threshold and classifying accounts with scores exceeding that threshold as high-risk. Setting the threshold level comes with a trade-off. A low level threshold leads to many accounts being classified as high risk, and even though this approach may accurately capture customers who are actually high-risk and about to default on their payments, it can also give rise to many low-risk accounts incorrectly classified as high-risk. By contrast, a high threshold can result in too many high-risk accounts being classified as low-risk.       

This type of trade-off is inherent in any classification problem, and involves trading off Type-I (false positives) and Type-II (false negatives) errors in a classical hypothesis testing context. In the  credit risk management context, a cost/benefit analysis can be formulated contrasting false positives to false negatives to make this trade-off explicit, and applying the threshold that will optimize an objective function in which costs and benefits associated with false positives and false negatives are inputs. 

A commonly used performance metric in the machine learning and statistics literature is a $2 \times 2$ contingency table, often referred to as the \emph{confusion matrix}, that describes the statistical behavior of any classification algorithm. In our application, the two rows correspond to ex post realizations of the two types of accounts in our sample, \emph{no default}  and \emph{default}. We define \emph{no default} accounts as those who do not become 90+ days delinquent during the forecast period, and \emph{default} accounts as those who do. The two columns correspond to ex ante classifications of the accounts into these categories. If a predictive model is applied to a set of accounts, each account falls into one of the four cells in the confusion matrix, thus the performance of the model can be assessed by the relative frequencies of the entries.  In the Neymann-Pearson hypothesis-testing framework, the lower-left entry is defined as Type-I error and the upper right as Type-II error, while the objective of the researcher is to minimize Type-II error (i.e., maximize "power") subject to a fixed level of Type-I error (i.e., "size").       

\begin{figure}[htbp]
\begin{center}
\bsk
\subfloat[Confusion Matrix]{\includegraphics[scale=0.55]{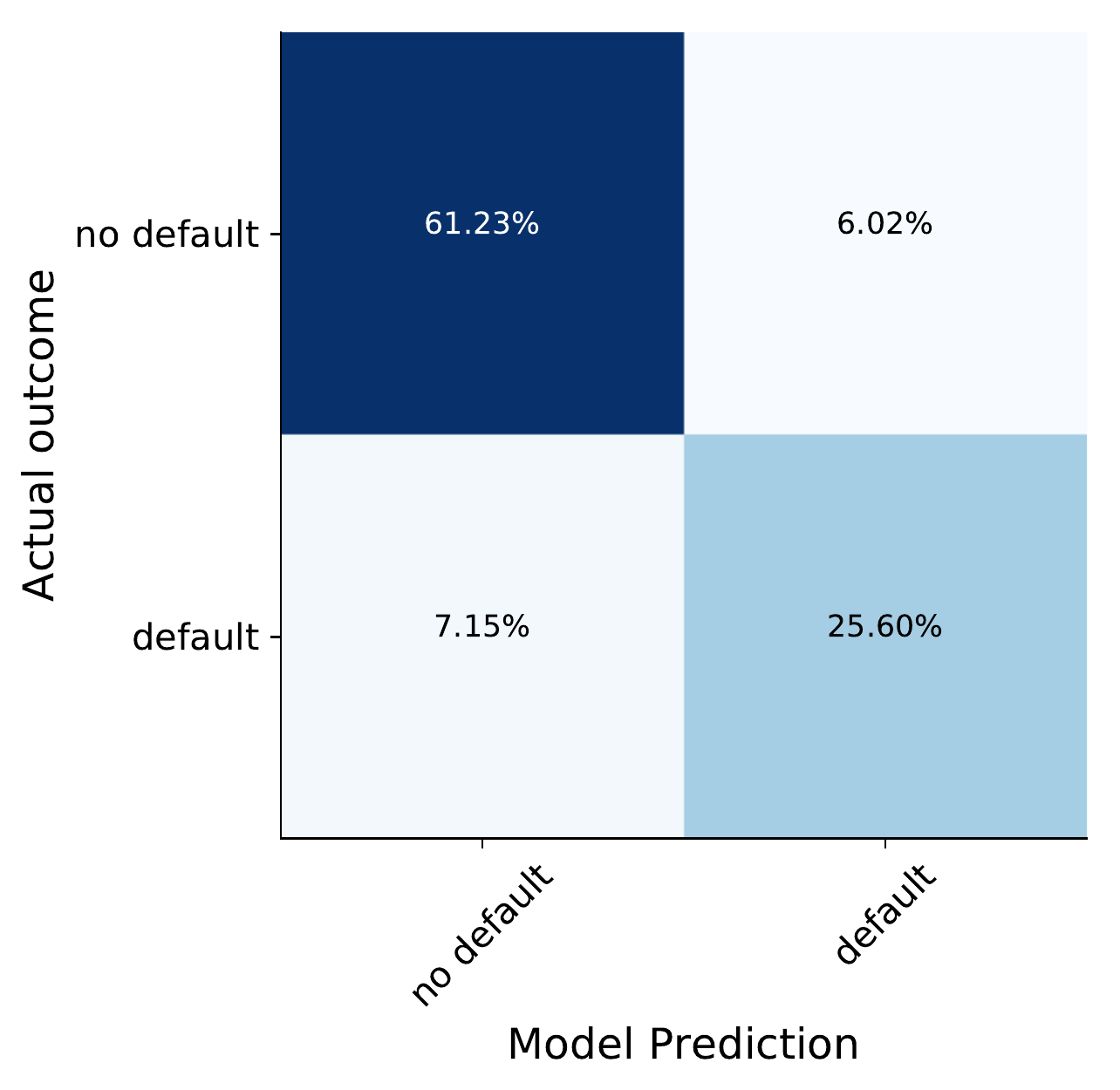}}
\subfloat[ROC Curve]{\includegraphics[scale=0.55]{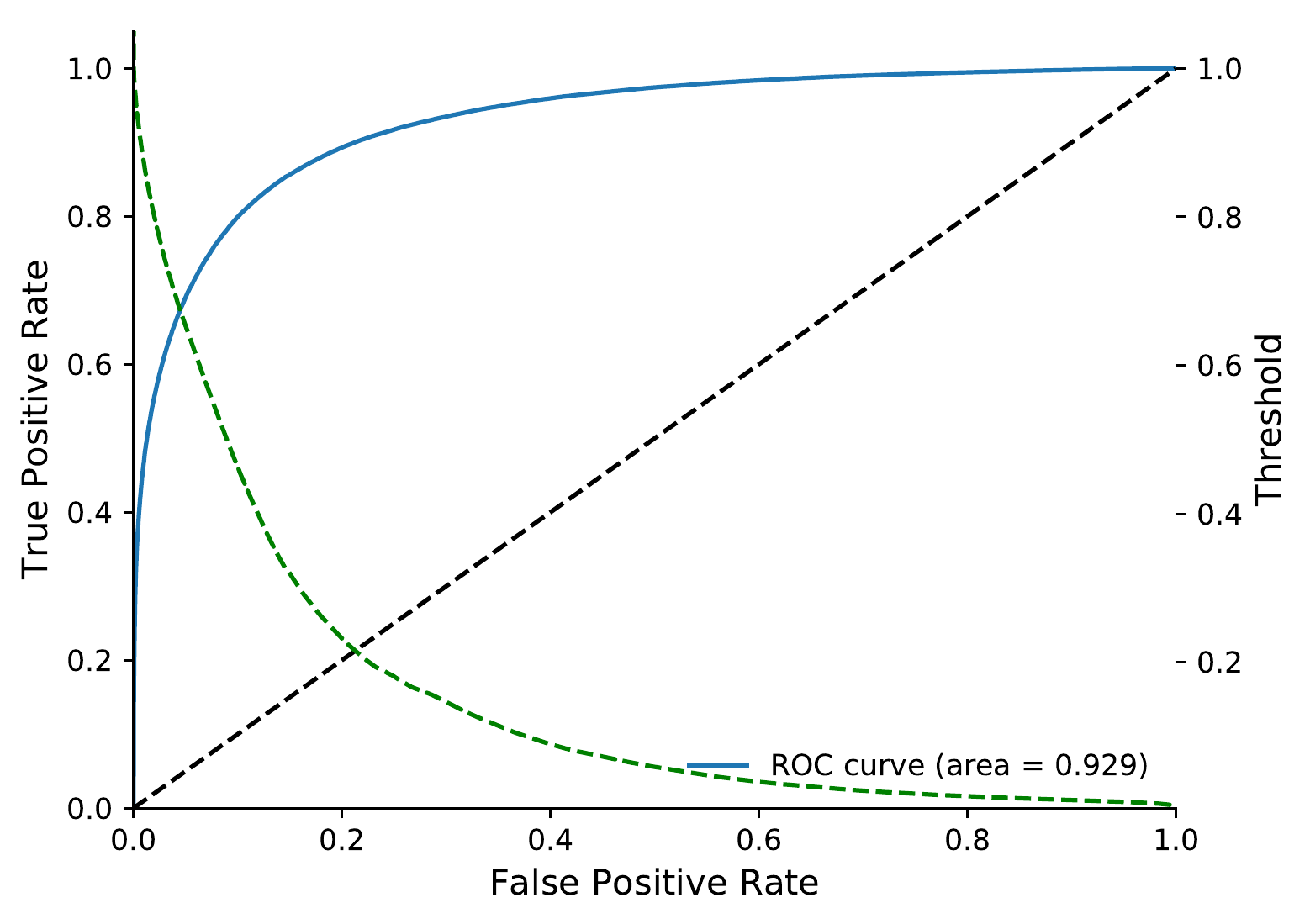}}
\end{center}
\caption{Confusion matrix and Receiver Operating Characteristic (ROC) curve of out-of-sample forecasts of 90+ days delinquencies over the 8Q forecast horizon based on our model of default risk. In Panel (a), rows correspond to actual states, with default defined as 90+ days delinquent, no default otherwise. Classifier threshold: 50\%. The numerical example is based on the model calibrated on 2011Q4 data and applied to 2013Q4 to generate out-of-sample predictions. Source: Authors' calculations based on Experian Data.}\label{fig:conf_roc}
\end{figure}

As an illustration, Figure \ref{fig:conf_roc} Panel (a) shows the confusion matrix for our hybrid DNN-GBT model calibrated using 2011Q4 data and evaluated on 2013Q4 data and a threshold of 50\%. This means that accounts with estimated delinquency probabilities greater than 50\% are classified as default and 50\% or below as no default. For this quarter, the model classified 61.23\% + 7.15\% = 68.38\% of the accounts as no default, of which 61.23\% did indeed not default and 7.15\% actually defaulted, that is, they were 90+ days delinquent in the subsequent 8 quarters. By the same token, of the 6.02\% + 25.60\% = 31.62\% borrowers who defaulted, the model accurately classified  25.60\%. Thus, the model's accuracy, defined as the percent of instances correctly classified, is the sum of the entries on the diagonal of the confusion matrix, that is, 61.23 \% + 25.60\% = 86.83\%.     


We can compute three additional performance metrics from the entries of the confusion matrix, which we describe heuristically here and define formally  in the appendix. \emph{Precision} measures the model's accuracy in instances that  are classified as default. \emph{Recall} refers to the number of accounts that defaulted as identified by the model divided by the actual number of defaulting accounts. Finally, the \emph{F-measure} is simply the harmonic mean of  precision and  recall. In an ideal scenario, we would have very high precision and recall. 
 
We can track the trade-off between true and false positives by varying the classification threshold of our model, and this trade-off is plotted in Figure \ref{fig:conf_roc} Panel (b). 
The blue line, called the Receiver Operating Characteristic (ROC) curve, is the pairwise plot of true and false positive rates for different classification thresholds (green line), and as the threshold decreases, the figure shows that the true positive rate increases, but so does the false positive rate. The ROC curve illustrates the non-linear nature of the trade-offs, implying that increase in true positive rates is not always proportionate with the increase in false positive rates. The optimal threshold then considers the cost of false positives with respect to the gain of true positives. If these are equal, the optimal threshold will correspond to the tangent point of the ROC curve with the 45 degree line. 

The last performance metric we consider is the area under the ROC curve, known as  AUC score, which is a widely used measure in the machine-learning literature for comparing models. It can be interpreted as the probability of the classifier assigning a higher  probability of being in default to an account that is actually in default. The ROC area of our model ranges from 0.9239 to 0.9305, demonstrating that our machine-learning classifiers have strong predictive power in separating the two classes. 

\begin{table}[!h]
\caption{Performance Metrics using Hybrid DNN-GBT, Full Sample}\label{tab:perf_metrics}
\footnotesize
\begin{center}

\begin{tabular}{cccccccc}
Training Window & Testing Window & AUC score & Precision & Recall & F-measure & Accuracy & Loss   \\ \hline 
& & & & & & & \\
2004Q1-2013Q4 & 2004Q1-2013Q4  &  0.9527 & 0.8662 & 0.8061 & 0.8351 & 0.8918 & 0.2609 \\
 &  &  &  &  &  &  &  \\
2004Q1 & 2006Q1 & 0.9244 & 0.8551 & 0.6988 & 0.7691 & 0.8637 & 0.3236 \\
2004Q2 & 2006Q2 & 0.9254 & 0.8488 & 0.7178 & 0.7779 & 0.8657 & 0.3181 \\
2004Q3 & 2006Q3 & 0.9262 & 0.8494 & 0.7253 & 0.7824 & 0.8667 & 0.3164 \\
2004Q4 & 2006Q4 & 0.9251 & 0.8499 & 0.7255 & 0.7828 & 0.8653 & 0.3203 \\
2005Q1 & 2007Q1 & 0.9257 & 0.8583 & 0.7209 & 0.7836 & 0.8642 & 0.3211 \\
2005Q2 & 2007Q2 & 0.9256 & 0.8595 & 0.7220 & 0.7848 & 0.8636 & 0.3221 \\
2005Q3 & 2007Q3 & 0.9249 & 0.8624 & 0.7169 & 0.7829 & 0.8621 & 0.3254 \\
2005Q4 & 2007Q4 & 0.9239 & 0.8558 & 0.7278 & 0.7866 & 0.8616 & 0.3273 \\
2006Q1 & 2008Q1 & 0.9252 & 0.8522 & 0.7371 & 0.7905 & 0.8619 & 0.3263 \\
2006Q2 & 2008Q2 & 0.9247 & 0.8570 & 0.7286 & 0.7876 & 0.8607 & 0.3272 \\
2006Q3 & 2008Q3 & 0.9255 & 0.8604 & 0.7270 & 0.7881 & 0.8609 & 0.3265 \\
2006Q4 & 2008Q4 & 0.9261 & 0.8564 & 0.7386 & 0.7931 & 0.8618 & 0.3248 \\
2007Q1 & 2009Q1 & 0.9279 & 0.8528 & 0.7489 & 0.7975 & 0.8635 & 0.3207 \\
2007Q2 & 2009Q2 & 0.9281 & 0.8487 & 0.7569 & 0.8002 & 0.8647 & 0.3195 \\
2007Q3 & 2009Q3 & 0.9289 & 0.8467 & 0.7617 & 0.8020 & 0.8656 & 0.3170 \\
2007Q4 & 2009Q4 & 0.9305 & 0.8524 & 0.7640 & 0.8058 & 0.8678 & 0.3129 \\
2008Q1 & 2010Q1 & 0.9302 & 0.8401 & 0.7802 & 0.8091 & 0.8678 & 0.3141 \\
2008Q2 & 2010Q2 & 0.9299 & 0.8345 & 0.7845 & 0.8087 & 0.8676 & 0.3147 \\
2008Q3 & 2010Q3 & 0.9297 & 0.8326 & 0.7859 & 0.8086 & 0.8676 & 0.3145 \\
2008Q4 & 2010Q4 & 0.9295 & 0.8339 & 0.7818 & 0.8070 & 0.8675 & 0.3158 \\
2009Q1 & 2011Q1 & 0.9302 & 0.8406 & 0.7773 & 0.8077 & 0.8689 & 0.3135 \\
2009Q2 & 2011Q2 & 0.9289 & 0.8332 & 0.7790 & 0.8051 & 0.8676 & 0.3163 \\
2009Q3 & 2011Q3 & 0.9294 & 0.8397 & 0.7719 & 0.8043 & 0.8686 & 0.3142 \\
2009Q4 & 2011Q4 & 0.9296 & 0.8365 & 0.7736 & 0.8038 & 0.8684 & 0.3135 \\
2010Q1 & 2012Q1 & 0.9301 & 0.8313 & 0.7802 & 0.8049 & 0.8689 & 0.3120 \\
2010Q2 & 2012Q2 & 0.9290 & 0.8312 & 0.7724 & 0.8008 & 0.8680 & 0.3136 \\
2010Q3 & 2012Q3 & 0.9288 & 0.8275 & 0.7735 & 0.7996 & 0.8683 & 0.3123 \\
2010Q4 & 2012Q4 & 0.9280 & 0.8177 & 0.7775 & 0.7971 & 0.8671 & 0.3146 \\
2011Q1 & 2013Q1 & 0.9288 & 0.8143 & 0.7815 & 0.7976 & 0.8674 & 0.3123 \\
2011Q2 & 2013Q2 & 0.9284 & 0.8139 & 0.7787 & 0.7959 & 0.8675 & 0.3120 \\
2011Q3 & 2013Q3 & 0.9288 & 0.8097 & 0.7820 & 0.7956 & 0.8675 & 0.3109 \\
2011Q4 & 2013Q4 & 0.9292 & 0.8095 & 0.7817 & 0.7954 & 0.8683 & 0.3085 \\ \hline 
\end{tabular}
\end{center}

\msk
\footnotesize{Performance metrics for our model of default risk. The model calibrations are specified by the training and testing windows. The results of classifications versus actual outcomes over the following 8Q  are used to calculate these performance metrics for 90+ days delinquencies within 8Q. Source: Authors' calculations based on Experian Data.}
\end{table}

Table \ref{tab:perf_metrics} reports the performance metrics widely used in the machine-learning literature for each of the 32+1 models discussed. Our models exhibit strong predictive power across the various performance metrics. For instance, the 85.70\% precision implies that when our classifier predicts that someone is going to default, there is an 85.70\% chance this person will actually default; while the 72.86\% recall means that we accurately identified 72.86\% of all the defaulters.     
Our approach of using only one quarter of data to train the model is rather restrictive. Using more quarters usually increases model performance, so since most credit scoring applications will use a training data that exceeds one quarter, performance metrics are likely to improve relative to what we report in our exercise.

\begin{table}[!h]
\caption{Performance Metrics using Hybrid DNN-GBT, Current}\label{tab:perf_metrics_current}
\footnotesize
\begin{center}
\begin{tabular}{cccccccc}
Training Window & Testing Window & AUC score & Precision & Recall & F-measure & Accuracy & Loss   \\ \hline 
& & & & & & & \\
2004Q1-2013Q4 & 2004Q1-2013Q4  &  0.9263 & 0.7974 & 0.5465 & 0.6485 & 0.9007 & 0.2423 \\
 &  &  &  &  &  &  &  \\
2004Q1 & 2006Q1 & 0.8774 & 0.7744 & 0.3960 & 0.5240 & 0.8673 & 0.3174 \\
2004Q2 & 2006Q2 & 0.8657 & 0.7288 & 0.3567 & 0.4790 & 0.8680 & 0.3156 \\
2004Q3 & 2006Q3 & 0.8642 & 0.7232 & 0.3564 & 0.4775 & 0.8678 & 0.3168 \\
2004Q4 & 2006Q4 & 0.8642 & 0.7256 & 0.3643 & 0.4851 & 0.8664 & 0.3206 \\
2005Q1 & 2007Q1 & 0.8642 & 0.7394 & 0.3611 & 0.4853 & 0.8617 & 0.3289 \\
2005Q2 & 2007Q2 & 0.8616 & 0.7331 & 0.3499 & 0.4737 & 0.8590 & 0.3329 \\
2005Q3 & 2007Q3 & 0.8606 & 0.7439 & 0.3349 & 0.4618 & 0.8571 & 0.3370 \\
2005Q4 & 2007Q4 & 0.8601 & 0.7270 & 0.3585 & 0.4802 & 0.8566 & 0.3380 \\
2006Q1 & 2008Q1 & 0.8621 & 0.7207 & 0.3787 & 0.4965 & 0.8548 & 0.3403 \\
2006Q2 & 2008Q2 & 0.8614 & 0.7264 & 0.3628 & 0.4839 & 0.8532 & 0.3414 \\
2006Q3 & 2008Q3 & 0.8606 & 0.7280 & 0.3479 & 0.4708 & 0.8536 & 0.3420 \\
2006Q4 & 2008Q4 & 0.8577 & 0.7115 & 0.3522 & 0.4712 & 0.8564 & 0.3366 \\
2007Q1 & 2009Q1 & 0.8602 & 0.7050 & 0.3714 & 0.4865 & 0.8604 & 0.3286 \\
2007Q2 & 2009Q2 & 0.8596 & 0.6920 & 0.3833 & 0.4934 & 0.8621 & 0.3256 \\
2007Q3 & 2009Q3 & 0.8597 & 0.6826 & 0.3922 & 0.4981 & 0.8646 & 0.3208 \\
2007Q4 & 2009Q4 & 0.8578 & 0.6859 & 0.3736 & 0.4838 & 0.8675 & 0.3163 \\
2008Q1 & 2010Q1 & 0.8609 & 0.6794 & 0.4175 & 0.5172 & 0.8688 & 0.3151 \\
2008Q2 & 2010Q2 & 0.8617 & 0.6662 & 0.4364 & 0.5273 & 0.8695 & 0.3135 \\
2008Q3 & 2010Q3 & 0.8620 & 0.6595 & 0.4483 & 0.5338 & 0.8699 & 0.3126 \\
2008Q4 & 2010Q4 & 0.8638 & 0.6710 & 0.4376 & 0.5298 & 0.8723 & 0.3092 \\
2009Q1 & 2011Q1 & 0.8665 & 0.6871 & 0.4381 & 0.5351 & 0.8725 & 0.3086 \\
2009Q2 & 2011Q2 & 0.8671 & 0.6784 & 0.4517 & 0.5424 & 0.8728 & 0.3085 \\
2009Q3 & 2011Q3 & 0.8685 & 0.6899 & 0.4377 & 0.5356 & 0.8733 & 0.3068 \\
2009Q4 & 2011Q4 & 0.8658 & 0.6802 & 0.4223 & 0.5211 & 0.8760 & 0.3025 \\
2010Q1 & 2012Q1 & 0.8669 & 0.6723 & 0.4332 & 0.5269 & 0.8752 & 0.3024 \\
2010Q2 & 2012Q2 & 0.8690 & 0.6894 & 0.4241 & 0.5252 & 0.8756 & 0.3018 \\
2010Q3 & 2012Q3 & 0.8683 & 0.6861 & 0.4200 & 0.5211 & 0.8766 & 0.3001 \\
2010Q4 & 2012Q4 & 0.8672 & 0.6784 & 0.4187 & 0.5178 & 0.8772 & 0.2988 \\
2011Q1 & 2013Q1 & 0.8700 & 0.6734 & 0.4462 & 0.5367 & 0.8763 & 0.2996 \\
2011Q2 & 2013Q2 & 0.8705 & 0.6746 & 0.4452 & 0.5364 & 0.8767 & 0.2985 \\
2011Q3 & 2013Q3 & 0.8695 & 0.6721 & 0.4374 & 0.5299 & 0.8775 & 0.2970 \\
2011Q4 & 2013Q4 & 0.8679 & 0.6668 & 0.4343 & 0.5260 & 0.8789 & 0.2951 \\ \hline 
\end{tabular}
\end{center}

\msk
\footnotesize{Performance metrics for our model of default risk for the current population. Borrowers who are current do not have any delinquencies. The model calibrations are specified by the training and testing windows. The results of classifications versus actual outcomes over the following 8Q  are used to calculate these performance metrics for 90+ days delinquencies within 8Q. Source: Authors' calculations based on Experian Data.}
\end{table}

Table \ref{tab:perf_metrics_current} reports the same performance metrics for the population of borrowers who are current, that is, they do not have any delinquencies in the quarter they are assessed. As previously noted, this is a smaller population with a lower probability of default. Performance metrics drop marginally relative to those for the model applied to the population of all borrowers but they are still very strong. For example, the AUC score drops from 92-93\% to 86-88\%, accuracy mostly remain in the same range and the loss increases by 1-2 percentage points.

\subsection{Model Interpretation}\label{sec:interpretation}
We use our hybrid DNN-GBT  model to uncover associations between the explanatory variables and default behavior. Since we do not identify causal relationships, our goal is simply to find covariates that have an important impact on default outcomes. Our findings can be used to better understand default behavior, further refine model specification and possibly aid in the formulation of theoretical models of consumer default. For this exercise, we mainly use the pooled model, which uses all available data. This allows us to assess factors that are critical in default behavior throughout the sample period with the best performing model. We also consider time variation in the factors influencing the default decision in subsets of our sample.

\subsubsection{Explanatory Power of Variables}
We start by examining the explanatory power of each of our features. We follow an approach similar to \citeN{sirignano}, which amounts to a perturbation analysis on the pooled sample using our hybrid model. First, we draw a random sample of 100,000 observations from the testing sample. Then, for each variable, we re-shuffle the feature, keeping the distribution intact and the model's loss function is evaluated with the changed covariate. We repeat this step 10 times, and report the average of the loss and accuracy. Then, the variable is replaced to its original values, and a perturbation test is performed on a new variable. Perturbing the variable of course reduces the accuracy of the model, and the test loss becomes larger. If a particular variable has strong explanatory power, the test loss will significantly increase. The test loss for the complete model when no variables are perturbed is the Baseline value. Features that have large explanatory power, and whose information is not contained in the other remaining variables will increase the loss significantly if they are altered. Table \ref{tab_shuffle} reports the results. Features relating to credit history, debt balances, and the number and credit available on revolving trades dominate the list.
Specifically, months since the most recent 90+ days delinquency and months since the oldest trade was opened each increase the loss by 13\%,  the number of open credit card trades and credit amount on revolving trades increase the loss by 11\%, while  the balance on first mortgage trades and monthly payment on first mortgages increase the loss by 10\%. These results suggest that revolving debt, length of credit history and temporal proximity to a delinquency are all important factors in default behavior. Based on publicly available information, length of the credit history is also an important determinant of standard credit scoring models, though payment history rather than balances or number of trades is understood as the most critical. 
This approach to assessing the importance of different features for the predicted probability of default has two major shortcomings. First, when features are highly correlated, the interpretation of feature importance can be biased by unrealistic data instances. To illustrate this problem, consider  two  highly correlated features. As we perturb  one of the features, we create instances that are unlikely or even impossible. For example, mortgage balances are highly correlated with and lower than total debt balances, yet this perturbation approach could create instances in which total debt balances are smaller than mortgage balances. Since many of the features are strongly correlated, care must be taken with  interpretation of  feature importance. We list the highly correlated features in Appendix \ref{sec:features}. An additional concern with this perturbation approach is that the distribution of some features are highly skewed, which implies that the probability of their value being different than where the mass of their distribution is concentrated is quite low. Moreover, skewness varies substantially across features, therefore the informativeness of the perturbation may differ across variables. In the next section, we examine a more robust approach that is less susceptible to these limitations.

\begin{table}[!h]
\caption{Explanatory Power of Variables}\label{tab_shuffle}
\begin{center}
\footnotesize 
\begin{tabular}{lll}
Feature & Accuracy & Loss \\ \hline 
 &  &  \\ 
Months since the oldest trade was opened & 0.8762 & 0.2941 \\
Months since the most recent 90 or more days delinquency & 0.8782 & 0.2936 \\
Open credit card trades & 0.8793 & 0.2895 \\
Credit amount on revolving trades & 0.8806 & 0.2876 \\
Balance on first mortgage trades & 0.8741 & 0.2876 \\
Monthly payment on open first mortgage trades & 0.8727 & 0.2866 \\
Open bankcard revolving, and charge trades & 0.8804 & 0.2865 \\
Total credit amount on open trades & 0.8803 & 0.2851 \\
Total debt balances & 0.8814 & 0.2844 \\
Monthly payment on all debt & 0.8797 & 0.2840 \\
Credit amount on open credit card trades & 0.8819 & 0.2812 \\
Worst ever status on any trades in the last 24 months & 0.8852 & 0.2784 \\
Months since the most recently opened first mortgage & 0.8835 & 0.2780 \\
Balance on collections & 0.8818 & 0.2779 \\
Monthly payment on credit card trades & 0.8831 & 0.2776 \\
Balance on bankcard revolving and charge trades & 0.8843 & 0.2769 \\
Balance on credit card trades & 0.8839 & 0.2769 \\
Credit amount on open mortgage trades & 0.8846 & 0.2765 \\
Monthly payment on open auto loan trades & 0.8837 & 0.2758 \\
Months since the most recent 30-180 days delinquency & 0.8849 & 0.2756 \\
Worst present status on any trades & 0.8856 & 0.2751 \\
Months since the most recently opened credit card trade & 0.8850 & 0.2746 \\
Credit amount on unsatisfied derogatory trades & 0.8825 & 0.2743 \\
Balance on revolving trades & 0.8848 & 0.2743 \\
Months since the most recently opened auto loan trade & 0.8848 & 0.2736 \\
Credit amount on open installment trades & 0.8846 & 0.2731 \\
Credit amount paid down on open first mortgage trades & 0.8870 & 0.2721 \\
\vdots & \vdots & \vdots  \\ 
Mortgage inquiries made inthe last 3 months & 0.8913 & 0.2605 \\
First mortgage trades opened in the last 6 months & 0.8914 & 0.2605 \\
Bankcard revolving and charge inquiries made in the last 3 months & 0.8914 & 0.2604 \\
Auto loan or lease inquiries made in the last 3 months & 0.8912 & 0.2603 \\
Balance on open bankcard revolving, and charge trades with credit line suspended & 0.8914 & 0.2603 \\
Baseline & 0.8914 & 0.2603 \\ \hline 
\end{tabular}
\end{center}

\msk
\footnotesize{This table reports a perturbation analysis on the pooled sample using our hybrid model. For each variable, we re-shuffle the feature, keeping the distribution intact in the test dataset and the model's loss function is evaluated on the test dataset with the changed covariate. We repeat this step 10 times, and report the average of the loss and accuracy. Then, the variable is replaced to its original values, and a perturbation test is performed on a new variable. Perturbing the variable of course reduces the accuracy of the model, and the test loss becomes larger. If a particular variable has strong explanatory power, the test loss will significantly increase. The test loss for the complete model when no variables are perturbed is the Baseline value.}

\end{table}

\subsubsection{Economic Significance of Variables}
We now turn to analyzing the economic significance of our features for default behavior. We adopt SHapley Additive exPlanations (SHAP), a unified framework for interpreting predictions, to explain the output of our hybrid deep learning model (for a detailed description of the approach see \citeN{SHAP}). SHAP uses a game theoretical concept to  assign each feature a local importance value for a given prediction. Though Shapley values are local by design, they can be combined into global explanations by averaging the absolute Shapley values featurewise. Then, we can compare features based on their absolute average Shapley values, with higher values implying higher feature importance. Similarly to permutation feature importance, SHAP is a feature importance measure. The main difference between the two is that while permutation feature importance is based on the decrease in model performance, SHAP is based on the magnitude of feature attributions.    

We first compute the Shapley values for the Deep Neural Network model and the Gradient Boosted Trees model separately, then simply average them for each individual and for each feature.\footnote{We implement Deep SHAP, a high-speed approximation algorithm for SHAP values in deep learning models to compute the Shapley values for our 5 hidden layer neural network. For GBT, we implement TreeExplainer, a high-speed exact algorithm for tree ensemble methods. Because our dataset is fairly large with many features, we pass a random sample of 100 observations, referred to as background observations, to compute the expected value for both models.} We use a random sample of 100,000 observations for explaining the model. By the Shapley efficient property, the SHAP values for an observation sum up to the difference between the predicted value of that observation and the expected value, computed using the background dataset:
\begin{equation}
f(x) = E_X[\widehat{f}(X)] + \sum_{j=1}^{M} \phi_j
\end{equation}
where f is the model prediction, M is the number of features, and $\phi_j \in R$ is the feature attribution for feature j (i.e., the Shapley values). Thus, we can interpret the Shapley value as the contribution of a feature value to the difference between the model's prediction and the mean prediction, given the current set of feature values. As an illustration, a SHAP value of 0.1 implies that the feature's value for that particular instance contributed to an increase of 0.1 to the predicted probability compared to the mean prediction. Features that are highly correlated can decrease the importance of the associated feature by splitting the importance between both features. We account for the effect of feature correlation on interpretability by grouping features with a correlation larger than 0.7, and summing the SHAP values within each groups. We denote these groups with an asterisk for the rest of the analysis and report the composition of feature groups in Table \ref{tab:feature_corr} in the appendix.

Figure \ref{fig:shap} sorts features by the sum of absolute SHAP value magnitudes, and plots the distribution of the impact each feature has on the model output for the twelve most important features or groups of correlated features. The color represents the feature value (red: high, blue: low), whereas the position on the horizontal axis denotes the contribution of the feature. The charts plot the distribution of SHAP values for individual instances in the 100,000 testing sample. The most important feature in terms of SHAP value magnitude is the worst status on any trades. High values of this variable tend to increase predicted default risk, whereas low values tend to decrease it, though the distribution of instances is dispersed. Features capturing credit history, such as length of credit history and recent delinquencies, also have high SHAP values, specifically, high values of these variables lower predicted default risk, with a much more dispersed distribution. Additionally, credit card utilization, credit amount on derogatory trades and outstanding collections are typically associated with an increase in predicted default probability. Higher total debt balances are also associated with a lower than expected predicted default risk, reiterating the notion that the borrowers with the most credit are also associated with lower predicted probability of default, which suggests that credit allocation decisions are made to minimize default probabilities. As in the perturbation exercise, we find that number of trades and balances seem to have the strongest association with variation in the predicted probability of default, whereas credit inquiries do not play a sizable role. 

\begin{figure}[!h]
\begin{center}
    \includegraphics[scale=0.55]{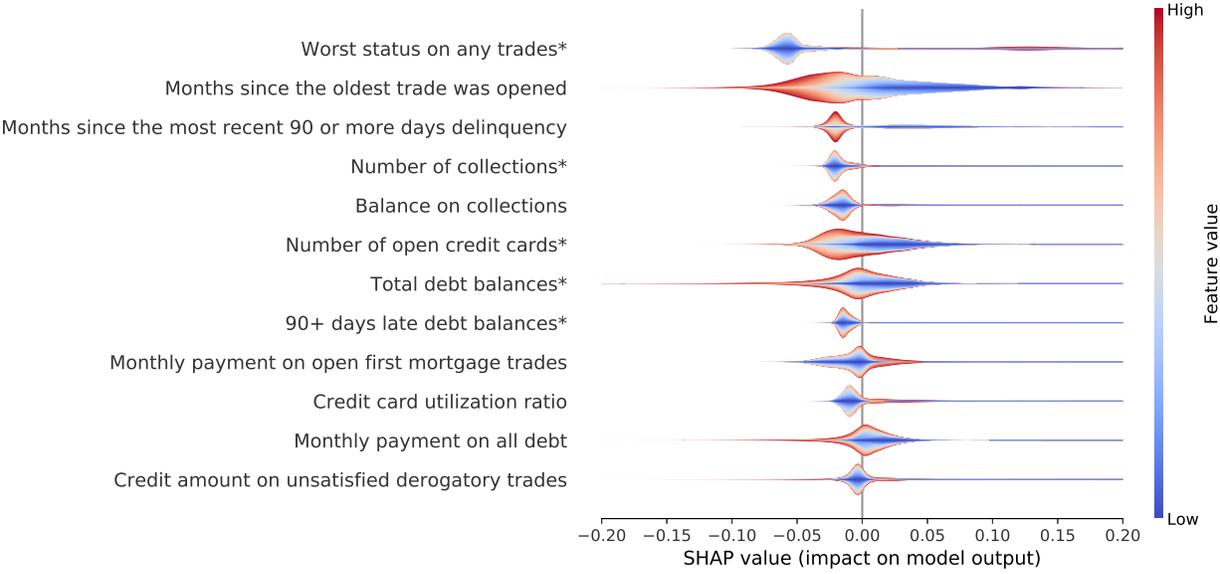}
     \ssk    
     
\caption{SHAP applied to predicted 90+ days delinquency within 8Q. Source: Authors' calculations based on Experian Data.}
\label{fig:shap}
\end{center}
\end{figure}

These results only point to correlations between the features and the predicted outcome and should not be interpreted causally. Yet, they can be used as a point of departure for a causal analysis of default and theoretical modeling. They are also important to comply with legal disclosure requirements. Both the Fair Credit Reporting Act ad the Equal Opportunity in Credit Access Act require lenders and developers of credit scoring models to reveal the most important factors leading to a denial of a credit application and for credit scores. The SHAP value provides an individualized assessment of such factors that can be used for making credit allocation decisions and communicating them to the borrower.  


\subsubsection{Temporal Determinants of Default}
We next look at the changing dynamics of default behavior by comparing models that are trained in different periods of time. For this analysis, we use our hybrid model. Specifically, we target the following time periods: 2006Q1, 2008Q1 and 2011Q1 as time periods before, during and after the 2007-2009 crisis, and compute default predictions for them with data trained in the same quarter and two years prior, that is in 2004Q1, 2006Q1, 2007Q1 and 2009Q1, respectively. We then calculate Shapley values for the two models.\footnote{We do this for both the Deep Neural Network and the Gradient Boosted Trees and similarly to how we obtain the output, we simply take the average of the Shapley values. For both our models, we use a random sample of 100 observations of the testing data scaled by the mean and standard deviation of the corresponding training data for reference value.} The first exercise provides an in-sample assessment for feature importance, while the second exercise can be used to assess feature importance out-of-sample. In both exercises, the model is the same, so comparing the results from the two exercises can help uncover which features are important for default prediction for a given period from an ex ante perspective and from an in-sample perspective.      
 
\begin{sidewaystable}[htpb]
\caption{SHAP Values over Time}\label{tab:shap_time} 
\scriptsize
\begin{center}
\begin{tabular}{lllllll}
 & \multicolumn{6}{c}{Prediction Date} \\
 & \multicolumn{2}{c}{2006Q1} & \multicolumn{2}{c}{2008Q1} & \multicolumn{2}{c}{2011Q1} \\
 & \multicolumn{6}{c}{Model} \\
 & 2004Q1 & 2006Q1 & 2006Q1 & 2008Q1 & 2009Q1 & 2011Q1 \\
Features &  &  &  &  &  &  \\ \hline
 &  &  &  &  &  &  \\
Total debt balances* & 0.072 (1) & 0.063 (1) & 0.064 (1) & 0.054 (1) & 0.047 (1) & 0.055 (1) \\
Months since the most recent 30-180 days delinquency & 0.036 (2) & 0.024 (5) & 0.02 (5) & 0.004 (34) & 0.004 (36) & 0.009 (15) \\
Number of open credit cards* & 0.031 (3) & 0.024 (4) & 0.023 (4) & 0.037 (2) & 0.022 (5) & 0.031 (3) \\
Number of mortgages* & 0.03 (4) & 0.049 (2) & 0.044 (2) & 0.033 (3) & 0.042 (3) & 0.05 (2) \\
Worst status on any trades* & 0.024 (5) & 0.027 (3) & 0.027 (3) & 0.032 (4) & 0.025 (4) & 0.02 (5) \\
Credit card debt* & 0.017 (6) & 0.009 (14) & 0.009 (12) & 0.02 (7) & 0.017 (8) & 0.018 (6) \\
Student debt* & 0.016 (7) & 0.012 (8) & 0.014 (8) & 0.01 (14) & 0.012 (12) & 0.016 (7) \\
Months since the most recent 30-180 days delinquency on auto loan or lease trades & 0.013 (8) & 0.009 (15) & 0.008 (16) & 0.009 (17) & 0.005 (29) & 0.008 (19) \\
Fraction of credit card debt to total debt & 0.012 (9) & 0.01 (11) & 0.011 (9) & 0.024 (6) & 0.018 (7) & 0.031 (4) \\
Inquiries made in the last 12 months & 0.012 (10) & 0.011 (9) & 0.011 (10) & 0.009 (18) & 0.019 (6) & 0.007 (21) \\
Installment trades & 0.011 (11) & 0.015 (6) & 0.015 (7) & 0.027 (5) & 0.045 (2) & 0.01 (13) \\
Auto loan* & 0.01 (14) & 0.01 (12) & 0.008 (14) & 0.006 (25) & 0.007 (19) & 0.006 (25) \\
Months since the oldest trade was opened & 0.011 (13) & 0.009 (13) & 0.008 (13) & 0.011 (10) & 0.006 (22) & 0.002 (49) \\
Months since the most recently closed, transferred, or refinanced first mortgage & 0.011 (12) & 0.013 (7) & 0.017 (6) & 0.01 (13) & 0.008 (14) & 0.01 (12) \\
Worst present status on an installment trade & 0.008 (15) & 0.011 (10) & 0.01 (11) & 0.012 (9) & 0.013 (10) & 0.013 (9) \\
Months since the most recently opened first mortgage & 0.008 (16) & 0.005 (23) & 0.005 (24) & 0.009 (16) & 0.007 (18) & 0.008 (20) \\
Early payoff trades & 0.007 (17) & 0.008 (16) & 0.008 (15) & 0.004 (35) & 0.004 (32) & 0.002 (48) \\
Credit card trades opened in the last 12 months & 0.006 (19) & 0.003 (31) & 0.004 (32) & 0.002 (51) & 0.002 (64) & 0.001 (64) \\
Monthly payment on all debt & 0.006 (18) & 0.005 (24) & 0.004 (27) & 0.008 (21) & 0.002 (69) & 0.003 (44) \\
Months since the most recently opened home equity line of credit trade & 0.005 (20) & 0.004 (27) & 0.005 (20) & 0.004 (32) & 0.003 (41) & 0.011 (10) \\ \hline
\end{tabular}
\end{center}

\scriptsize{
This table reports the Shapley values for a selected 20 features for three out-of-sample models. For each prediction window, we compute the Shapley value for each of the observations and for each feature. We then calculate the average of the absolute value for each feature, and report the results for the selected features. Finally, we rank the results based on the feature's relative rank in the given prediction window in parentheses. Source: Authors' calculations based on Experian data.}
\end{sidewaystable}

Table \ref{tab:shap_time} reports the results.\footnote{The features are sorted by the sum of absolute SHAP value magnitudes over the first period.} For each period, it is interesting to compare the variation in SHAP values from an ex ante and contemporaneous perspective, and additionally we are interested in comparing variation in SHAP values for given features in the different time periods. In all testing windows, total debt balances has the highest SHAP value. Number of open credit cards, mortgages, and worst status on any trades are consistently among the five most influential features. Student debt and credit card debt, are generally between the sixth and fourteenth most significant in terms of SHAP values.  The SHAP value is quite stable over time for most features, but there are some variables for which it changes substantially. One example is months since the most recent 30-180 days delinquency which ranks second and fifth for 2004Q1 and 2006Q1 but moves down to fifteenth in sample and thirty-sixth out-of sample for 2011Q1.  The length of the credit history is never among the ten most important features, and features related to credit utilization do not register high SHAP values in any of the time periods. Overall these results confirm our findings from the pooled model, suggesting the balances and number of trades, in addition to delinquency status, have a strong association with default risk according to our model.

\subsection{Model Comparisons}\label{sec:comparison}
We examine the out-of-sample behavior of a collection of alternative machine learning techniques in this section. To motivate our choice of deep learning, leading to our hybrid DNN-GBT model, we begin by illustrating the importance of hidden layers that enable us to capture multi-level interactions between features by comparing how neural networks of different depth perform on the pooled sample. For this exercise, we fix the number of neurons per layer at 512, and build neural networks up to 5 hidden layers.\footnote{The architecture reported in Table \ref{tab7} was not optimized. We picked 512 for each layer randomly. Another variant of this exercise is where we use the optimal 5-layer neural network architecture with a SELU activation function and remove the neurons layer by layer. We report the results in Table \ref{tab:pooled_leave}.}    

We benchmark our results against  the logistic regression, which is a commonly used technique in credit scoring and can be interpreted as a neural network with no hidden layers. Table \ref{tab7} reports the in- and out-of-sample behavior  for neural networks with 0-5 hidden layers. The number of hidden layers measure the complexity of the network, and we found that the marginal improvements in performance beyond 4 layers are small. Depth, however, is not directly proportional to out-of-sample performance due to higher model capacity. Table \ref{tab7} also shows that applying dropout improves the out-of-sample fit for networks of higher depths. This demonstrates that dropout serves as an effective regularization tool and addresses over-fitting for networks of greater depths.

 \begin{table}[!h]
 \caption{Neural networks comparison: Loss \& Accuracy}
 \label{tab7}
\small
\begin{center}
\begin{tabular}{lllll}
Model & \multicolumn{2}{c}{In-sample Loss}  & \multicolumn{2}{c}{Out-of-sample Loss}   \\
 & w/o Dropout & Dropout & w/o Dropout & Dropout \\ \hline
 & & & & \\ 
Logistic Regression & 0.3451 & 0.3451 & 0.3449 & 0.3449 \\
1 layer & 0.3094 & 0.3104 & 0.3118 & 0.3117 \\
2 layers & 0.2961 & 0.3013 & 0.3080 & 0.3065 \\
3 layers & 0.2847 & 0.2934 & 0.3053 & 0.3021 \\
4 layers & 0.2757 & 0.2870 & 0.3031 & 0.2987 \\
5 layers & 0.2794 & 0.2860 & 0.3028 & 0.2982 \\ \hline 
 & & & & \\ 
Model & \multicolumn{2}{c}{In-sample Accuracy}   & \multicolumn{2}{c}{Out-of-sample Accuracy}   \\
 & w/o Dropout & Dropout & w/o Dropout & Dropout \\ \hline
 & & & & \\ 
Logistic Regression & 0.8564 & 0.8564 & 0.8566 & 0.8566 \\
1 layer & 0.8694 & 0.8690 & 0.8682 & 0.8685 \\
2 layers & 0.8754 & 0.8733 & 0.8700 & 0.8708 \\
3 layers & 0.8808 & 0.8769 & 0.8720 & 0.8727 \\
4 layers & 0.8847 & 0.8794 & 0.8735 & 0.8739 \\
5 layers & 0.8829 & 0.8800 & 0.8732 & 0.8741 \\ \hline 
\end{tabular}
\msk
\end{center}

\footnotesize{In-sample and out-of-sample loss (categorical cross-entropy) and accuracy for neural networks of different depth and for logistic regression. Models are calibrated and evaluated on the pooled sample (2004Q1 - 2013Q4). Source: Authors' calculations based on Experian Data.}

\end{table}

\begin{figure}[htbp]
\begin{center}
\includegraphics[scale=0.6]{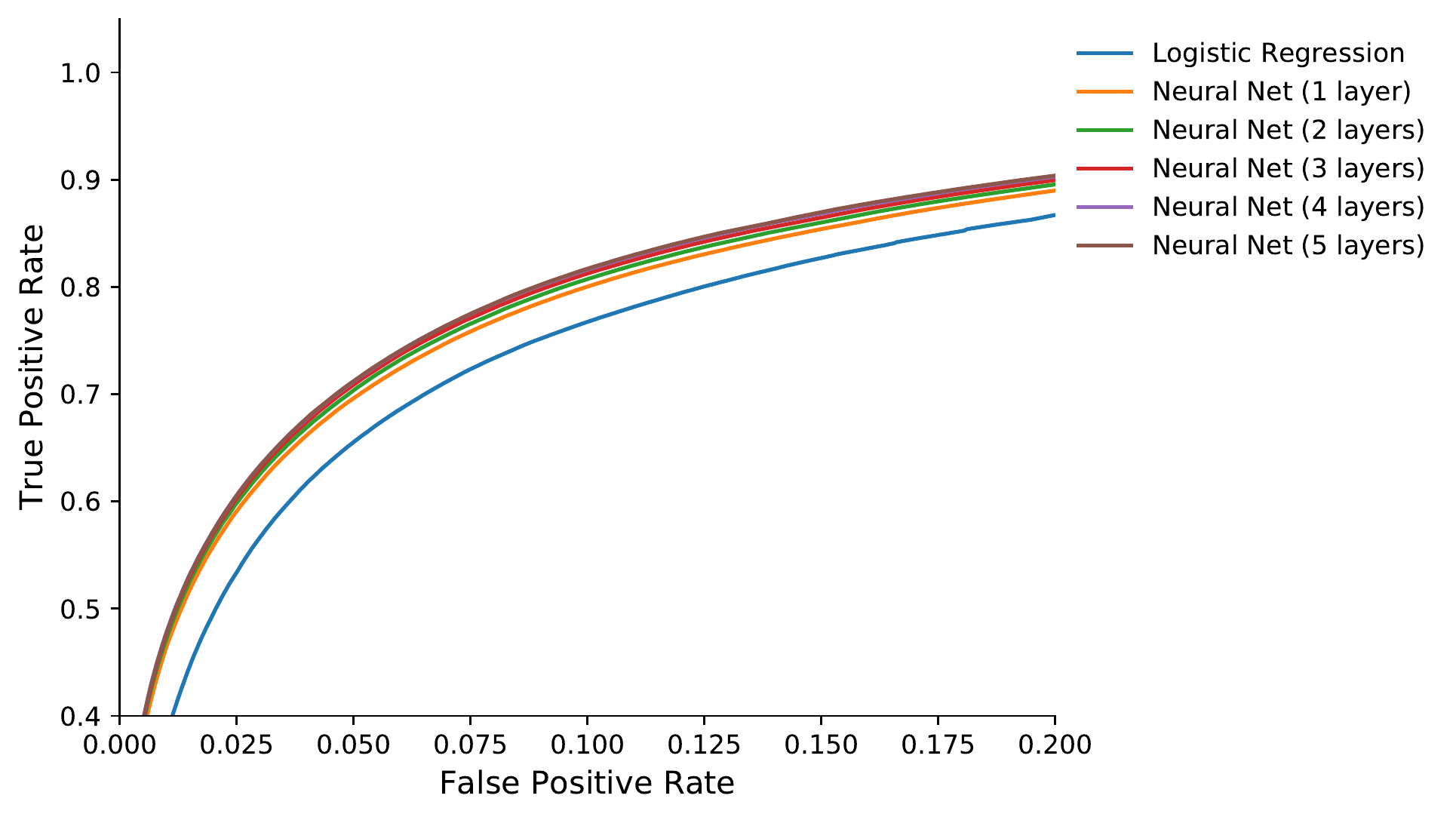}
 
\msk     
\caption{Out-of-sample ROC curves for various models with dropout. Models are calibrated and evaluated on the pooled sample (2004Q1 - 2013Q4).  
Source: Authors' calculations based on Experian Data.}\label{fig:ROC_comp}
\end{center}
\end{figure}

The results in Table \ref{tab7} suggest that there are complex non-linear relationships among the features used as inputs in the model. This is further supported by the fact that permitting non-linear relationships between default behavior and explanatory variables produces the largest model improvement. Going from a linear model (0 layers) to the simplest non-linear model (1 layer) generates the most sizable reduction in out-of-sample loss. To see this from another angle, we plot the ROC curves for our neural networks considered in Figure \ref{fig:ROC_comp}. We can see that the logistic regression is dominated by all models that allow for non-linear relationships, while the improvements for deeper models are marginal.      

We next analyze a number of machine learning techniques that are possible alternatives to our hybrid model. These algorithms  have been used in other credit scoring applications, and include decision trees (CART, see \citeN{khandani}), random forests (RF, see \citeN{butaru}), neural networks (see \citeN{West}), gradient boosting (GBT, see \citeN{xia}) and logistic regression. 
We use the out of sample loss as our main comparison metric,  with lower loss values corresponding to better model performance.  We tune the hyper-parameters for each model and present the results in Table \ref{tab:model_comparison} for our baseline 1 quarter training/validation samples.  Our hybrid model performs the best, with gradient boosting coming second, while deep neural network outperforms both random forest and decision trees in all samples. We repeat this comparison with an expanding  training window and find that deep neural networks benefit the most from more training data. 

\clearpage 

\begin{table}[!h]
\caption{Model Comparison: Out-of-Sample Loss}
\label{tab:model_comparison}
\footnotesize
\begin{center}
\begin{tabular}{cccccccc}
Training Window & Testing Window & Combined & GBT & DNN &  RF    & CART & Logistic   \\ \hline 
& & & & & & & \\ 
2004Q1 & 2006Q1 & 0.3236 & 0.3235 & 0.3290 & 0.3274 & 0.3432 & 0.3499 \\
2004Q2 & 2006Q2 & 0.3181 & 0.3185 & 0.3229 & 0.3231 & 0.3372 & 0.3465 \\
2004Q3 & 2006Q3 & 0.3164 & 0.3161 & 0.3228 & 0.3210 & 0.3361 & 0.3445 \\
2004Q4 & 2006Q4 & 0.3203 & 0.3198 & 0.3262 & 0.3245 & 0.3396 & 0.3464 \\
2005Q1 & 2007Q1 & 0.3211 & 0.3209 & 0.3270 & 0.3259 & 0.3427 & 0.3491 \\
2005Q2 & 2007Q2 & 0.3221 & 0.3215 & 0.3279 & 0.3272 & 0.3440 & 0.3512 \\
2005Q3 & 2007Q3 & 0.3254 & 0.3246 & 0.3315 & 0.3301 & 0.3455 & 0.3525 \\
2005Q4 & 2007Q4 & 0.3273 & 0.3275 & 0.3324 & 0.3327 & 0.3498 & 0.3560 \\
2006Q1 & 2008Q1 & 0.3263 & 0.3264 & 0.3315 & 0.3316 & 0.3477 & 0.3558 \\
2006Q2 & 2008Q2 & 0.3272 & 0.3277 & 0.3317 & 0.3321 & 0.3495 & 0.3563 \\
2006Q3 & 2008Q3 & 0.3265 & 0.3256 & 0.3328 & 0.3299 & 0.3474 & 0.3565 \\
2006Q4 & 2008Q4 & 0.3248 & 0.3259 & 0.3296 & 0.3293 & 0.3462 & 0.3547 \\
2007Q1 & 2009Q1 & 0.3207 & 0.3211 & 0.3267 & 0.3244 & 0.3425 & 0.3503 \\
2007Q2 & 2009Q2 & 0.3195 & 0.3202 & 0.3246 & 0.3236 & 0.3427 & 0.3496 \\
2007Q3 & 2009Q3 & 0.3170 & 0.3174 & 0.3224 & 0.3205 & 0.3383 & 0.3464 \\
2007Q4 & 2009Q4 & 0.3129 & 0.3127 & 0.3189 & 0.3170 & 0.3348 & 0.3425 \\
2008Q1 & 2010Q1 & 0.3141 & 0.3141 & 0.3201 & 0.3182 & 0.3360 & 0.3453 \\
2008Q2 & 2010Q2 & 0.3147 & 0.3144 & 0.3205 & 0.3188 & 0.3355 & 0.3463 \\
2008Q3 & 2010Q3 & 0.3145 & 0.3148 & 0.3200 & 0.3187 & 0.3353 & 0.3468 \\
2008Q4 & 2010Q4 & 0.3158 & 0.3156 & 0.3215 & 0.3196 & 0.3365 & 0.3459 \\
2009Q1 & 2011Q1 & 0.3135 & 0.3138 & 0.3192 & 0.3174 & 0.3320 & 0.3453 \\
2009Q2 & 2011Q2 & 0.3163 & 0.3163 & 0.3218 & 0.3202 & 0.3346 & 0.3472 \\
2009Q3 & 2011Q3 & 0.3142 & 0.3148 & 0.3191 & 0.3186 & 0.3322 & 0.3457 \\
2009Q4 & 2011Q4 & 0.3135 & 0.3137 & 0.3185 & 0.3176 & 0.3313 & 0.3446 \\
2010Q1 & 2012Q1 & 0.3120 & 0.3118 & 0.3175 & 0.3166 & 0.3298 & 0.3440 \\
2010Q2 & 2012Q2 & 0.3136 & 0.3140 & 0.3189 & 0.3186 & 0.3313 & 0.3453 \\
2010Q3 & 2012Q3 & 0.3123 & 0.3127 & 0.3172 & 0.3176 & 0.3310 & 0.3440 \\
2010Q4 & 2012Q4 & 0.3146 & 0.3142 & 0.3207 & 0.3195 & 0.3325 & 0.3455 \\
2011Q1 & 2013Q1 & 0.3123 & 0.3124 & 0.3174 & 0.3167 & 0.3313 & 0.3440 \\
2011Q2 & 2013Q2 & 0.3120 & 0.3123 & 0.3166 & 0.3169 & 0.3320 & 0.3432 \\
2011Q3 & 2013Q3 & 0.3109 & 0.3108 & 0.3164 & 0.3152 & 0.3315 & 0.3416 \\
2011Q4 & 2013Q4 & 0.3085 & 0.3088 & 0.3134 & 0.3136 & 0.3281 & 0.3406 \\
 &  &  &  &  &  & &  \\
Average &  & 0.3173 & 0.3175 & 0.3229 & 0.3230 & 0.3377 & 0.3476 \\ \hline 
\end{tabular}
\end{center}

\footnotesize{Performance comparison of machine learning classification models of consumer default risk. The model calibrations are specified by the training and testing windows. The results of predicted probabilities versus actual outcomes over the following 8Q testing period are used to calculate the loss metric for 90+ days delinquencies within 8Q. Combined refers to the hybrid DNN-GBT model, DNN refers to deep neural network, RF refers to random forest, GBT refers to gradient boosted trees, while CART refers to decision tree.  Source: Authors' calculations based on Experian Data.}
\end{table}

Empirically, ensembles perform better when there is a significant diversity among the models (see \citeN{kuncheva2003measures}). Table \ref{tab:shap_comparison} in Appendix \ref{app:model_comp} shows the SHAP values for our hybrid DNN-GBT model in comparison to GBT and DNN models for the pooled sample. The results suggest there are significant differences between the DNN and GBT. For instance, monthly payment on first mortgage trades is the sixth most important feature for GBT, while only fiftysecond for DNN. Even more striking perhaps is that the number of collections is ranked most important for DNN, while only sixteenth for GBT. The ensemble approach  can thus be thought of as providing diversification, which can reduce the variance of any of the two models. If one of the models puts too high of a weight on a feature, the other model may mitigate this effect. 

It is important to emphasize that these results do not imply that there does not exist a random forest or CART model that cannot outperform our hybrid model. The best model will depend on the specific  sample. The  exercise is intended to illustrate that the complexity of the model is proportional to its accuracy to a certain degree, and that deep neural networks improve substantially on shallow models, such as logistic regression.

\section{Applications}\label{sec:applications}
In this section, we use our model in a number of applications. We first provide a comparison between the performance of our model and conventional credit score. Second, we show that our model can score more borrowers. Finally, we show that our model is able to accurately predict variations in aggregate default risk.
\subsection{Comparison with Credit Score}\label{sec:credit_score_comp}
In this section, we compare the performance of our deep neural networks to a conventional credit score.\footnote{The deep-learning forecasts for each quarter are obtained using  out-of-sample input data, as reported in Table \ref{tab:full_sample}.} The credit score is a summary indicator intended to predict the risk of default by the borrower and it is widely used by the financial industry.  For most unsecured debt, lenders typically verify a perspective borrower's credit score at the time of application and sometimes a short recent sample of their credit history. For larger unsecured debts, lenders also typically require some form of income verification, as they do for secured debts, such as mortgages and auto loans. Still, the credit score is often a key determinant of crucial terms of the borrowing contract, such as the interest rate, the downpayment or the credit limit. We have access to a widely used conventional credit score that uses information from the three credit bureaus.

Table \ref{tab:rank_corr} shows the relationship between credit score, predicted probability and realized default rate, where default is defined as usual as 90+ days delinquency in the subsequent 8 quarters. 
The calculation proceeds as follows. We first compute the number of unique credit scores in the data. We create the same number of bins of equal size in our predicted probability distribution, and calculate the realized frequency of 90+ days delinquencies in the subsequent 8 quarters for each of these bins. Since higher credit scores correspond to lower probability of default, we present the negative of the rank correlation with realized defaults for the credit score.  The results  indicate that even though credit score is successful in rank-ordering customers by their future default rates, with rank correlations between 0.980 and 0.994, our deep neural network performs better, with rank correlations always at 0.999. Figure \ref{fig:rank_gini_comp} in Appendix \ref{app:score} plots the time series of these rank correlations by quarter for the entire sample period. The figure shows that the rank correlation for the predicted probability of default generated by our model is remarkably stable over time, while for the credit score it fluctuates from lows of around 0.975 before 2012 to a peak go 0.995 in 2013Q2 with notable quarter by quarter variation. This property of credit scores may be due to the fact that the credit score is an ordinal ranking and its distribution is designed to be stable over time, even if default risk at an individual or aggregate level may change substantially. Figure \ref{fig:vs_hist} in Appendix \ref{app:score} displays the histogram of credit score distributions in our sample for selected years, and show that these distributions are virtually identical over time.     

A common way to measure the accuracy of conventional credit scoring models is the Gini coefficient, which measures the dispersion of the credit score distribution and therefore its ability to separate borrowers by their default risk. The Gini coefficient is related to a key performance metric for machine learning algorithm, the AUC score, with $Gini=2*AUC-1$, so we can compare the performance of the credit score to our model along this dimension. Figure \ref{fig:rank_gini_comp} plots the Gini coefficient for the credit score and our predicted default probability by quarter. The Gini coefficient for our model is about 0.85 between 2006Q1 and 2008Q3, and then rises to 0.86. For the credit score, the Gini coefficient is close to 0.81 until 2012Q3 when it drops to approximately 0.79 until the end of the sample, suggesting a drop in performance of the credit score in the aftermath of the Great Recession.

\begin{table}[htbp]
\caption{Credit Score-Predicted Probability of Default Comparison}\label{tab:rank_corr}
\footnotesize
\begin{center}
\begin{tabular}{lllllllll}
 Metric & 2006 & 2007 & 2008 & 2009 & 2010 & 2011 & 2012 & 2013 \\ \hline 
 & &  &  &  &  &  &  &  \\
\textbf{Rank Correlation}  &  &  &  &  &  &  &  \\
Credit Score & 0.9881 & 0.9804 & 0.9882 & 0.9816 & 0.9825 & 0.9861 & 0.9906 & 0.9944 \\
Predicted Probability & 0.9991 &	0.9994 &	0.9993 &	0.9992 &	0.9991 &	0.9992 &	0.9992 &	0.9992 \\
 & &  &  &  &  &  &  &  \\
\textbf{GINI Coefficient}  &  &  &  &  &  &  &  \\
Credit Score & 0.8108 & 0.8142 & 0.8137 & 0.8143 & 0.8078 & 0.8008 & 0.7942 & 0.7898 \\
Predicted Probability & 0.8533 &	0.8528 &	0.8531 &	0.8604 &	0.8615 &	0.8608 &	0.8598 &	0.8579\\ \hline 
\end{tabular}

\end{center}
\ssk

\footnotesize{Rank correlation between credit score, predicted probability of default according to our model and subsequent realized default frequency by year. For the credit score, report the rank correlation between each unique value of the score and the default frequency. For predicted probability of default based on our hybrid DNN-GBT model, we first generate a number of bins equal to the number of unique credit score realizations in the data  and then calculate the realized default frequency for each bin. Source: Authors' calculations based on Experian Data.}

\end{table}

\begin{figure}[h]
\begin{center}
 \subfloat[Predicted Default Probability]{\includegraphics[scale=0.5]{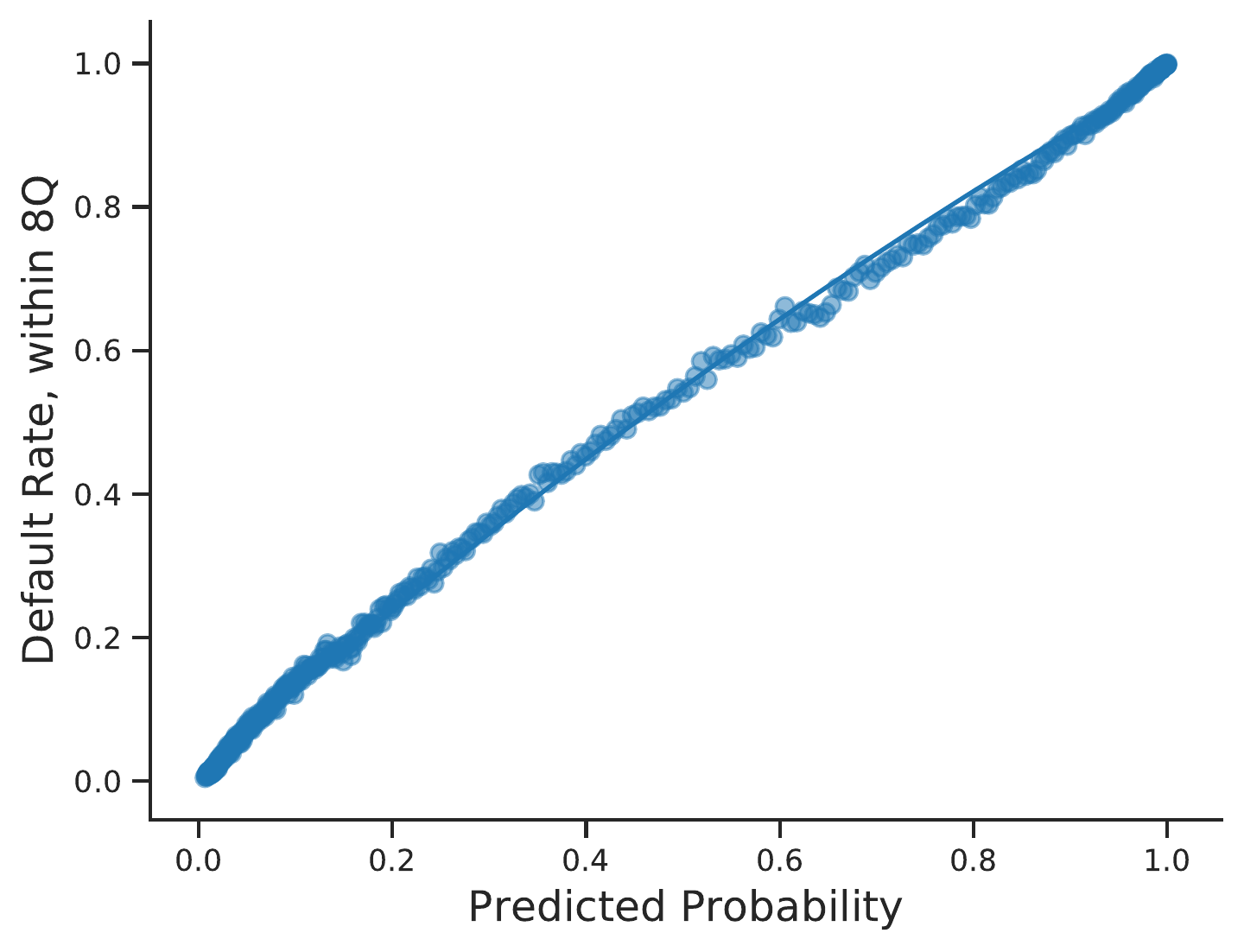}} \hskip 0.25truein
 \subfloat[Credit Score]{\includegraphics[scale=0.5]{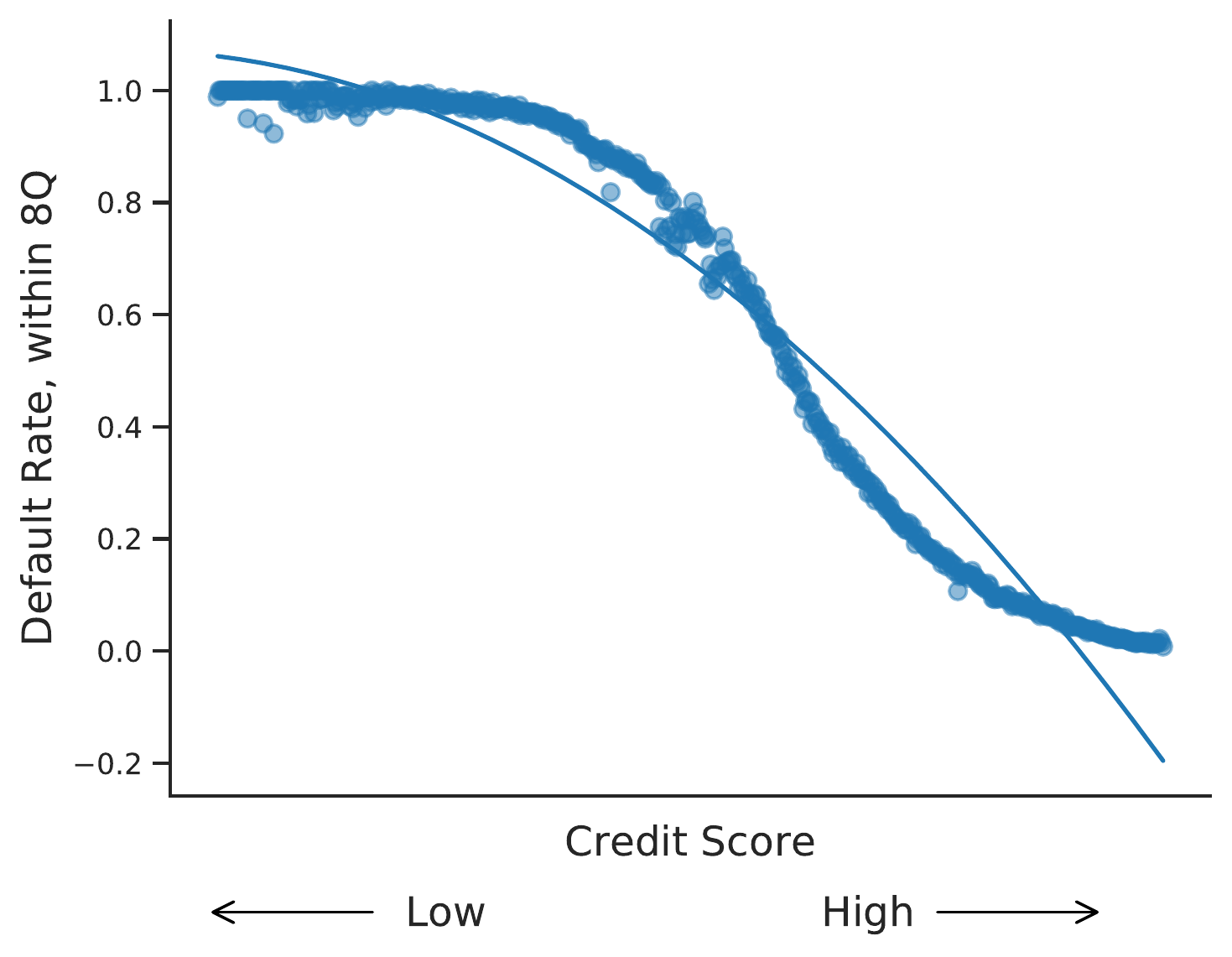}}
\caption{Scatter plot of realized default rates against model predicted default probability (a) and the credit score (b), with associated second-order polynomial fitted approximations for the year 2008. Source: Authors' calculations based on Experian Data.}\label{fig:ranking}
\end{center}
\end{figure}

Figure \ref{fig:ranking} plots a scatterplot of the realized default rate against the credit score (left panel) and our predicted probability (right panel) for all quarters in the year 2008. In addition to the raw data, we also plot second-order polynomial-fitted curves to approximate the relationship. The scatter plots of realized default rates against the predictions from our hybrid model lay mostly on the 45 degree line, consistent with the very high rank correlations reported in Table \ref{tab:rank_corr}. By contrast, the relation between realized default rates and credit scores has an inverted S-shape, with the realized default rate equal to one for a large range of low credit scores and equal to zero for a large range of low credit scores, and a large variation only for intermediate credit scores.

\begin{figure}[!h]
\begin{center}

 \subfloat[Predicted Default Probability]{\includegraphics[scale=0.45]{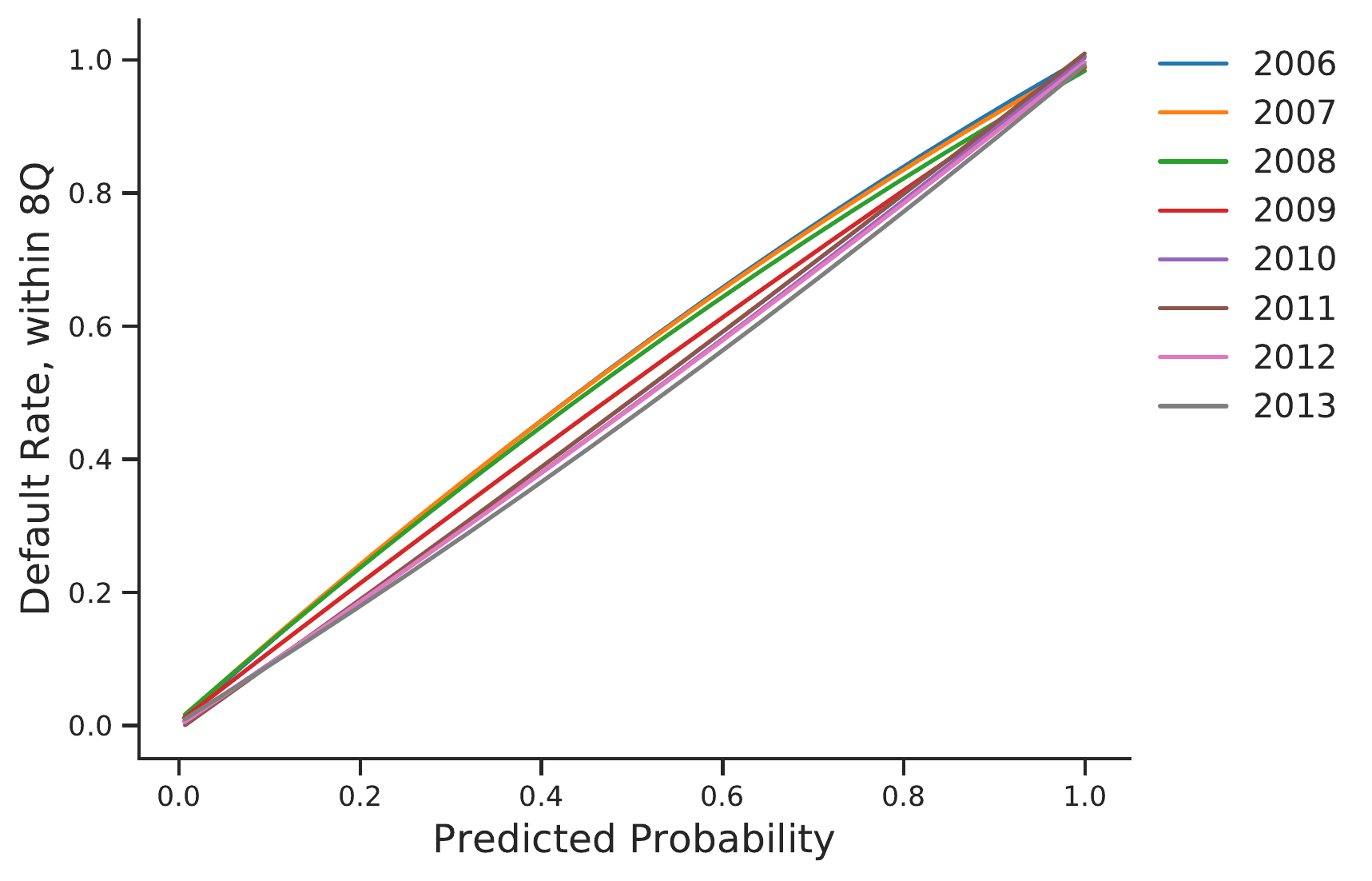}} \hskip 0.25truein
  \subfloat[Credit Score]{\includegraphics[scale=0.45]{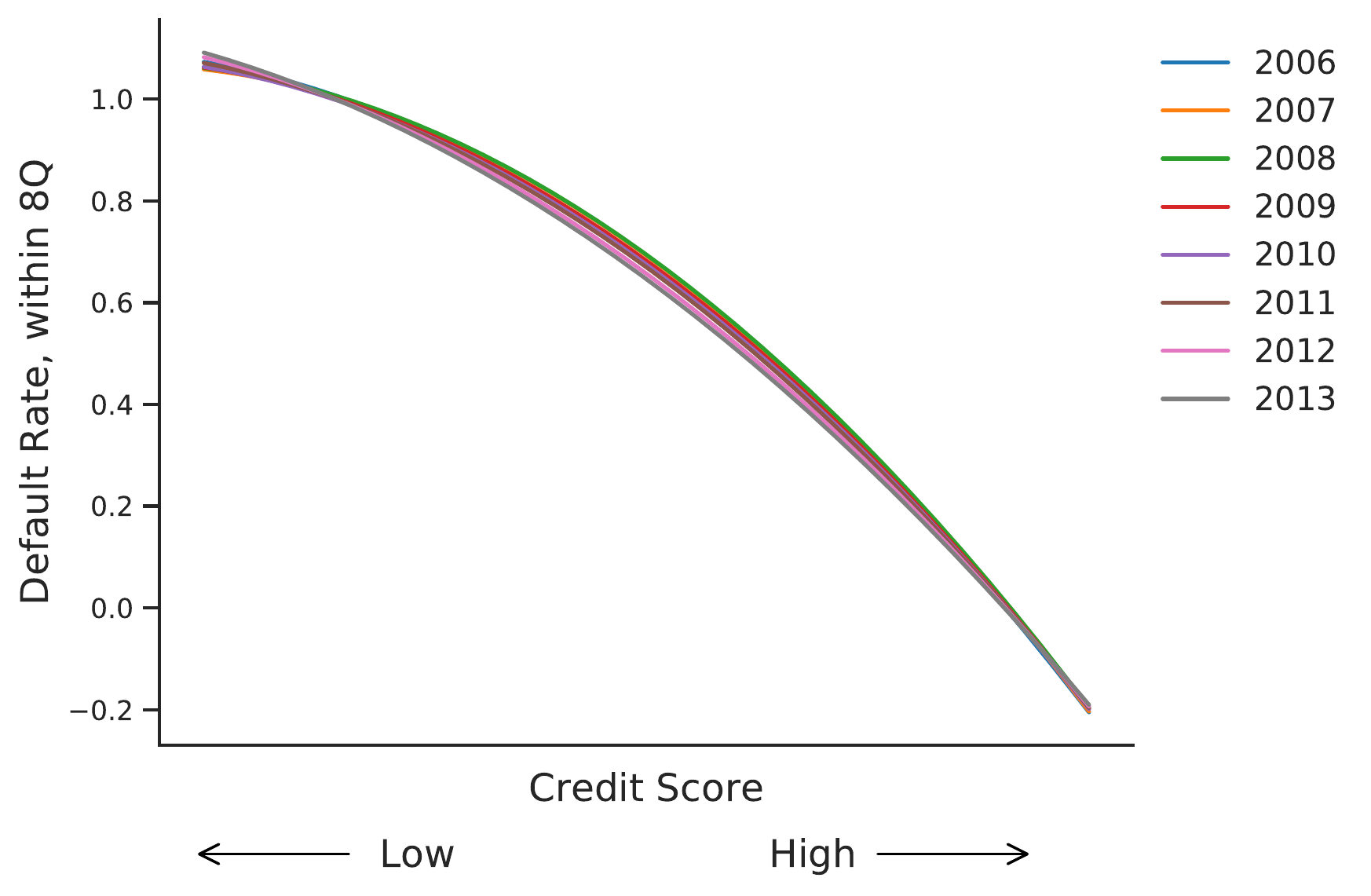}}
  
\caption{Second-order polynomial approximation of the relationship between realized default rates against model predicted default probability (a) and the credit score (b) for selected years. Source: Authors' calculations based on Experian Data.}\label{fig:polynomial}
\end{center}
\end{figure}

Figure \ref{fig:polynomial} plots second-order polynomial-fitted curves approximating the relation between realized default rates and those predicted by our model and the credit score for all  years in which our model prediction is available, starting in 2006 until 2013, to examine how the relation between realized and predicted defaults varies with aggregate economic conditions. For the years at the height of the Great Recession, the default rate seems to be somewhat higher than our model prediction, but in all years the relation is very close to a 45 degree line. By contrast, there is virtually no change in the relation between the realized default rate and the credit score. This is by construction, since the distribution of credit scores is designed to only provide a relative ranking of default risk across borrowers.\footnote{Credit scores are specifically designed to provide a stable ranking by using multiple years of data.}    This property of the credit score  implies that it is unable to forecast variations in aggregate default risk. In Section \ref{sec:macro}, we will show that our model is able to capture variations in aggregate default risk while retaining a consistent ability to separate borrowers by their individual default risk.

Table \ref{tab:rank_corr_current} reports the rank correlation with the realized default rate and the Gini coefficients by year for the credit score and the probability of default predicted by our model restricting attention to the current population, that is those borrowers who do not have any outstanding delinquencies in the quarter of interest. The rank correlation between the credit score and the realized default rate drops by 1-3 percentage points for these borrowers, whereas for our model it drops by less than a quarter of 1 percent. The Gini coefficient drops from 80-81\% to 68-69\% for the credit score and from 85-86\% to 72-74\%. These results suggest that when measured on the population of current borrowers, the performance advantage of our model relative to a conventional credit score grows.\footnote{Appendix \ref{app:score} also plots the time series of the rank correlation and the Gini coefficient for the credit score and our model for the current population. The credit score shows a large drop in these statistics for the credit score during the Great Recession whereas for our model they are both stable over time. This is consistent with the notion that the performance of the credit score dropped during the 2007-2009 period.}

\begin{table}[htbp]
\caption{Credit Score-Predicted Probability of Default Comparison-- Current Borrowers}\label{tab:rank_corr_current}
\footnotesize
\begin{center}
\begin{tabular}{lllllllll}
 Metric & 2006 & 2007 & 2008 & 2009 & 2010 & 2011 & 2012 & 2013 \\ \hline 
 & &  &  &  &  &  &  &  \\
\textbf{Rank Correlation}  &  &  &  &  &  &  &  \\
Credit Score & 0.9670 & 0.9613 & 0.9445 & 0.9653 & 0.9683 & 0.9585 & 0.9806 & 0.9559 \\
Predicted Probability & 0.9980 &	0.9983 &	0.9980 &	0.9977 &	0.9969 &	0.9974 &	0.9972 &	0.9971 \\
 &  &  &  &  &  &  &  &  \\
Credit Score & 0.6933 & 0.6935 & 0.6908 & 0.6807 & 0.6777 & 0.6833 & 0.6795 & 0.6810 \\
Predicted Probability & 0.7360 &	0.7232 &	0.7207 &	0.7183 &	0.7233 &	0.7334 &	0.7356 &	0.7389 \\ \hline 
\end{tabular}

\end{center}
\ssk

\footnotesize{Rank correlation between the credit score, predicted probability of default according to our model and subsequent realized default frequency by year for the population of current borrowers. Current borrowers do not have any delinquencies. For the credit score, report the rank correlation between each unique value of the score and the default frequency. For predicted probability of default based on our hybrid model, we first generate a number of bins equal to the number of unique credit scores in the data  and then calculate the realized default frequency for each bin. Source: Authors' calculations based on Experian Data.}

\end{table}

We examine how the ranking of borrowers varies under credit score and our model to understand the differences in performance under the two approaches. To do so, we consider the industry classification of borrowers into five risk categories Deep Subprime, Subprime,  Near Prime, Prime and Super Prime.\footnote{The threshold levels for these categories are: 1) Deep Subprime: up to 499 credit score; 2) Subprime: 500-600  credit score; 3) Near Prime: 601-660 credit score; 4) Prime: 661-780 credit score; 5) Super Prime: higher than 781 credit score.} As shown in Table \ref{tab:cs_comp}, these categories account for respectively, 6.5\%, 21.2\%, 14.1\%, 33.3\% and 24.9\% of all borrowers. We then create 5 correspondingly sized bins in our predicted probability of default with bin 1 corresponding to the 6.5\% of borrowers with the highest predicted default risk and bin 5 to the 24.9\% of all borrowers with the lowest predicted default risk. Finally, we calculate the fraction of borrowers in each credit score category that is in each of the 5 predicted default risk categories and their realized and predicted default rate.  The results are displayed in Table \ref{tab:cs_comp}. We also report the average realized and predicted default rate for each credit score category overall (columns 7 and 8) and for each predicted default risk category  for all credit score (last 5 rows). 

\begin{table}[htpb]
\caption{Comparison with Credit Score}\label{tab:cs_comp}
\begin{center}
\begin{tabular}[c]{p{1.125in}p{0.5in}p{0.25in}p{0.5in}p{0.6in}p{0.6in}p{0.6in}p{0.65in}}
\multicolumn{2}{c}{ Credit Score }  & \multicolumn{2}{c}{Predicted Default  } &  \multicolumn{2}{c}{Default Rate} &  \multicolumn{2}{c}{Average Default Rate}    \\
\multicolumn{2}{c}{  }  & \multicolumn{2}{c}{Probability  } & & &  \multicolumn{2}{c}{}   \\ 
 	&	Share 	&	 	&	 Share 	&	 Realized 	&	 Predicted 	&	 Realized	&	 Predicted	\\ 

	(1) & (2) & (3) & (4) & (5) & (6) & (7) & (8) \\
\hline

Deep Subprime & 6.48\% & 1 & 48.11\% & 99.53\% & 99.44\% & 95.45\% & 94.56\% \\
 &  & 2 & 50.93\% & 92.21\% & 90.78\% &  &  \\
 &  & 3 & 0.90\% & 64.10\% & 52.30\% &  &  \\
 &  & 4 & 0.04\% & 42.94\% & 13.02\% &  &  \\
 &  & 5 & 0.01\% & 33.82\% & 1.90\% &  &  \\ 	\hline
Subprime & 21.22\% & 1 & 14.82\% & 99.32\% & 99.29\% & 78.64\% & 77.30\% \\
 &  & 2 & 64.48\% & 84.31\% & 83.91\% &  &  \\
 &  & 3 & 16.68\% & 52.01\% & 46.43\% &  &  \\
 &  & 4 & 4.00\% & 21.97\% & 18.25\% &  &  \\
 &  & 5 & 0.02\% & 15.97\% & 2.09\% &  &  \\ 	\hline
Near Prime & 14.09\% & 1 & 1.33\% & 99.03\% & 99.14\% & 43.71\% & 42.66\% \\
 &  & 2 & 24.16\% & 76.59\% & 78.78\% &  &  \\
 &  & 3 & 42.86\% & 41.41\% & 40.43\% &  &  \\
 &  & 4 & 31.02\% & 19.74\% & 16.02\% &  &  \\
 &  & 5 & 0.63\% & 3.30\% & 2.51\% &  &  \\ 	\hline
Prime & 33.31\% & 1 & 0.08\% & 98.90\% & 99.11\% & 14.31\% & 14.44\% \\
 &  & 2 & 2.45\% & 74.60\% & 78.36\% &  &  \\
 &  & 3 & 13.20\% & 33.46\% & 36.45\% &  &  \\
 &  & 4 & 70.93\% & 10.65\% & 10.32\% &  &  \\
 &  & 5 & 13.35\% & 3.22\% & 2.41\% &  &  \\ 	\hline
Super Prime & 24.90\% & 1 & 0.00\% & 100.00\% & 99.26\% & 2.56\% & 2.64\% \\
 &  & 2 & 0.09\% & 80.55\% & 81.25\% &  &  \\
 &  & 3 & 0.22\% & 31.98\% & 34.64\% &  &  \\
 &  & 4 & 17.93\% & 5.44\% & 5.89\% &  &  \\
 &  & 5 & 81.76\% & 1.76\% & 1.75\% &  &  \\ 	\hline
All &  & 1 & 100\% & 99.41\% & 99.36\% &  &  \\
 &  & 2 &  & 83.93\% & 83.94\% &  &  \\
 &  & 3 &  & 41.65\% & 40.72\% &  &  \\
 &  & 4 &  & 11.44\% & 10.67\% &  &  \\
 &  & 5 &  & 2.03\% & 1.87\% &  &	\\ 	\hline
\end{tabular}

\end{center}

\ssk
\footnotesize{Borrowers are classified into 5 categories of default risk standard in the industry named in column (1). The threshold levels for these categories are: 1) Deep Subprime: up to 499 credit score; 2) Subprime: 500-600  credit score; 3) Near Prime: 601-660 credit score; 4) Prime: 661-780 credit score; 5) Super Prime: higher than 781 credit score. The fraction of borrowers in each category is reported in column (2). Borrowers are also assigned to 5 categories of predicted default risk based on our hybrid model from the highest default risk (1) to the lowest (5), where the share of borrowers in each predicted default risk category is the same as for the credit score categories Deep Subprime to Super Prime. For each credit score risk category the share of borrowers in each predicted default risk category is reported in column (4). Columns (5) and (6) report the corresponding realized and predicted default probability for each credit score category interacted with predicted default risk category. Columns (7) and (8) report the average realized and predicted default probability for each credit score category. All rates, fractions and shares in percentage. Total \# of observations: 17,732,772. Time period 2006Q1-2013Q4. Source: Authors' calculations based on Experian Data.}
\ssk

\end{table}

These results suggest that our model does well in predicting the default probability of borrowers in all categories, with a slight tendency to under-predict the probability of default by 1-2 percentage points for the Deep Subprime and Subprime borrowers. The majority of Deep Subprime borrowers fall in the two lowest categories of predicted default risk. For Subprime borrowers, 64\% fall into the corresponding second category of default risk, while 15\% fall in the first and 17\% into the third. This corresponds to a sizable discrepancy as the average realized default probability for Subprime borrowers is 79\%, whereas it is 99\% for those in the first category and only 52\% for those in the third. By contrast, the predicted default risk is very close to the realized default risk for Subprime borrowers in all categories of the predicted default risk distribution, with a discrepancy under 1 percentage point for predicted default risk category 1 and 2, and around 5 percentage points for category 3 and 4.
Near Prime borrowers also display a wide dispersion across predicted default risk categories with only 43\% falling into the corresponding third category, 24\% falling in category 2 (higher default risk) and 31\% falling into category 4 (lower default risk).  Again, the realized default rates vary substantially for Near Prime borrowers by predicted default risk category, from 77\% in category 2, to 41\% and 20\% in category 3 and 4, respectively, while the predicted default risk in much closer to the realized, with a maximum 4 percentage point discrepancy.
The discrepancy in classification for the credit score are lower for Prime and Super Prime borrowers. 13\% of Prime borrowers fall into category 3 (higher default risk), 13\% in category 5 (lower default risk) and 71\% in the corresponding category 4. The realized default rates are 11\% for Prime borrowers in category 4, and 34\% and 3\% respectively for Prime borrowers in category 3 and 5. Only 18\% of Super Prime borrowers fall in category 4 of predicted default risk (higher risk) and 82\% fall in the corresponding category 5. Moreover, the differences in realized default risk between these categories are minor, with a realized default rate of 5\% and 2\% for categories 4 and 5, respectively.
These results suggest that credit scores misclassify borrowers across risk categories with very different realized default rates. By contrast, as shown in the bottom 5 rows of Table  \ref{tab:cs_comp} and by columns (6) and (8), our model is very successful at predicting the default rate for borrowers irrespective of their credit score.

We next investigate feature attributions differences across credit scores and our hybrid model. We grouped our features into five categories to correspond to information we obtained from marketing resources for the credit score, and aggregated the absolute value of the SHAP values for each instance across each categories across our testing dataset for the pooled model. These categories are, payment history, amount owed, length of credit history, credit mix and new credit. Their contribution towards credit scores is reported in Table \ref{tab:cs_vs_pp}. 

Contrary to credit scores, features relating to credit inquiries and debt products account for 10\% of the total variation in predicted probabilities for our hybrid model. The aggregate impact of these two factors is exactly half of the variation they explain of credit scores, which can partly be attributed to the low number of features we include in our models pertaining to new credit and credit mix. However, even their per feature contribution is lower for our hybrid model.\footnote{The per feature contribution for credit inquiries and debt products are 0.004 and 0.005 respectively, contrasted with 0.007, 0.008 for payment history and amounts owed.} Next, payment history and length of credit account 30\% and 10\% of the variation in predicted probabilities, each being 5\% short of their contribution towards credit scores. Perhaps most strikingly, accounts owed explain 50\% of the variation for our hybrid model, while only 30\% for credit scores. This exercise once again illustrates that features relating to debt balances are the most important determinants for our model's output, contrasted with credit scores, where payment history is registered as the most important predictor.

\begin{table}[!h]
\caption{Credit Score vs. Hybrid Model: Model Explanation}\label{tab:cs_vs_pp}
\begin{center}
\begin{tabular}{lccc}
 & & \multicolumn{2}{c}{Model} \\ 
 & &  Hybrid & Credit Score \\ 
Feature Group & \# of Features &  \\ \hline
& & & \\
Payment History & 42 & 0.30 & 0.35 \\
Amounts Owed & 65 &  0.50 & 0.30 \\
Length of Credit & 11 &  0.1 & 0.15 \\
Credit Mix & 16 & 0.08 & 0.10 \\
New Credit & 5 &  0.02 & 0.10 \\ \hline 
\end{tabular}
\end{center}

\ssk
\footnotesize{
This table reports the Shapley values for five feature groups across four models. For each prediction window, we compute the Shapley value for each of the observations and for each feature. We then calculate the sum of the absolute value for each feature, aggregate it across the feature groups and report the results for the group. We normalized the results so that for each model the four groups sum up to 1. Source: Authors' calculations based on Experian data.}
\end{table}

\subsubsection{Coverage}\label{sec:unscored}
Consumers with limited credit histories encounter severe difficulties in accessing credit markets. As explained earlier in the paper, lenders often rely on credit scores to make lending decisions. If a borrower's credit report does not have sufficient information to evaluate their default risk, lenders are unlikely to grant credit. Consequently, consumers with limited credit histories have a hard time accessing credit markets.      
These consumers can be divided into two groups. The first group consists of individuals without credit records, often referred to in the industry as "credit invisibles." The second group are those consumers who do have credit records, but are considered "unscorable," that is they have insufficient credit histories to generate a credit score. \citeN{CFPB_2016_unscored} find that 11\% of the US population lacks credit records and an additional 8.3\% have a credit record but do not have a credit score.

A credit record may be considered unscorable for two reasons:  (1) it contains insufficient information to generate a score, meaning the record either has too few accounts or has accounts that are too new to contain sufficient payment history to calculate a reliable credit score; or (2) it has become stale in that it contains no recently reported activity. The exact definition of what constitutes insufficient or stale information differs across credit scoring models, as each model uses its own proprietary definition.\footnote{The FICO score has the most restrictive requirements and does not score borrowers who show no updates or reports on  credit file in past 6 months, or no accounts at least 6 months old. The rest of the industry has been trying to expand the universe of scorable borrowers and typically adopts a more flexible approach to increase the ranks of borrowers who are scored. Vantage Score has been particularly pushing the need to expand the universe of scorable customers and they have implemented several successful changes to the most recent version of their scoring model that have indeed substantially decreased the number of unscored consumers. For more information, see \url{https://www.vantagescore.com/resource/174/scoring-credit-invisibles-using-machine-learning-techniques}.}

The challenges that credit invisibles and unscored consumers  face in accessing credit markets has generated considerable attention from researchers and industry participants. \citeN{CFPB_2016_unscored} show that young, minority and low income borrowers are disproportionately represented among the unscored. Several studies have explored the potential of various types of alternative data to supplement the information contained in credit reports and allow credit scores to be generated for these consumers.\footnote{See for example \citeN{jagtiani2018roles}. For an industry perspective, see \citeN{Oliver_Wyman_2017alternative}.} 
Our model generates a predicted probability of default for every individual in our sample without an empty credit record, so effectively there are no active borrowers that we do not score. Our ability to score every active borrower  is partly due to the fact that our model does not use any lagged observations. As previously discussed, many of the features in our model have a temporal dimension which renders the use of lags unnecessary. This constitutes an additional advantage of our model when compared to traditional credit scores. 

\subsection{Predicting Systemic Risk}\label{sec:macro}
\begin{figure}[htbp]
\begin{center}
\subfloat[Predicted and realized aggregate default rate]{\includegraphics[scale=0.5]{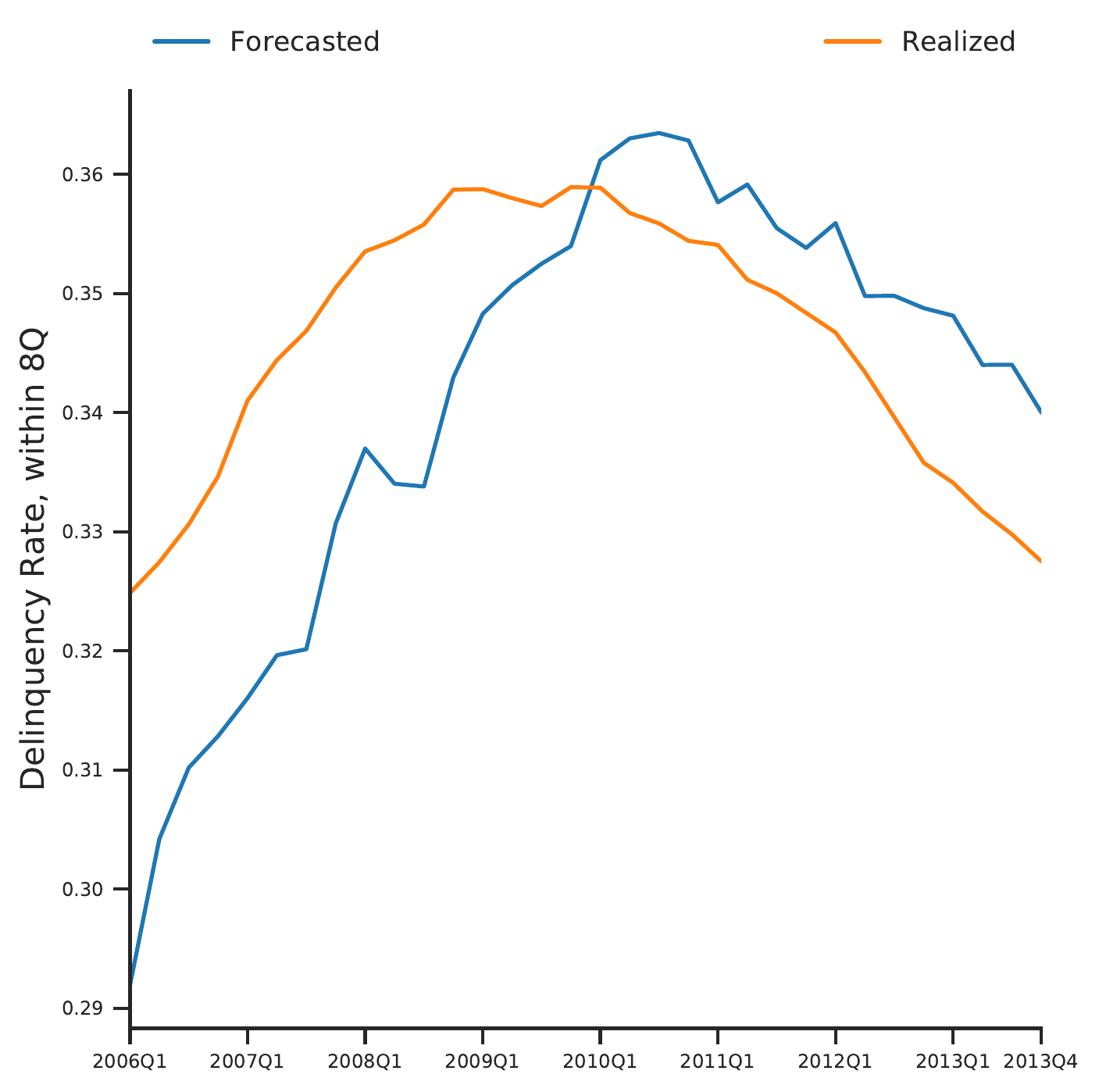}}
\subfloat[Correlation between predicted and realized]{\includegraphics[scale=0.525]{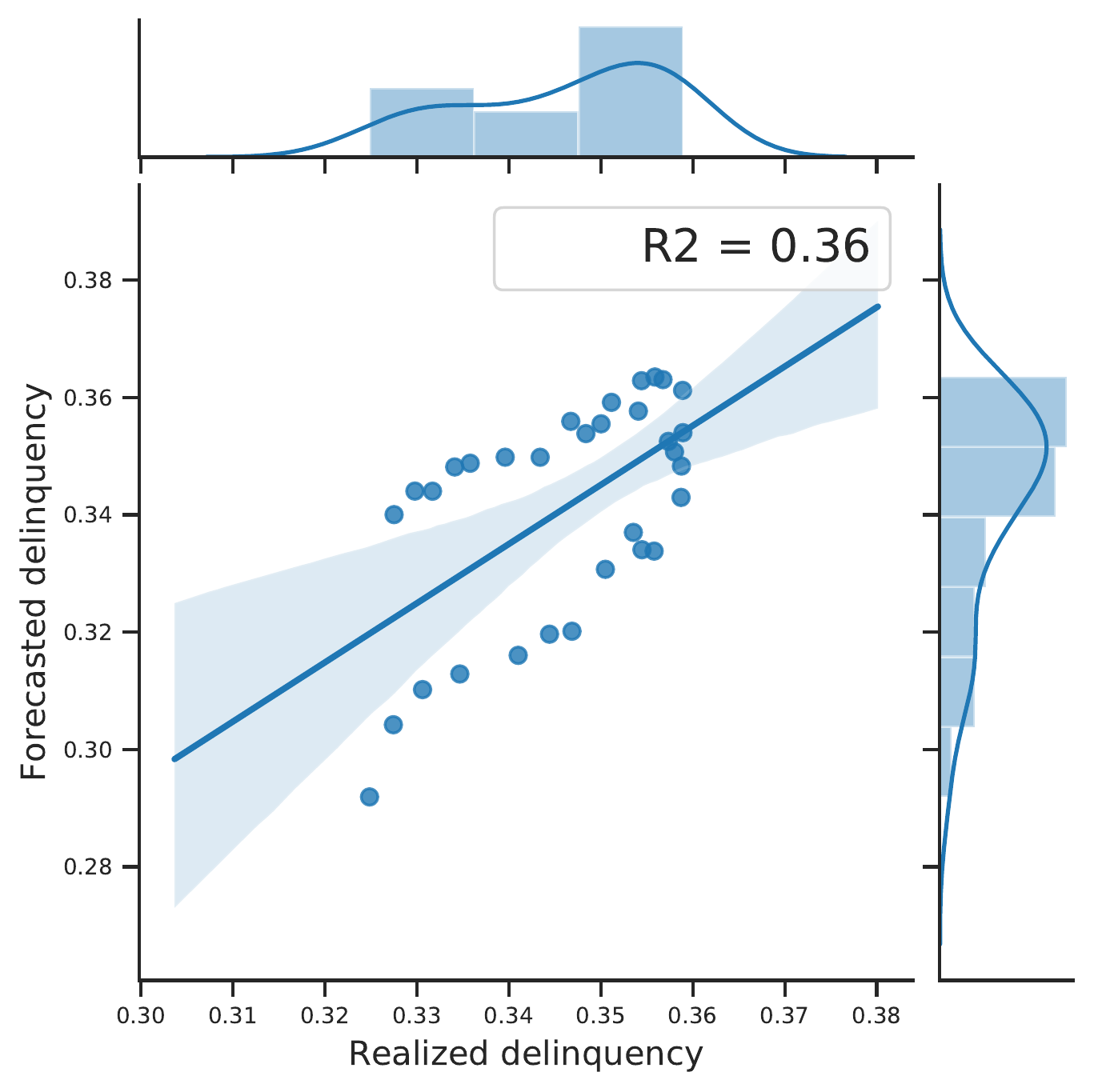}}

\end{center}

\caption{Fraction of consumers with 90+ days delinquency within the subsequent 8 quarters, predicted by our hybrid DNN-GBT model and realized. Aggregate default rates are obtained by averaging across all consumers in each period. Source: Authors' calculations based on Experian Data.}\label{fig:macro}

\end{figure}

We next turn to analyze the aggregate forecasting power of our hybrid model. We aggregate the deep-learning forecasts for individual accounts to generate macroeconomic forecasts of credit risk by taking the average of the predicted probabilities over a given forecast period. Since our sample of consumers in nationally representative in each quarter, this will provide an unbiased estimate of the aggregate default risk predicted by our model.   We calculate the aggregate default probability for 2006Q1-2013Q4, and show that our model is able to predict the spike in delinquencies during the 2007-2009 financial crisis and also the reduction in delinquencies since then. This estimate of aggregate default risk could be used as a proxy of systemic risk in the household sector. 
The results are displayed in Figure \ref{fig:macro}.
Panel (a) plots the aggregate predicted default rate from our hybrid model and compares it to the aggregate realized default rate. While our predicted aggregate default rate is approximately 2 percentage points lower than the realized in 2006 and 2007, it rises at a similar speed as the realized default rate. It peaks in 2010Q2, approximately 2 quarters after the peak in the realized rate and then declines in the ensuing period, again reflecting the behavior of the realized rate, though it overestimates it by about 1 percentage point. Panel (b) shows a scatter plot of the predicted aggregate default rate against the realized for the different quarters in our sample period. The correlation between the predicted and realized aggregate default rate is 36\%. 

\subsection{Value Added}\label{sec:value_added}
We assess the economic salience of our hybrid DNN-GBT model by analyzing its value added for lenders and borrowers. For lenders, we examine the role our model can play in minimizing the losses from default. For borrowers, we calculate the interest savings for borrowers who are misclassified as having an excessively high probability of default based on the credit score compared to our model.  

\subsubsection{Lenders}\label{sec:value_added_lenders}
\begin{figure}[htbp]
\begin{center}
\subfloat[Hybrid vs. No Forecast]{\includegraphics[scale=0.33]{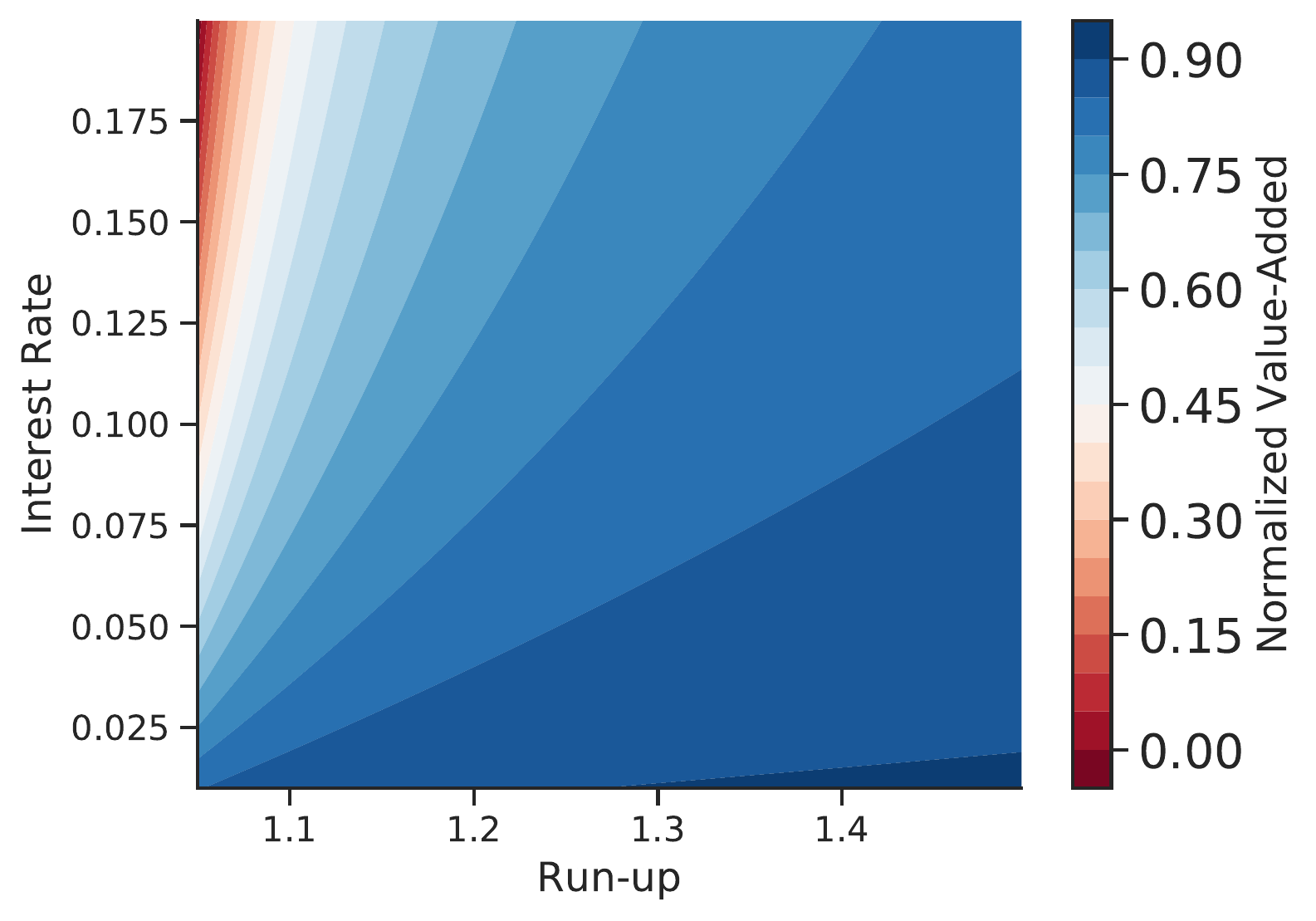}}
\subfloat[Hybrid vs. Logistic]{\includegraphics[scale=0.33]{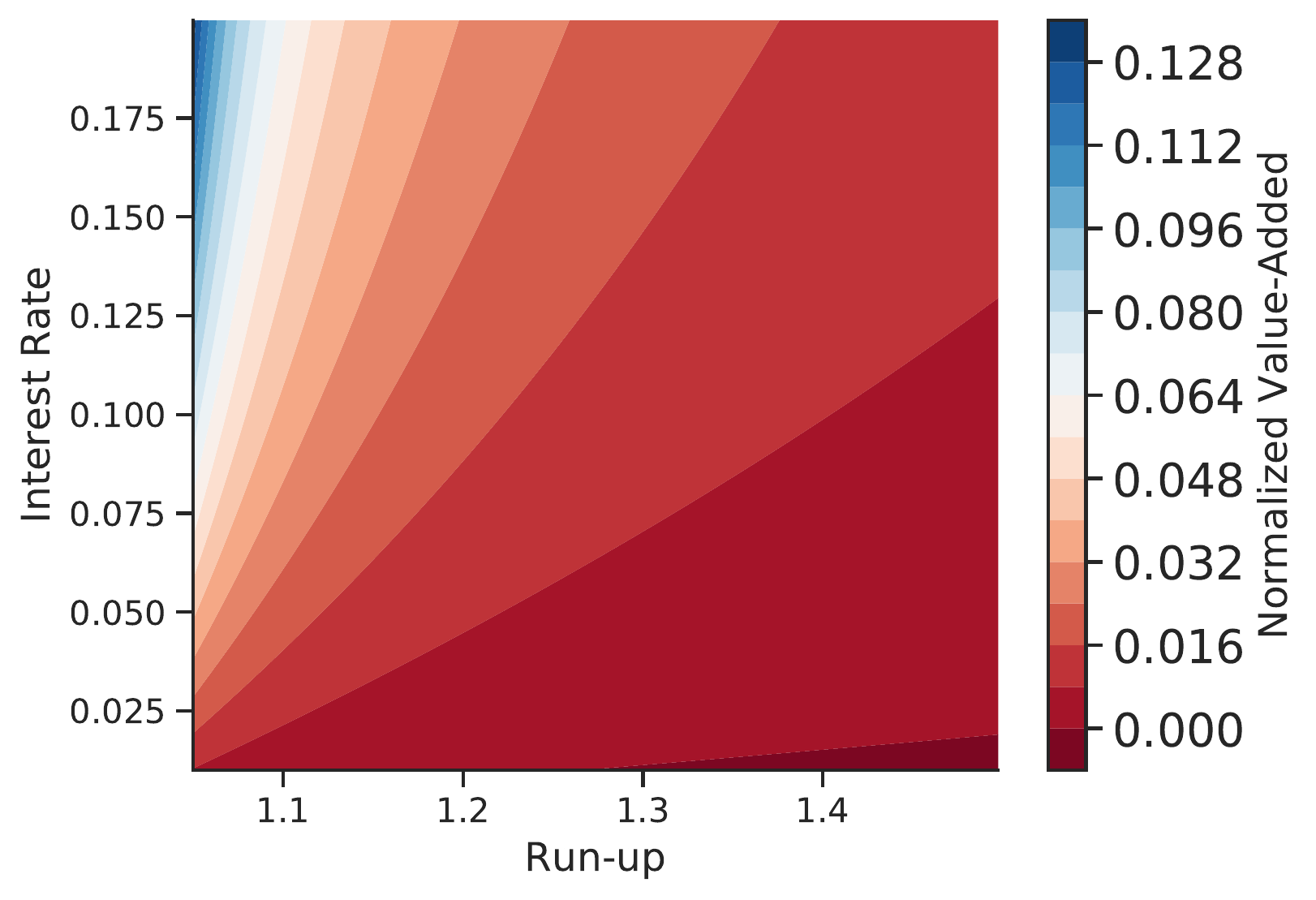}}
\subfloat[Hybrid vs. RF]{\includegraphics[scale=0.33]{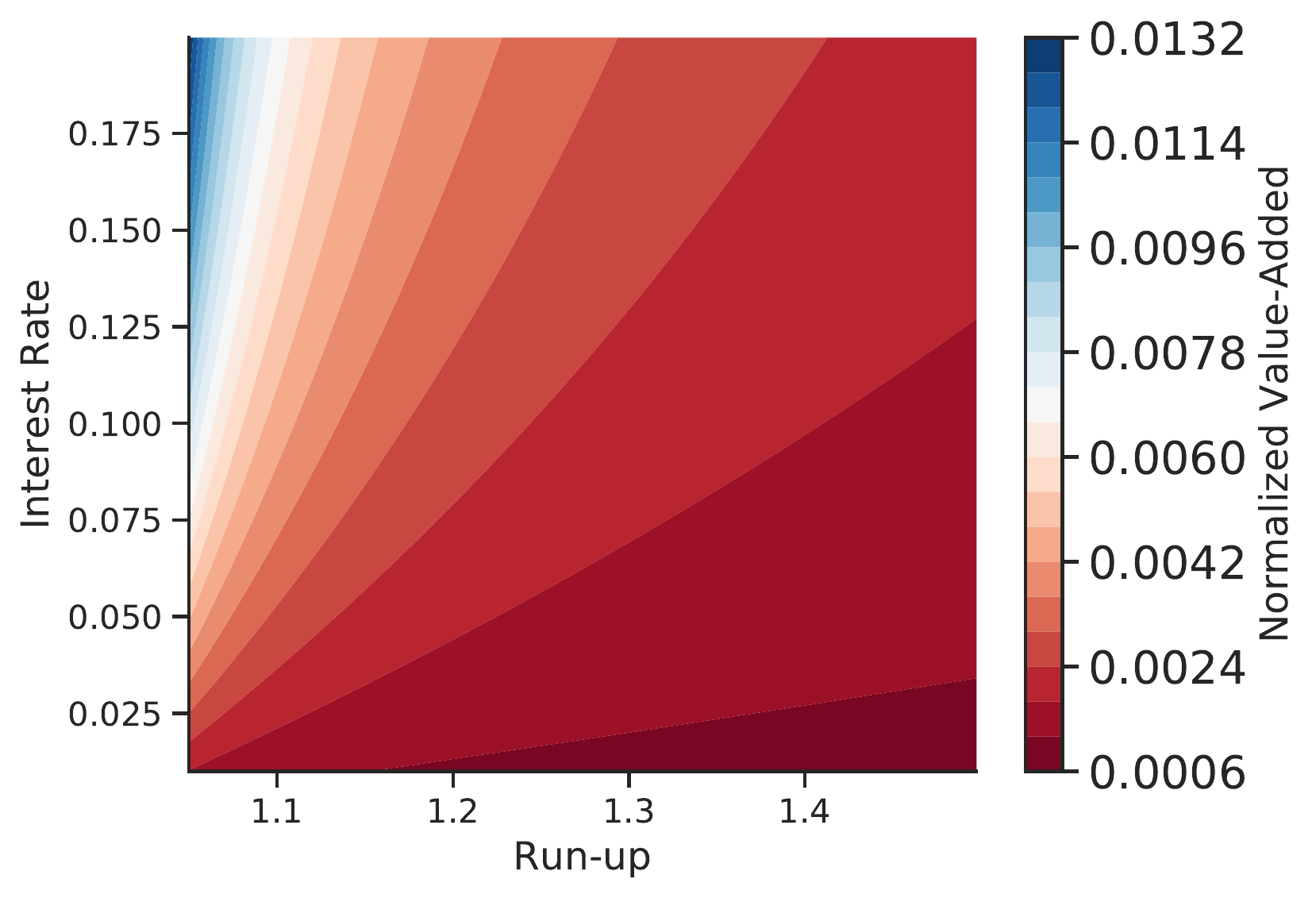}}
\end{center}

\caption{Value-added of machine-learning forecasts of 90+ days delinquency over
8Q forecast horizons on data from 2012Q4. VA values are calculated with amortization period N = 3 years and a 50\% classification threshold. Source: Authors' calculations based on Experian Data.}\label{fig:value_added}

\end{figure}

We follow the framework proposed by \citeN{khandani}, which compares the value of having a prediction of default risk to having none, and we make the same simplifying assumptions with respect to the revenues and costs of the consumer lending business. Specifically, in absence of any forecasts, it is assumed a lender will take no action regarding credit risk, implying that customers who default will generate losses for the lender, and customers who are current on their payments will generate positive revenues from  financing fees on their running balances. To simplify, we assume that all defaulting and non-defaulting customers have the same running balance, $B_r$, but defaulting customers increase their balance to $B_d$ prior to default. We refer to the ratio between $B_d$ and $B_r$ as "run-up." It is assumed that with a model to predict default risk, a lender can avoid losses of defaulting customers by cutting their credit line and avoiding run-up. Then, the value added as proposed by \citeN{khandani} can be written as follows:
\begin{equation}
    VA(r,N,TN,FN,FP) = \frac{TN - FN \big[1-(1+r)^{-N}\big] \big[\frac{B_d}{B_r}-1\big]^{-1}}{TN+FP}
\end{equation}
where $r$ refers to the interest rate, $N$ the loan's amortization period, and $TN,FN,FP$ refer to true negatives, false negatives and false positives respectively. Panel (a) of Figure \ref{fig:value_added} plots the Value Added (VA) as a function of interest rate and the ratio of run-up balance for our out-of-sample forecasts of 90+ days delinquencies over the subsequent 8 quarters for 2012Q4. These  estimates imply cost savings of over 60\% of total losses when compared to having no forecast model for a run-up of 1.2 at a 10\% interest rate for an amortization period for 3 years.      

We next compare the value added of our hybrid model with default predictions generated by a logistic regression. This exercise illustrates the gains from adopting a better technology for credit allocation. Panel (b) of Figure \ref{fig:value_added} shows more modest, but substantial cost savings in the range of 1-9\% and approximately 5\% for a 1.2 run-up at a 10\% interest rate with a 3 year amortization period. Panel (c) calculates the cost savings associated to using our hybrid model in comparison to random forest. In this case the cost savings range from 0.1-1.3\%. This exercise then confirms the advantages of using deep learning over other technologies in predicting default.

\subsubsection{Borrowers}\label{sec:value_added_consumers}
We now examine the potential cost savings for consumers who would be offered credit according to the predicted default probability implied by our model instead of a conventional credit score. Following our approach in Section \ref{sec:credit_score_comp}, we create credit score categories based on common industry standards  and corresponding predicted probability bins with the same number of observations, and we place customers in these bins based on their credit score at account origination for each of their credit cards.\footnote{We look at the months since most recently opened credit card account to infer account origination. We drop customers with months since the most recently opened credit card greater than 72.} The distribution of customers is summarized in Table \ref{tab:value_consumer_shares} in Appendix \ref{app:cost_savings}. We then follow the information on interest rates by credit score category in Table 2 in \citeN{Stroebel_2015} to obtain the cost of credit on credit card balances. \footnote{Credit card interest rates are notoriously invariant to overall changes in interest rates, so the calculations reported in this section apply irrespective of the time period. See \citeN{ausubel1991failure} and \citeN{calem1995consumer}.} 
  To obtain the cost savings for consumers, we use the difference in interest rates by credit score category based on how they would be classified according to our model.\footnote{We draw interest rates from a truncated normal distribution with mean and standard deviation as in \citeN{Stroebel_2015}.} For customers who are placed in higher risk categories by the credit score compared to our predicted probability of default, interest rates on credit cards are higher than they would have been if they had been classified according to our model. Thus, using our model to score consumers rather than the credit score would generate the cost savings for them. For customers placed in risk categories by the credit score that are too low relative to the default risk predicted by our model, interest rates will be higher under our model. The calculation is made for each individual consumer. The average  for each credit score category is then computed and reported in Table \ref{tab:value_consumer_desc}. The information on interest rates and balances, and the dollar value of cost savings for different credit card categories is reported in Table \ref{tab:value_consumer_savings}. We report this in percentage of credit card balances in the top panel and in current USD terms in the bottom panel. The largest gains accrue to customers with Subprime and Near Prime credit scores. As we showed in Section \ref{sec:credit_score_comp}, they are more likely to be attributed a probability of default by the credit score that is too low compared to our model predictions. Additionally, the biggest variation in credit card interest rates occurs across Subprime and Near Prime borrowers in comparison to Prime based on \citeN{Stroebel_2015}. The cost savings for these borrowers average out to 4-5\% of total credit card balances of \$1,088-1465. Gains for Prime and Super Prime borrowers who are attributed a lower default probability by our model are very modest, as credit card interest rates vary little by credit score for Prime and Superprime borrowers. On the other hand, Prime and Superprime borrowers who are placed in group 1, corresponding to the highest predicted default probability based on our model face, substantial losses in the order of 4-5\% of total credit card balances or \$269-412. The cumulated interest rate cost savings across all consumers in our sample is \$754,527,819, which amounts to \$43 per capita.

\begin{table}[!h]
\caption{Cost of Credit Risk Misclassification}
\small 
\label{tab:value_consumer_savings}
\begin{center}
\begin{tabular}[c]{p{1.5in}p{0.75in}p{0.75in}p{0.75in}p{0.75in}p{0.85in}}
 &  & \multicolumn{4}{c}{Credit Score} \\ \hline
   &  &  &  &  &  \\
& & Subprime, Near Prime & Prime Low & Prime Mid &  Prime High, Superprime \\  \hline
  &  &  &  &  &  \\
 & \multicolumn{5}{c}{\textbf{Annual Interest Rate Savings}} \\
    &  &  &  &  &  \\
Predicted Default Bin & 1 & 0.00\% & -5.13\% & -4.28\% & -4.92\% \\
 & 2 & 5.12\% & 0.00\% & 0.85\% & 0.21\% \\
 & 3 & 4.29\% & -0.85\% & 0.00\% & -0.65\% \\
 & 4 & 4.93\% & -0.21\% & 0.64\% & 0.00\% \\ 
  &  &  &  &  &  \\
 & \multicolumn{5}{c}{\textbf{Annual Average Cost Saving (\$)}} \\
    &  &  &  &  &  \\
Predicted Default Bin & 1 & 0 & -412 & -269 & -301 \\
 & 2 & 1088 & 0 & 81 & 14 \\
 & 3 & 1465 & -178 & 0 & -52 \\
 & 4 & 1279 & -47 & 118 & 0 \\ \hline
\end{tabular}
\end{center}

\ssk
\footnotesize{
This table reports the average cost savings for consumers across credit score and predicted default probability bins for our sample. The cumulative savings for consumers on both credit card and bankcard debt adds up to \$754,527,819. Time period 2006Q1-2013Q4. Source: Authors' calculations based on Experian Data. }
\end{table}

This calculation provide us with a lower bound for the cost savings of being classified according to our model in comparison to the credit score, as they do not take into account the higher credit limits and potential behavioral responses of customers faced with higher borrowing capacity and lower interest rates. As shown in \citeN{Stroebel_2015}, changes in the cost of funds for lenders mainly translate into changes in credit limits and exclusively for higher credit score borrowers. Therefore, being placed in a higher risk category for consumers also inhibits their ability to benefit from expansionary monetary policy. Additionally, we do not take into account the fact that more expensive credit in the form of higher interest rate costs makes it more likely that the consumer will incur missed payments in response to temporary changes in income. Fees for missed payments constitute a substantial component of credit card costs for consumers, and the ability to avoid these fees would contribute to substantial cost savings for consumers (see \citeN{agarwal2014regulating}).

\section{Conclusion}
We have proposed to use deep learning to develop a model to predict consumer default. Our model uses the same data used by conventional scoring models and abides with all legislative restrictions in the United States. We show that our model compares favorably to conventional credit scoring models in ranking individual consumers by their default risk, and is also able to capture variations in aggregate default risk. Our model is interpretable and allows to identify the factors that are most strongly associated with default. Whereas conventional credit scoring models emphasize utilization rates, our analysis suggests that the number and balances on open trades are the factors which associate more strongly to higher default probabilities. Our model is able to provide a default prediction for all consumers with a non-empty credit record. Additionally, we show that our hybrid DNN-GBT model performs better than standard machine learning models of default based on logistic regression and can accrue cost saving to lenders in the order of 1-9\% compared to default predictions based on logistic regression, as well as interest rate cost savings for consumers of up to \$1,465 per year.

\bibliography{main}{}
\bibliographystyle{achicago}

\clearpage

\end{onehalfspacing}

\newpage
\appendix
\renewcommand{\baselinestretch}{1}

\clearpage

\appendix
\section*{Appendix}
\addcontentsline{toc}{section}{Appendices}
\renewcommand{\thesubsection}{\Alph{subsection}}
\section{Performance Metrics}\label{app:performance_defs}
Suppose a binary classifier is given and applied to a sample of N observations. For each instance i, let $y_i$ denote the true outcome.  For each observation, the model generates a probability that an observation with feature vector $x_i$ belongs to class 1. This predicted probability, $f(x_i)$ is then evaluated based on a threshold to classify observations into class 1 or 0. Given a threshold level (c), let True Positive (TP) denote the number of observations that are correctly classified as type 0, True Negative (TN) be the number of observations that are correctly classified as type 1, False Positive (FP) be the number of observations that are type 1 but incorrectly classified as type 0, and, finally, False Negative (FN) be the number of observations that are actually of type 0 but incorrectly classified as type 0. Based on these definitions, one can define the following metrics to assess the performance of the classifier:
\begin{equation}
    \textrm{True Negative Rate (TNR)} \equiv \frac{\textrm{TN}}{\textrm{TN+FP}}
\end{equation}
\begin{equation}
    \textrm{False Positive Rate (FPR)} \equiv \frac{\textrm{FP}}{\textrm{FP+TN}}
\end{equation}
\begin{equation}
    \textrm{Precision}   \equiv \frac{\textrm{TP}}{\textrm{TP + FP}}
\end{equation}
\begin{equation}
    \textrm{Recall}   \equiv \frac{\textrm{TP}}{\textrm{TP + FN}}
\end{equation}
\begin{equation}
    \textrm{F-measure}   \equiv \frac{\textrm{2 $\times$ Recall $\times$ Precision}}{\textrm{Precision + Recall}}
\end{equation}
\begin{equation}
    \textrm{Accuracy}   \equiv \frac{\textrm{TP+TN}}{\textrm{TP+TN+FP+FN}}
\end{equation}
\begin{equation}
    \textrm{Youden's J statistic}   \equiv \frac{\textrm{TP}}{\textrm{TP + FN}} + \frac{\textrm{TN}}{\textrm{TN + FP}} - 1
\end{equation}
\begin{equation}
    \textrm{ROC AUC}   = \int_{\infty}^{-\infty} \textrm{TPR}(c) \textrm{FPR}'(c) \textrm{dc}
\end{equation}
\begin{equation}
   \textrm{Cross-entropy loss} = -\frac{1}{N} \sum_{i=1}^{N} (y_i \cdot \log(f(x_i)) + (1 - y_i) \cdot  \log(1 - f(x_i))
\end{equation}
\begin{equation}
    \textrm{Brier score}   = \frac{1}{N} \sum_{i=1}^{N} (f(x_i) - y_i)^2
\end{equation}

\section{Data Pre-Processing}\label{sec:data_processing}
Our original dataset contains 33,600,000 observations. We discard observations of individuals with missing birth information, deceased individuals and restrict our analysis to individuals aged between 18 and 85, residing in one of the 50 states or the District of Columbia, with 8 consecutive quarters of non-missing default behavior. This leaves us with 22,004,753 data points. Our itemized sample restrictions are summarized in Table \ref{tab:restrictions} below.      

\begin{table}[htbp]
\caption{Itemized Sample Restrictions}\label{tab:restrictions}
\normalsize
\begin{center}
\begin{tabular}{lll}
 &  & Observations \\ \hline 
   &  &  \\
 Credit Report Data &  & 33,600,000 \\ \hline
     &  &  \\
  Remove & &  \\ \hline 
   &  &  \\
Deceased &  & - 513,270 \\
Age &  & - 4,718,804 \\
Residence &  & - 953,215 \\ 
Prediction Window & & - 5,409,958 \\ \hline 
   & & \\
Prediction Sample & & 22,004,753 \\ \hline \hline 
\end{tabular}
\end{center}
\end{table}

\subsection*{Feature Scaling}
We normalize all explanatory variables by their means and standard deviations:
\begin{equation}
    z_i = \dfrac{x_i-\mu_x}{\sigma_x}
\end{equation}
where  $\mathbf{x} = (x_1,x_2,\ldots,x_k)$, and $\mathbf{z_i}$ is the $i^{\textrm{th}}$ normalized data.

\subsection*{Train-Test Split}
For most of our analysis we split the data to account for look-ahead bias, i.e., the training set consists of data 8Q prior to the testing data. Then, we scale the testing data by the mean and standard deviation of the training data.
In an alternative specification, we split our pooled data into three chunks: training set (60\%), holdout set (20\%), and testing set (20\%). We report each specifications in Table \ref{tab:full_sample} - Table \ref{tab:current}. Except for parts of Section \ref{sec:interpretation}, we used the predictions generated by our models on the temporal splits.    

In each specifications, we randomly shuffled the data to ensure that the mini-batch gradients are unbiased. If gradients are biased, training may not converge and accuracy may be lost.

\section{Model Estimation}\label{app:estimation}
Our estimation consists of four steps. First, we specify the loss function. Second, we choose the optimization algorithm. Third, we optimize the hyperparameters of the model. Fourth, we train our models.

\subsection*{Loss Function}
Suppose $\mathbf{y}$ is the ground truth vector of default, and $\mathbf{\hat{y}}$ is the estimate obtained directly from the last layer given input vector $\mathbf{x} = (x_1,x_2,\ldots,x_k)$. By construction, $y_i = \{0,1\}$ and $\hat{y_i} \in [0,1]$. We minimize the categorical cross-entropy loss function\footnote{Loss function measures the inconsistency between the predicted and the actual value. The performance of a model increases as the loss function decreases. There are several other types of loss functions, including mean squared error, hinge, and Poisson. The categorical cross-entropy is often used for classification problems.} to estimate the parameter specified in (7). We do this by choosing $\theta$ that minimizes the distance between the predicted $\mathbf{\hat{y}}$ and the actual $\mathbf{y}$ values. Given N training examples, the categorical cross-entropy loss can be written as:

\begin{equation}
   L(\hat{y}, y) = -\frac{1}{N} \sum_{i=1}^{N} (y_i \cdot log(\hat{y_i}) + (1 - y_i) \cdot  log(1 - \hat{y_i})
\end{equation}

We apply an iterative optimization algorithm to find the minimum of the categorical cross-entropy loss function. We next describe this algorithm.

\subsection*{DNN Optimization Algorithm}
Deep learning models are computationally demanding due to their high degree of non-linearity, non-convexity and rich parameterization. Given the size of the data, gradient descent is impractical. We follow the standard approach of using stochastic gradient descent (SGD) to train our deep learning models (see \citeN{goodfellow}). Stochastic gradient descent is an iterative algorithm that uses small random subsets of the data to calculate the gradient of the objective function. Specifically, a subset of the data, referred to as a mini-batch (the size of the mini-batch is called the batch size), is loaded into memory and the gradient is computed on this subset. The gradient is then updated, and the process is repeated until convergence.      

We adopt the Adaptive Moment Estimation (Adam), a computationally efficient variant of the SGD introduced by (see \citeN{kingma}) to train our neural networks. The Adam optimization algorithm can be summarized as follows: 

\begin{enumerate}
    \item Fix the learning rate $\alpha$, the exponential decay rates for the moment estimates: $\beta_1$,$\beta_2 \in [0,1)$, and the objective function. Initialize the parameter vector $\theta_0$, the first and second moment vector $m_0$ and $v_0$ respectively, and the timestep t.
    \item While $\theta_t$ does not converge, do the following:
    \begin{enumerate}
        \item Compute the gradients with respect to the objective function at timestep t:
        \begin{equation}
            g_t = \nabla_\theta f_t(\theta_{t-1})
        \end{equation}
        \item Update the first and second moment estimates:
        \begin{equation}
            m_t = \beta_1 \cdot m_{t-1} + (1 - \beta_1) \cdot g_t
        \end{equation}
        \begin{equation}
            v_t = \beta_2 \cdot v_{t-1} + (1 - \beta_2) \cdot g_t^2
        \end{equation}
        \item Compute the bias-corrected first and second moment estimates:
                \begin{equation}
            \hat{m_t} = \frac{m_t}{1-\beta_1^t}
        \end{equation}
        \begin{equation}
            \hat{v_t} = \frac{v_t}{1-\beta_2^t}
        \end{equation}
        \item Update the parameters:
        \begin{equation}
            \theta_t = \theta_{t-1} - \dfrac{\alpha \cdot \hat{m_t}}{\sqrt{\hat{v_t}} + \epsilon}
        \end{equation}
    \end{enumerate}
    
\end{enumerate}

The hyperparameters have intuitive interpretations and typically require little tuning. We apply the default setting suggested by the authors of \citeN{kingma}, these are $\alpha = 0.001$, $\beta_1 = 0.9$, $\beta_2 = 0.999$ and $\epsilon = 10^{-7}$.    

\subsection*{GBT Algorithm}
Fit a shallow tree (e.g., with depth L = 1). Using the prediction residuals from the first tree, fit a second tree with the same shallow depth L. Weight the predictions of the second tree by $\nu \in (0,1)$ to prevent the model from overfitting the residuals, and then aggregate the forecasts of these two trees. Until a total of K trees is reached in the ensemble, at each step k, fit a shallow tree to the residuals from the model with k-1 trees, and add its prediction to the forecast of the ensemble with a shrinkage weight of $\nu$.

\subsection*{Regularization}
Neural networks are low-bias, high-variance models (i.e., they tend to overfit to their training data). We implement three routines to mitigate this. First, we apply dropout to each of the layers (see \citeN{srivastava}). During training, neurons are randomly dropped (along with their connections) from the neural network with probability p (referred to as the dropout rate), which prevents complex co-adaptations on training data.     

Second, we implement "early stopping", a general machine learning regularization tool. After each time the optimization algorithm passes through the training data (i.e., referred to as an epoch), the parameters are gradually updated to minimize the prediction errors in the training data, and predictions are generated for the validation sample. We terminate the optimization when the validation sample loss has not decreased in the past 50 epochs. Early stopping is a popular substitute to l2 regularization, since it achieves regularization at a substantially lower computational cost.     

Last, we use batch normalization (see \citeN{ioffe}), a technique for controlling the variability of features across different regions of the network and across different datasets. It is motivated by the internal covariate shift, a phenomenon in which inputs of hidden layers follow different distributions than their counterparts in the validation sample. This problem is frequently encountered when fitting deep neural networks that involve many parameters and rather complex structures. For each hidden unit in each training step, the algorithm cross-sectionally de-means and variance standardizes the batch inputs to restore the representation power of the unit.

\subsection*{Hyperparameter selection}
Deep learning models require a number of hyperparameters to be selected. We follow the standard approach by cross-validating the hyperparameters via a validation set. We fix a training and validation set, and then train neural networks with different hyperparameters on the training set and compare the loss function on the validation set.
We cross-validate the number of layers, the number of units per layer, the dropout rate, the batch size, and the activation function (i.e., the type of non-linearity) via Tree-structured Parzen Estimator (TPE) approach (see \citeN{bergtpe}),\footnote{We use TPE since it outperformed random search (see \citeN{bergtpe}), which was shown to be both theoretically and empirically more efficient than standard techniques such as trials on a grid. Other widely used strategies are grid search and manual search.} and select the hyperparameters with the lowest validation loss.     

The training set for our out-of-sample hyperparameter optimization comes from 2004Q3, while the validation set is from 2006Q3. Table \ref{tab:hyper-ml} summarizes our machine learning model hyperparameters. For our neural network, we used 5 hidden layers, with 150-600-1000-600-400 neurons per layer, RELU activation function, a batch size of 4096, a learning rate of 0.003, and a dropout rate of 50\%. For our GBT, We found that a learning rate of 0.05, a max tree depth of 6, a max bin size of 64, with 1000 trees gave us good performance. All GBT models were run until their validation accuracy was non-improving for a hundred rounds and were trained on CPUs.    

\begin{table}[htbp]
\caption{Hyperparameters for Machine Learning Models: Out-of-sample Exercise}\label{tab:hyper-ml}
\normalsize
\begin{center}
\begin{tabular}{lll}
Model & Tree Depth  & \# of Trees  \\ \hline 
   &  &   \\
CART & 7 &  \\
RF & 20  & 900 \\
GBT & 6 & 1000 \\ \hline  \hline 
\end{tabular}
\end{center}
\end{table}

For the pooled sample prediction, we increased the number of neurons per layers to 512,1024,2048,1024,512 and decreased the dropout rate to 20\%, keeping the activation function, the batch size, and the learning rate unchanged. We instituted early stopping with a patience of 1,000 for GBT, and trained a model of depth 6 with up to 10,000 trees and a learning rate of 0.3. We report the results of the best performing GBT. 

\subsection*{Implementation}
We include 139 features for each individual. Since we work with panel data, there is a sample for each quarter of data. We train roughly 20 million samples, which takes up around 20 gigabytes of data. Our deep learning models are made up of millions of free parameters. Since the estimation procedure relies on computing gradients via backpropagation, which tends to be time and memory intensive, using conventional computing resources (e.g., desktop) would be impractical (if not infeasible). We address the time and memory intensity with two methods. First, to save memory, we use single precision floating point operations, which halves the memory requirements and results in a substantial computational speedup. Second, to accelerate the learning, we parallelized our computations and trained all of our models on a GPU cluster\footnote{1 node with 4 NVIDIA GeForceGTX1080 GPUs. The pooled model trains within 36 hours.}. In our setting, GPU computations were over 40X faster than CPU for our deep neural networks. For a discussion on the impact of GPUs in deep learning see \citeN{schmidhuberdeep}.       

We conduct our analysis using Python 3.6.3 (Python Software Foundation), building on the packages numpy (\citeN{walt2011numpy}), pandas (\citeN{mckinney2010data}) and matplotlib (\citeN{hunter2007matplotlib}). We develop our deep neural networks with keras (\citeN{chollet2015keras}) running on top of Google TensorFlow, a powerful library for large-scale machine learning on heterogenous systems (\citeN{abadi2016tensorflow}). We run our machine learning algorithms using sci-kit learn (\citeN{pedregosa2011scikit}) and (\citeN{xgboost}).

\subsection*{Features}\label{sec:features}
Table \ref{tab:features} lists our model inputs. Table \ref{tab:sum_stats} provides summary statistics for selected features. For the SHAP value analysis, we grouped features that had a correlation higher than 0.7. These groups are presented in Table \ref{tab:feature_corr}. 

\subsection*{Weighting}\label{sec:weighting}
We have investigated alternative weighting schemes, and the results are reported in Table \ref{tab:weighting}. The sample corresponds to Table \ref{tab:model_comparison}, where the benchmark to this exercise is the average reported in the last row. Based on this exercise, the optimal weight on the DNN would be between 0.2 and 0.4. We selected an equally weighted model, given that the decision maker would not have access to this data ex-ante. 

\begin{table}[htbp]
\caption{Weighting Schemes and Average Loss}\label{tab:weighting}
\begin{center}
\begin{tabular}{ll}
Weight on DNN & Average Loss \\ \hline
& \\ 
0.1 & 0.3170 \\
0.2 & 0.3168 \\
0.3 & 0.3168 \\
0.4 & 0.3170 \\
0.5 & 0.3173 \\
0.6 & 0.3180 \\
0.7 & 0.3188 \\
0.8 & 0.3198 \\
0.9 & 0.3211 \\ \hline 
\end{tabular}
\end{center}
\msk

\footnotesize{Performance comparison of our hybrid DNN-GBT model under different weighting schemes. The results of predicted probabilities versus actual outcomes over the following 8Q (testing period) are used to calculate the loss metric for 90+ days delinquencies within 8Q. DNN refers to deep neural network, Source: Authors' calculations based on Experian Data.}

\end{table}

\begin{table}[htbp]
\footnotesize
\caption{Model Inputs}\label{tab:features}
\begin{tabular}{p{3.5in}p{3.5in}}
\hline
& \\
90 day delinquencies in the last 36 months & Credit amount paid down on open first mortgage trades \\
90 days delinquencies in the last 12 months & Credit amount paid down on open second mortgage trades \\
90 days delinquencies in the last 24 months & Credit card trades opened in the last 12 months \\
90 days delinquencies in the last 6 months & Credit card utilization ratio \\
Auto loan or lease inquiries made in the last 3 months & Dismissed bankruptcies \\
Auto loan trades opened in the last 6 months & Early payoff trades \\
Balance on 30 days late bankcard trades & Fannie Mae first mortgage trades opened prior to June 2009 \\
Balance on 30 days late mortgage trades & First mortgage trades opened in the last 6 months \\
Balance on 60 days late bankcard trades & Fraction of 30 days delinquent debt to total debt \\
Balance on 60 days late mortgage trades & Fraction of 60 days delinquent debt to total debt \\
Balance on 90-180 days late bankcard trades & Fraction of 90 days delinquent debt to total debt \\
Balance on 90-180 days late mortgage trades & Fraction of auto loan to total debt \\
Balance on authorized user trades & Fraction of credit card debt to total debt \\
Balance on bankcard trades & Fraction of HELOC to total debt \\
Balance on collections & Fraction of mortgage to total debt \\
Balance on credit card trades & Freddie Mac first mortgage trades opened prior to June 2009 \\
Balance on derogatory bankcard trades & HELOC trades ever 90 or more days delinquent or derogatory\\
Balance on derogatory mortgage trades & HELOC utilization ratio \\
Balance on first mortgage trades & Inquiries made in the last 12 months \\
Balance on HELOC trades & Installment trades \\
Balance on mortgage trades & Installment utilization ratio \\
Balance on open 30 days late installment trades & Judgments with amount \textgreater \$1000 \\
Balance on open 30 days late revolving trades & Monthly payment on all debt \\
Balance on open 60 days late installment trades & Monthly payment on credit card trades \\
Balance on open 60 days late revolving trades & Monthly payment on HELOC trades \\
Balance on open 90-180 days late installment trades & Monthly payment on open auto loan trades \\
Balance on open 90-180 days late revolving trades & Monthly payment on open first mortgage trades \\
Balance on open auto loan trades & Monthly payment on open second mortgage trades \\
Balance on open bankcard trades with credit line suspended & Monthly payment on student trades \\
Balance on open derogatory installment trades & Mortgage trades \\
Balance on open derogatory revolving trades & Mortgage type \\
Balance on open installment trades & Mortgage inquiries made inthe last 3 months \\
Balance on open personal liable business loans & Open auto loan trades \\
Balance on open revolving trades & Open bankcard trades \\
Balance on revolving trades & Open bankcard trades opened in the last 6 months \\
Balance on second mortgage trades & Open credit card trades \\
Balance on student trades & Open first mortgage trades \\
Bankcard inquiries made in the last 3 months & Open HELOC trades \\
Bankruptcies filed within the last 24 months & Open mortgage trades \\
Chapter 13 bankruptcies & Open personal liable business loans \\
Chapter 7 bankruptcies & Open second mortgage trades \\
Charge-off amount on unsatisfied charge-off trades & Petitioned bankruptcies \\
Charge-off trades & Public record bankruptcies \\
Collections placed in the last 12 months & Public records filed in the last 24 months \\
Credit amount on deferred student trades & Total 30 days late debt balances \\
Credit amount on non-deferred student trades & Total 60 days late debt balances \\
Credit amount on open credit card trades & Total 90 or more days delinquent debt balances \\
Credit amount on open HELOC trades & Total 90-180 days late debt balances \\
Credit amount on open installment trades & Total credit amount on open trades \\
Credit amount on open mortgage trades & Total debt balances \\
Credit amount on revolving trades & Total derogatory debt balances \\
Credit amount on unsatisfied derogatory trades & Trades legally paid in full for less than the full balance \\
\vdots & \vdots \\ \hline 
\end{tabular}
\end{table}

\begin{table}[htbp]
\footnotesize
\begin{tabular}{l}
\hline
\vdots \\
 Months since the most recently closed transferred or refinanced first mortgage \\
Months since the most recently opened auto loan trade \\
Months since the most recently opened credit card trade \\
Months since the most recently opened first mortgage \\
Months since the most recently opened HELOC trade \\
Months since the most recently opened second mortgage \\
Months since the most recent 30-180 days delinquency on auto loan or lease trades \\
Months since the most recent 30-180 days delinquency on mortgage trade \\
Months since the most recent 30-180days delinquency on credit card trades \\
Months since the most recent foreclosure proceeding started on first mortgage \\
Months since the most recent public record bankruptcy filed \\
Months since the most recent 30-180 days delinquency \\
Months since the oldest trade was opened \\
Months since the most recent 90 or more days delinquency \\
Presence of outstanding governmental agency debts \\
Presently foreclosed first mortgage trades that occurred in the last 24 months \\
Student trades ever 90 or more days delinquent or derogatory occurred in the last 24 months \\
Trades ever 90 or more days delinquent or derogatory occurred in the last 24 months \\
Presently foreclosed first mortgages \\
Ratio of inquiries to trades opened in the last 6 months \\
Utility trades \\
Utilization ratio \\
Unsatisfied collections \\
Unsatisfied repossession trades \\
Worst ever status on a credit card trade in the last 24 months \\
Worst ever status on a mortgage trade in the last 24 months \\
Worst ever status on an auto loan or lease trade in the last 24 months \\
Worst ever status on any trades in the last 24 months \\
Worst present status on a mortgage trade \\
Worst present status on a revolving trade \\
Worst present status on an auto loan or lease trade \\
Worst present status on an installment trade \\
Worst present status on an open trade \\
Worst present status on any trades \\
Worst present status on bankcard trades \\ \hline 
\end{tabular}
\centering
\ssk

\footnotesize{List of features included in our model. }
\end{table}

\begin{table}[htbp]
\caption{Summary Statistics}\label{tab:sum_stats}
\footnotesize
\begin{tabular}{p{3.75in}lllll}
Feature                                                                                                   & Mean      & Std. Dev & 25\% & Median & 75\%  \\ \hline                                                                                            &      &     &           \\
Balance on collections & 724.74 & 3951.88 & 0 & 0 & 0 \\
Balance on credit card trades & 4573.20 & 9824.31 & 0 & 834 & 4591 \\
Balance on mortgage trades & 63408.03 & 160889.35 & 0 & 0 & 76301 \\
Balance on open auto loan trades & 4472.88 & 11608.48 & 0 & 0 & 3915 \\
Balance on open installment trades & 8754.12 & 32554.07 & 0 & 0 & 10583 \\
Balance on open personal liable business loan & 290.94 & 17183.32 & 0 & 0 & 0 \\
Balance on revolving trades & 4532.12 & 9804.11 & 0 & 761 & 4478 \\
Balance on student trades & 3523.54 & 15537.59 & 0 & 0 & 0 \\
Charge-off amount on unsatisfied charge-off trades & 1264.56 & 83747.09 & 0 & 0 & 0 \\
Collections placed in the last 12 months & 0.38 & 1.24 & 0 & 0 & 0 \\
Credit amount on open credit card trades & 21475.90 & 30662.29 & 0 & 8600 & 31947 \\
Credit amount on open installment trades & 12178.34 & 37332.78 & 0 & 0 & 17286 \\
Credit amount on revolving trades & 21382.80 & 30396.75 & 0 & 8641 & 31890 \\
Credit amount on unsatisfied derogatory trades & 11048.38 & 70187.96 & 0 & 0 & 1500 \\
Credit amount paid down on open first mortgage trades & 6109.98 & 164527.99 & 0 & 0 & 2255 \\
Credit card utilization ratio & 0.51 & 2.31 & 0 & 0.06 & 0.37 \\
Early payoff trades & 0.94 & 1.64 & 0 & 0 & 1 \\
Fraction of 30 days delinquent debt to total debt & 0.02 & 0.11 & 0 & 0 & 0 \\
Fraction of 60 days delinquent debt to total debt & 0.01 & 0.08 & 0 & 0 & 0 \\
Fraction of 90 days delinquent debt to total debt & 0.03 & 0.15 & 0 & 0 & 0 \\
Fraction of auto loan to total debt & 0.12 & 0.27 & 0.00 & 0.00 & 0.04 \\
Fraction of credit card debt to total debt & 0.27 & 0.40 & 0.00 & 0.03 & 0.44 \\
Fraction of home equity line of credit to total debt & 0.03 & 0.13 & 0.00 & 0.00 & 0.00 \\
Fraction of mortgage to total debt & 0.30 & 0.42 & 0.00 & 0.00 & 0.82 \\
Inquiries made in the last 12 months & 1.38 & 2.28 & 0 & 1 & 2 \\
Judgments with amount \textgreater \$1000 & 0.06 & 0.32 & 0 & 0 & 0 \\
Monthly payment on all debt & 907.95 & 11059.39 & 20 & 333 & 1232 \\
Monthly payment on credit card trades & 121.22 & 366.75 & 0 & 31 & 125 \\
Monthly payment on open auto loan trades & 132.70 & 313.24 & 0 & 0 & 227 \\
Monthly payment on student trades & 21.12 & 4628.97 & 0 & 0 & 0 \\
Months since the most recently opened credit card trade & 35.64 & 46.17 & 7 & 20 & 47 \\
Months since the oldest trade was opened & 196.45 & 126.35 & 98 & 178 & 271 \\
Open auto loan trades & 0.34 & 0.60 & 0 & 0 & 1 \\
Open credit card trades & 3.56 & 3.91 & 0 & 2 & 5 \\
Open mortgage trades & 0.50 & 0.83 & 0 & 0 & 1 \\
Total 60 days late debt balances & 619.09 & 13253.67 & 0 & 0 & 0 \\
Total 90 or more days delinquent debt balances & 3125.34 & 33186.26 & 0 & 0 & 0 \\
Total credit amount on open trades & 108480.31 & 259094.00 & 3000 & 33146 & 146535 \\
Total debt balances & 77126.36 & 170742.97 & 318 & 11738 & 95808 \\
Total derogatory debt balances & 1287.84 & 18422.40 & 0 & 0 & 0 \\
Trades ever 90 or more days delinquent or derogatory, last 24 months & 1.22 & 2.78 & 0 & 0 & 1 \\
Utilization ratio & 0.56 & 0.88 & 0.027 & 0.52 & 0.83 \\ \hline 
\end{tabular}
\end{table}

\begin{table}[htbp]
\footnotesize
\caption{Feature Groups}\label{tab:feature_corr} 
\begin{tabular}{ll}
{\ul Total debt balances*} & {\ul Number of collections*} \\
Total debt balances & Collections placed in the last 12 months \\
Total credit amount on open trades & Trades ever 90 or more days delinquent \\
Balance on mortgage trades & Unsatisfied collections \\
Credit amount on open mortgage trades &  \\
Balance on first mortgage trades & {\ul Number of open credit cards*} \\
 & Open credit card trades \\
{\ul 30 days late debt balances*} & Credit amount on open credit card trades \\
Total 30 days late debt balances & Open bankcard trades \\
Balance on 30 days late mortgage trades & Credit amount on revolving trades \\
 &  \\
{\ul 60 days late debt balances*} & {\ul Number of HELOC loans*} \\
Total 60 days late debt balances & Open home equity line of credit trades \\
Balance on 60 days late mortgage trades & Home equity line of credit utilization ratio \\
 &  \\
{\ul 90+ days late debt balances*} & {\ul Number of mortgages*} \\
Total 90 or more days delinquent debt balances & Fraction of mortgage to total debt \\
Total 90-180 days late debt balances & Mortgage trades \\
Balance on 90-180 days late mortgage trades & Open mortgage trades \\
Total derogatory debt balances & Mortgage type \\
Balance on derogatory mortgage trades & Open first mortgage trades \\
 &  \\
{\ul Balance on installment loans*} & {\ul Foreclosed first mortgages*} \\
Credit amount on open installment trades & Presently foreclosed first mortgages \\
Balance on open installment trades & Presently foreclosed first mortgage trades, last 24 months \\
 &  \\
{\ul Balance on HELOC loans*} & {\ul Auto loan*} \\
Credit amount on open home equity line of credit trades & Open auto loan trades \\
Balance on home equity line of credit trades & Monthly payment on open auto loan trades \\
Balance on open revolving trades & Balance on open auto loan trades \\
 &  \\
{\ul Student debt*} & {\ul Credit card debt*} \\
Balance on student trades & Balance on credit card trades \\
Credit amount on non-deferred student trades & Balance on revolving trades \\
 & Balance on bankcard trades \\
{\ul Worst status on any trades*} &  \\
Worst ever status on any trades in the last 24 months & {\ul Trades ever 90 or more days delinquent or derogatory*} \\
Worst present status on any trades & 90 days delinquencies in the last 6 months \\
 & 90 days delinquencies in the last 12 months \\
{\ul Worst status on credit cards*} & 90 days delinquencies in the last 24 months \\
Worst present status on a revolving trade & 90 days delinquencies in the last 36 months \\
Worst present status on bankcard trades &  \\
 & {\ul Bankruptcy history*} \\
{\ul Fraction of 90 days late debt to total debt*} & Public record bankruptcies \\
Worst present status on an open trade & Chapter 7 bankruptcies \\
Fraction of 90 days delinquent debt to total debt & Months since the most recent public record bankruptcy filed
\end{tabular}
\end{table}

\clearpage 
\section{Comparison with Credit Scores}\label{app:score}
The credit score is a summary indicator intended to predict the risk of default by the borrower and it is widely used by the financial industry.  For most unsecured debt, lenders typically verify a perspective borrower's credit score at the time of application and sometimes a short recent sample of their credit history. For larger unsecured debts, lenders also typically require some form of income verification, as they do for secured debts, such as mortgages and auto loans. Still, the credit score is often a key determinant of crucial terms of the borrowing contract, such as the interest rate, the downpayment or the credit limit.  

The most widely known credit score is the FICO score, a measure generated by the Fair Isaac Corporation, which has been in existence in its current form since 1989. Each of the three major credit reporting bureaus-- Equifax, Experian and TransUnion-- also have their own proprietary credit scores. Credit scoring models are not public, though they are restricted by the law, mainly the Fair Credit Reporting Act of 1970 and the Consumer Credit Reporting Reform Act of 1996. The legislation mandates that consumers be made aware of the 4 main factors that may affect their credit score. 
Based on available descriptive materials from FICO and the credit bureaus, these are payment history and outstanding debt, which account for more than 65\% of the variation in credit scores, followed by credit history, or the age of existing accounts, which explains  15\% of the variation, followed by new accounts and types of credit used (10\%) and new "hard" inquiries, that is credit report inquiries coming from prospective lenders after a borrower initiated credit application.

U.S. law prohibits credit scoring models from considering a borrower's race, color, religion, national origin, sex and marital status, age, address, as well as any receipt of public assistance, or the exercise of any consumer right under the Consumer Credit Protection Act. The credit score cannot be based on information not found in a borrower's credit report, such as salary, occupation, title, employer, date employed or employment history, or interest rates being charged on particular accounts. Finally, any items in the credit report reported as child/family support obligations are not permitted, as well as "soft" inquiries\footnote{These include "consumer-initiated" inquiries, such as requests to view one's own  credit report, "promotional inquiries,"  requests made by lenders in order to make pre-approved credit offers, or "administrative inquiries," requests made by lenders to review open accounts. Requests that are marked as coming from employers are also not counted.} and any information that is not proven to be predictive of future credit performance.

\begin{figure}[htbp]
\centering
\includegraphics[scale=0.425]{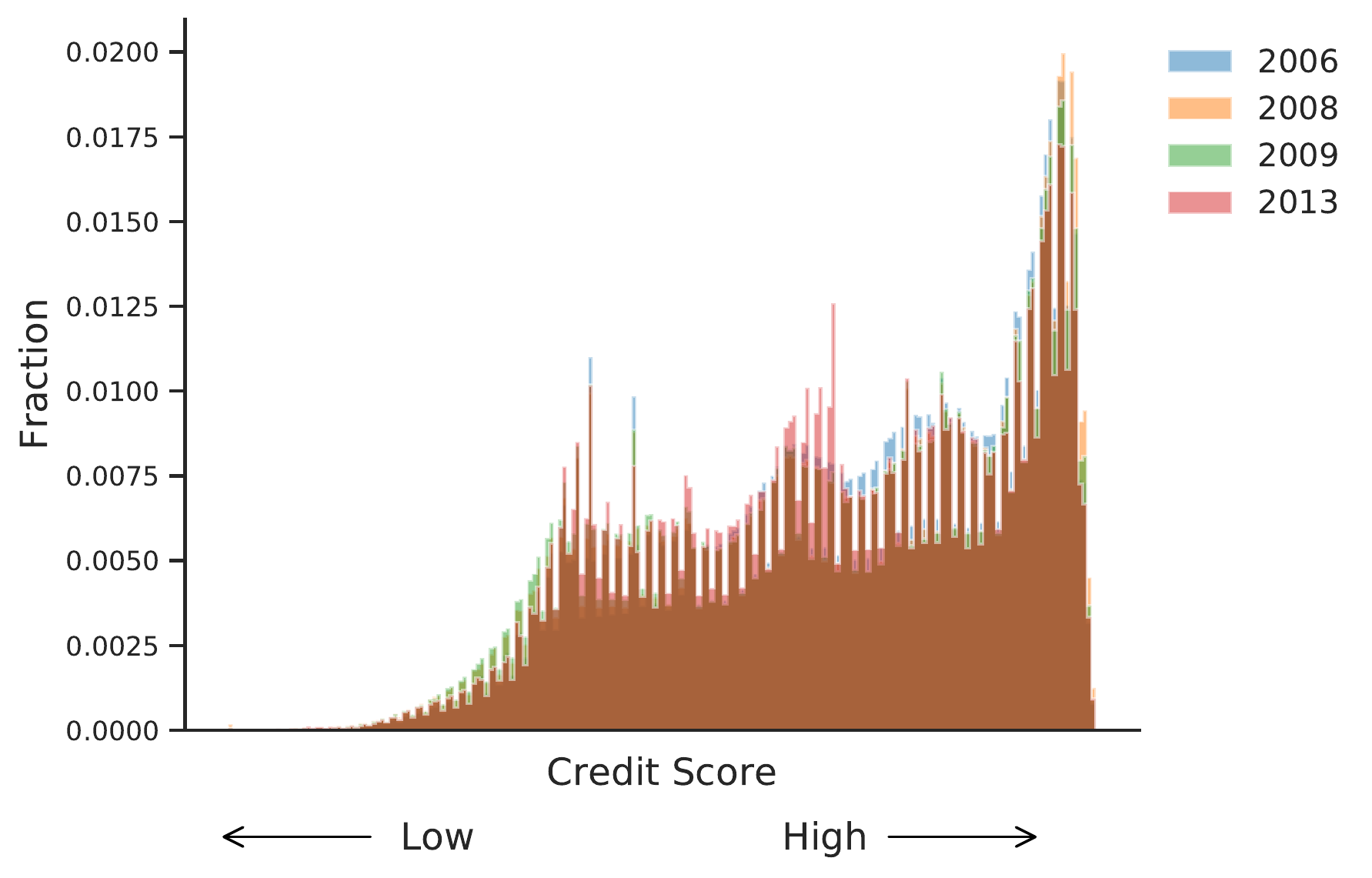}
\caption{Histogram of the credit score in our data by year for selected years.  Source: Source: Authors' calculations based on Experian data.}\label{fig:vs_hist}
\end{figure}

\begin{figure}[htbp]
\centering
\subfloat[Rank Correlation]{\includegraphics[scale=0.325]{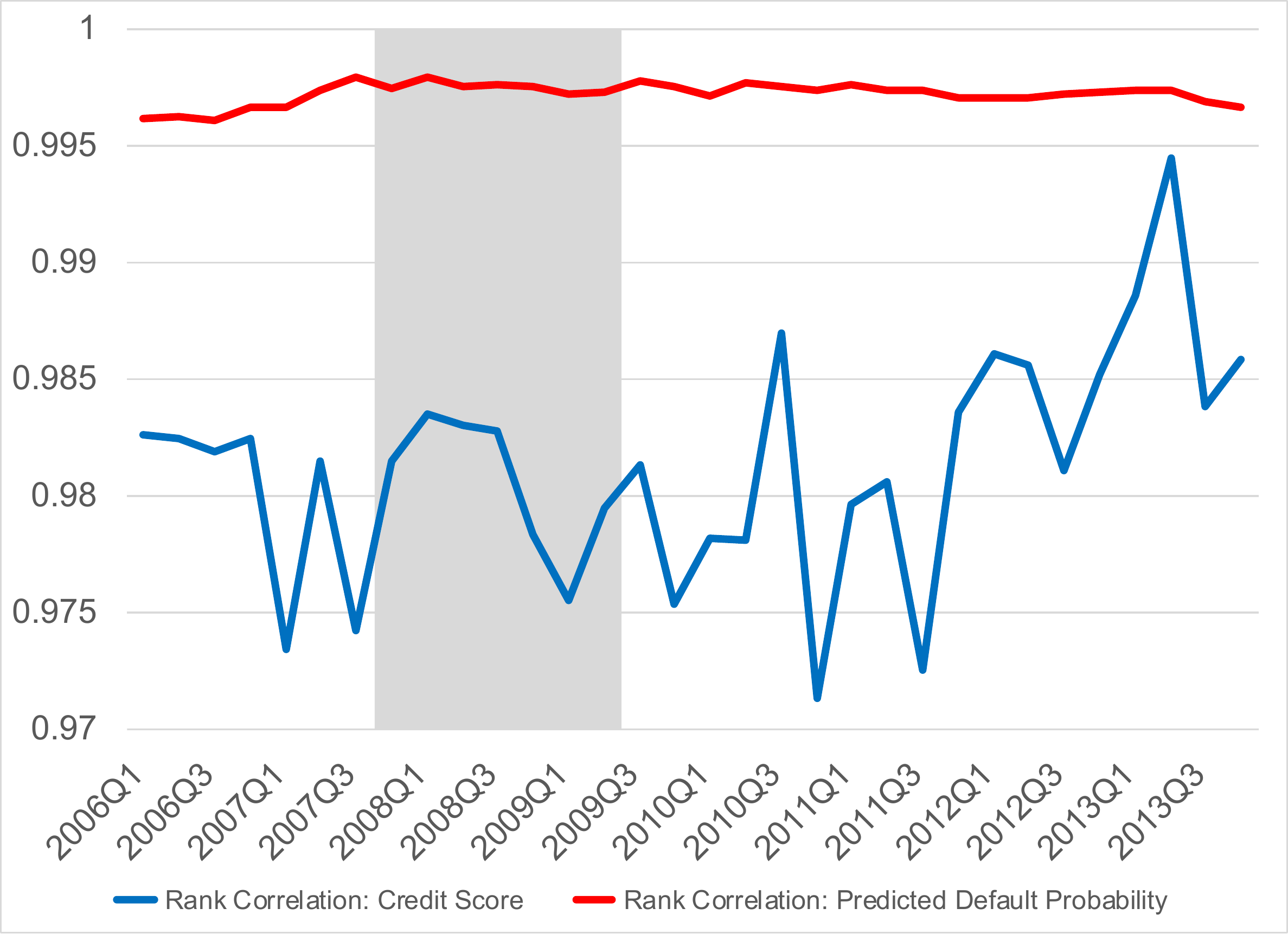}  } 
\subfloat[Gini Coefficient]{\includegraphics[scale=0.325]{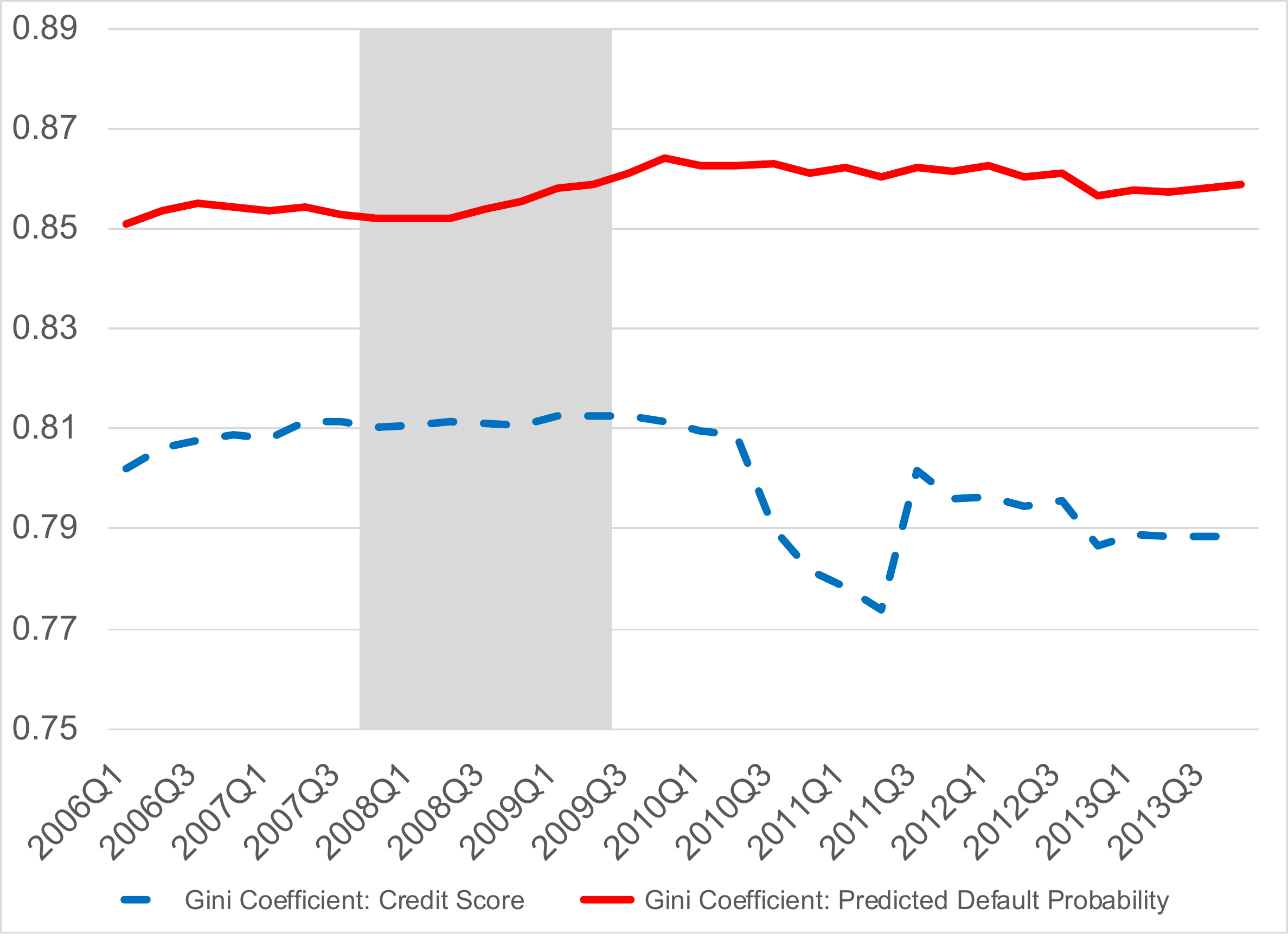}  }  \\  
\subfloat[Rank Correlation: Current]{\includegraphics[scale=0.325]{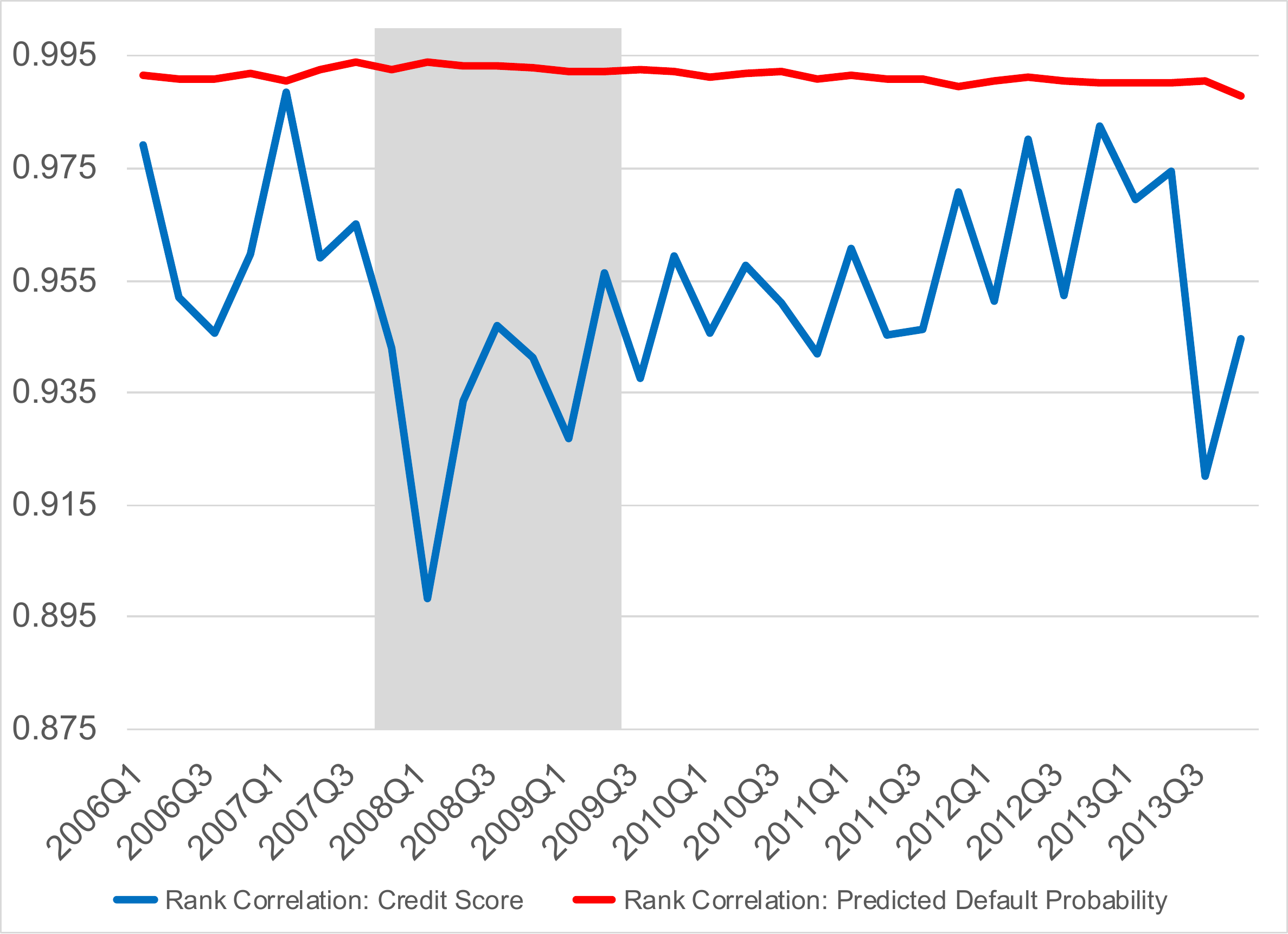}  }    
\subfloat[Gini Coefficient: Current]{\includegraphics[scale=0.325]{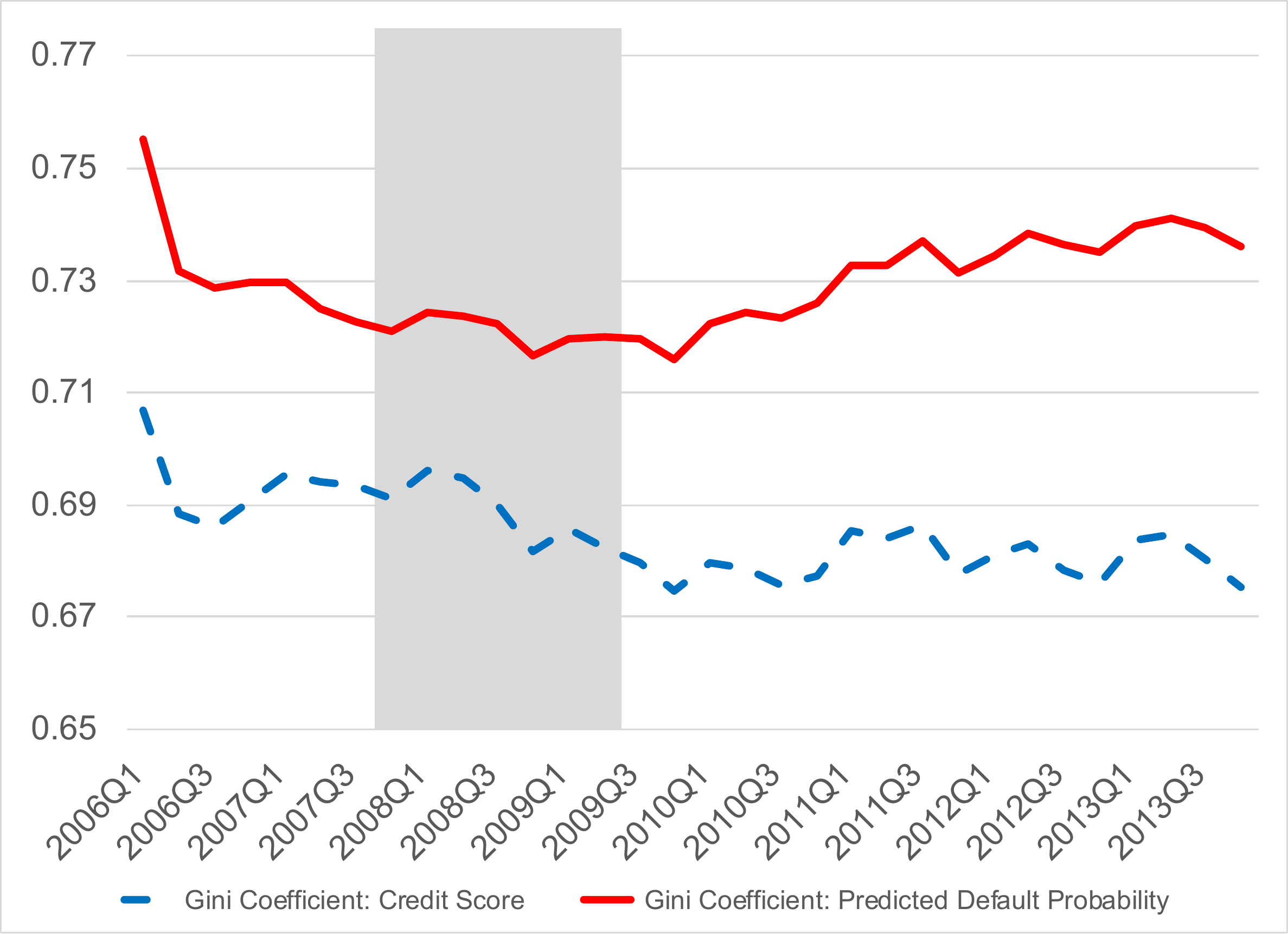}  }  
\caption{
Absolute value of rank correlation with realized default rate for the credit score and model predicted default probability for the full sample (a), for the current population (c), and Gini coefficients for the credit score and model predicted default probability by quarter for the full sample (b), and for the current population (d).  Source: Authors' calculations based on Experian data.}\label{fig:rank_gini_comp}
\end{figure}

\section{Model Comparison}\label{app:model_comp}
Table \ref{tab:model_comparison} compares the performance our our DNN to GBT by keeping our models' architecture the same, but expanding the training data by including observations up till the date specified by the training window. This exercise illustrates that while the performance of GBT remains similar, DNN benefits significantly from having more data to train on. Table \ref{tab:model_comparison3} looks at performance differences when we allow only the most recent 4 quarters for training. Next, Table \ref{tab:shap_comparison} examines the SHAP values across four different models on the pooled sample: (1) logistic, (2) DNN, (3) GBT, and (4) the hybrid model. Last, we computed the SHAP contributions across different debt categories, and the results are reported in Table \ref{tab:shap_group_comparison}. 

\begin{table}[htbp]
\caption{Model Comparison: DNN vs. GBT}\label{tab:model_comparison2}
\footnotesize
\begin{center}
\begin{tabular}{cccccc}
Training Window* & Testing Window & \multicolumn{2}{c}{AUC-score} & \multicolumn{2}{c}{Loss} \\ \hline 
 &  &  &  &  &  \\
 &  & DNN & GBT & DNN & GBT \\ \hline 
 &  &  &  &  &  \\
2004Q1 & 2006Q1 & 0.9221 & 0.9242 & 0.3283 & 0.3234 \\
2004Q2 & 2006Q2 & 0.9235 & 0.9251 & 0.3232 & 0.3190 \\
2004Q3 & 2006Q3 & 0.9245 & 0.9265 & 0.3221 & 0.3161 \\
2004Q4 & 2006Q4 & 0.9240 & 0.9257 & 0.3242 & 0.3192 \\
2005Q1 & 2007Q1 & 0.9249 & 0.9261 & 0.3247 & 0.3208 \\
2005Q2 & 2007Q2 & 0.9247 & 0.9259 & 0.3275 & 0.3226 \\
2005Q3 & 2007Q3 & 0.9240 & 0.9253 & 0.3288 & 0.3251 \\
2005Q4 & 2007Q4 & 0.9231 & 0.9243 & 0.3307 & 0.3283 \\
2006Q1 & 2008Q1 & 0.9239 & 0.9249 & 0.3322 & 0.3288 \\
2006Q2 & 2008Q2 & 0.9233 & 0.9243 & 0.3338 & 0.3307 \\
2006Q3 & 2008Q3 & 0.9243 & 0.9253 & 0.3305 & 0.3283 \\
2006Q4 & 2008Q4 & 0.9250 & 0.9257 & 0.3305 & 0.3280 \\
2007Q1 & 2009Q1 & 0.9266 & 0.9275 & 0.3258 & 0.3233 \\
2007Q2 & 2009Q2 & 0.9268 & 0.9277 & 0.3240 & 0.3222 \\
2007Q3 & 2009Q3 & 0.9282 & 0.9288 & 0.3196 & 0.3183 \\
2007Q4 & 2009Q4 & 0.9299 & 0.9307 & 0.3156 & 0.3133 \\
2008Q1 & 2010Q1 & 0.9299 & 0.9308 & 0.3167 & 0.3140 \\
2008Q2 & 2010Q2 & 0.9300 & 0.9308 & 0.3152 & 0.3128 \\
2008Q3 & 2010Q3 & 0.9300 & 0.9308 & 0.3142 & 0.3118 \\
2008Q4 & 2010Q4 & 0.9300 & 0.9307 & 0.3143 & 0.3123 \\
2009Q1 & 2011Q1 & 0.9311 & 0.9317 & 0.3115 & 0.3097 \\
2009Q2 & 2011Q2 & 0.9297 & 0.9302 & 0.3139 & 0.3125 \\
2009Q3 & 2011Q3 & 0.9298 & 0.9303 & 0.3134 & 0.3116 \\
2009Q4 & 2011Q4 & 0.9297 & 0.9302 & 0.3131 & 0.3114 \\
2010Q1 & 2012Q1 & 0.9303 & 0.9307 & 0.3112 & 0.3099 \\
2010Q2 & 2012Q2 & 0.9288 & 0.9293 & 0.3136 & 0.3124 \\
2010Q3 & 2012Q3 & 0.9286 & 0.9290 & 0.3120 & 0.3112 \\
2010Q4 & 2012Q4 & 0.9281 & 0.9286 & 0.3132 & 0.3121 \\
2011Q1 & 2013Q1 & 0.9287 & 0.9293 & 0.3111 & 0.3102 \\
2011Q2 & 2013Q2 & 0.9284 & 0.9289 & 0.3118 & 0.3103 \\
2011Q3 & 2013Q3 & 0.9290 & 0.9294 & 0.3098 & 0.3085 \\
2011Q4 & 2013Q4 & 0.9290 & 0.9295 & 0.3084 & 0.3075 \\ 
 & & &  &  &   \\
Average &        & 0.9271 & 0.9281 & 0.3195 & 0.3171 \\ \hline 
\end{tabular}
\end{center}

\msk

\footnotesize{Performance comparison of the two best performing machine learning classification models of consumer default risk. The model calibrations are specified by the training and testing windows. * implies that all data was used up to the quarter specified. The results of predicted probabilities versus actual outcomes over the following 8Q (testing period) are used to calculate the loss metric and the AUC-score for 90+ days delinquencies within 8Q. DNN refers to deep neural network, GBT refers to gradient boosted trees. Source: Authors' calculations based on Experian Data.}

\end{table}

\begin{table}[htbp]
\caption{Model Comparison: DNN vs. GBT}\label{tab:model_comparison3}
\footnotesize
\begin{center}
\begin{tabular}{ccccccc}
Training Window Start & Training Window End & Testing Window & \multicolumn{2}{c}{AUC-score} & \multicolumn{2}{c}{Loss} \\ \hline 
 &  &  &  & &  &  \\
 & & & DNN & GBT & DNN & GBT \\ \hline 
 & & &  &  &  &  \\
2004Q1 & 2004Q1 & 2006Q1 & 0.9221 & 0.9242 & 0.3305 & 0.3234 \\
2004Q1 & 2004Q2 & 2006Q2 & 0.9235 & 0.9251 & 0.3232 & 0.3190 \\
2004Q1 & 2004Q3 & 2006Q3 & 0.9246 & 0.9265 & 0.3212 & 0.3161 \\
2004Q1 & 2004Q4 & 2006Q4 & 0.9240 & 0.9257 & 0.3238 & 0.3192 \\
2004Q2 & 2005Q1 & 2007Q1 & 0.9248 & 0.9261 & 0.3238 & 0.3205 \\
2004Q3 & 2005Q2 & 2007Q2 & 0.9244 & 0.9258 & 0.3250 & 0.3220 \\
2004Q4 & 2005Q3 & 2007Q3 & 0.9245 & 0.9253 & 0.3270 & 0.3244 \\
2005Q1 & 2005Q4 & 2007Q4 & 0.9230 & 0.9243 & 0.3300 & 0.3273 \\
2005Q2 & 2006Q1 & 2008Q1 & 0.9233 & 0.9251 & 0.3307 & 0.3270 \\
2005Q3 & 2006Q2 & 2008Q2 & 0.9232 & 0.9244 & 0.3317 & 0.3287 \\
2005Q4 & 2006Q3 & 2008Q3 & 0.9242 & 0.9254 & 0.3296 & 0.3263 \\
2006Q1 & 2006Q4 & 2008Q4 & 0.9249 & 0.9257 & 0.3286 & 0.3261 \\
2006Q2 & 2007Q1 & 2009Q1 & 0.9264 & 0.9277 & 0.3249 & 0.3216 \\
2006Q3 & 2007Q2 & 2009Q2 & 0.9268 & 0.9278 & 0.3227 & 0.3209 \\
2006Q4 & 2007Q3 & 2009Q3 & 0.9276 & 0.9288 & 0.3205 & 0.3172 \\
2007Q1 & 2007Q4 & 2009Q4 & 0.9293 & 0.9307 & 0.3160 & 0.3124 \\
2007Q2 & 2008Q1 & 2010Q1 & 0.9293 & 0.9306 & 0.3171 & 0.3133 \\
2007Q3 & 2008Q2 & 2010Q2 & 0.9289 & 0.9304 & 0.3169 & 0.3129 \\
2007Q4 & 2008Q3 & 2010Q3 & 0.9286 & 0.9302 & 0.3168 & 0.3128 \\
2008Q1 & 2008Q4 & 2010Q4 & 0.9280 & 0.9299 & 0.3191 & 0.3142 \\
2008Q2 & 2009Q1 & 2011Q1 & 0.9294 & 0.9307 & 0.3156 & 0.3121 \\
2008Q3 & 2009Q2 & 2011Q2 & 0.9277 & 0.9291 & 0.3191 & 0.3150 \\
2008Q4 & 2009Q3 & 2011Q3 & 0.9276 & 0.9291 & 0.3181 & 0.3147 \\
2009Q1 & 2009Q4 & 2011Q4 & 0.9277 & 0.9294 & 0.3179 & 0.3138 \\
2009Q2 & 2010Q1 & 2012Q1 & 0.9291 & 0.9302 & 0.3144 & 0.3113 \\
2009Q3 & 2010Q2 & 2012Q2 & 0.9276 & 0.9290 & 0.3168 & 0.3133 \\
2009Q4 & 2010Q3 & 2012Q3 & 0.9278 & 0.9288 & 0.3145 & 0.3119 \\
2010Q1 & 2010Q4 & 2012Q4 & 0.9270 & 0.9284 & 0.3165 & 0.3132 \\
2010Q2 & 2011Q1 & 2013Q1 & 0.9278 & 0.9291 & 0.3140 & 0.3111 \\
2010Q3 & 2011Q2 & 2013Q2 & 0.9275 & 0.9286 & 0.3140 & 0.3112 \\
2010Q4 & 2011Q3 & 2013Q3 & 0.9278 & 0.9292 & 0.3128 & 0.3095 \\
2011Q1 & 2011Q4 & 2013Q4 & 0.9281 & 0.9293 & 0.3117 & 0.3084 \\
 &  &  &  &  &  &  \\
Average &  &  & 0.9265 & 0.9278 & 0.3208 & 0.3172 \\ \hline 
\end{tabular}
\end{center}

\msk

\footnotesize{Performance comparison of the two best performing machine learning classification models of consumer default risk. The model calibrations are specified by the training and testing windows. The results of predicted probabilities versus actual outcomes over the following 8Q (testing period) are used to calculate the loss metric and the AUC-score for 90+ days delinquencies within 8Q. DNN refers to deep neural network, GBT refers to gradient boosted trees. Source: Authors' calculations based on Experian Data.}

\end{table}

\begin{table}[htbp]
\caption{Shap Values across Models}\label{tab:shap_comparison}
\footnotesize
\begin{center}
\begin{tabular}{lcccc}
Feature                                                                   & Hybrid & Logistic & DNN & GBT \\ \hline 
& & & & \\ 
Worst status on any trades* & 0.074 (1) & 0.115 (1) & 0.027 (4) & 0.123 (1) \\
Months since the oldest trade was opened & 0.038 (2) & 0.038 (4) & 0.024 (5) & 0.055 (2) \\
Months since the most recent 90 or more days delinquency & 0.029 (3) & 0.019 (8) & 0.027 (3) & 0.036 (5) \\
Number of collections* & 0.028 (4) & 0.064 (2) & 0.044 (1) & 0.017 (16) \\
Balance on collections & 0.025 (5) & 0.004 (26) & 0.003 (41) & 0.048 (3) \\
Number of open credit cards* & 0.024 (6) & 0.042 (3) & 0.037 (2) & 0.031 (8) \\
Total debt balances* & 0.023 (7) & 0.005 (22) & 0.009 (9) & 0.042 (4) \\
90+ days late debt balances* & 0.02 (8) & 0.011 (11) & 0.01 (8) & 0.03 (9) \\
Monthly payment on open first mortgage trades & 0.017 (9) & 0.0 (99) & 0.002 (52) & 0.033 (6) \\
Credit card utilization ratio & 0.016 (10) & 0.002 (36) & 0.006 (17) & 0.027 (10) \\
Monthly payment on all debt & 0.016 (11) & 0.0 (97) & 0.002 (48) & 0.032 (7) \\
Credit amount on unsatisfied derogatory trades & 0.013 (12) & 0.001 (58) & 0.005 (22) & 0.024 (11) \\
Months since the most recent 30-180 days delinquency & 0.012 (13) & 0.002 (35) & 0.01 (6) & 0.018 (13) \\
Inquiries made in the last 12 months & 0.012 (14) & 0.017 (10) & 0.008 (10) & 0.017 (14) \\
Credit card debt* & 0.01 (15) & 0.018 (9) & 0.005 (20) & 0.018 (12) \\
Months since the most recently opened credit card trade & 0.009 (16) & 0.009 (14) & 0.006 (19) & 0.015 (18) \\
Credit amount paid down on open first mortgage trades & 0.009 (17) & 0.001 (60) & 0.003 (36) & 0.016 (17) \\
Utilization ratio & 0.009 (18) & 0.008 (16) & 0.007 (13) & 0.014 (19) \\
Balance on installment loans* & 0.008 (19) & 0.002 (41) & 0.005 (23) & 0.017 (15) \\
Monthly payment on credit card trades & 0.008 (20) & 0.008 (18) & 0.005 (21) & 0.013 (22)  \\ \hline     
\end{tabular}
\end{center}

\msk

\footnotesize{This table reports the Shapley values for four selected machine learning classification models of consumer default risk. We sorted the features based on the feature's relative rank (in parentheses) using the hybrid model. Source: Authors' calculations based on Experian Data.}

\end{table}

\begin{table}[htbp]
\begin{center}
\caption{Shap Values across Debt Categories}\label{tab:shap_group_comparison}
\begin{tabular}{lllll}
& \multicolumn{4}{c}{Model} \\
& Logistic & DNN & GBT & Hybrid \\ 
Feature Group  & & & & \\ \hline 
& & & & \\ 
Total Debt & 0.544 & 0.515 & 0.470 & 0.504 \\
Mortgage Debt & 0.150 & 0.119 & 0.176 & 0.156 \\
Credit Card Debt & 0.136 & 0.177 & 0.146 & 0.146 \\
Installment \& Revolving Debt & 0.085 & 0.084 & 0.095 & 0.086 \\
Auto Debt & 0.025 & 0.057 & 0.051 & 0.049 \\
Student Debt & 0.039 & 0.027 & 0.034 & 0.034 \\
HELOC Debt & 0.021 & 0.020 & 0.027 & 0.025 \\ \hline 
\end{tabular}
\end{center}

\msk

\footnotesize{This table reports the aggregate absolute Shapley values for four selected machine learning classification models of consumer default risk. We grouped our features into debt categories, and computed the sum of the absolute SHAP values. For ease of interpretability, we normalized our feature groups to 1 for each of our models. Source: Authors' calculations based on Experian Data.}

\end{table}

 \begin{table}[htbp]
 \caption{Neural networks comparison: Loss \& Accuracy}
 \label{tab:pooled_leave}
\small
\begin{center}
\begin{tabular}{lllll}
Model & \multicolumn{2}{c}{In-sample Loss}  & \multicolumn{2}{c}{Out-of-sample Loss}   \\
 & w/o Dropout & Dropout & w/o Dropout & Dropout \\ \hline
 & & & & \\ 
Logistic Regression & 0.3451 & 0.3451 & 0.3449 & 0.3450 \\
1 layer & 0.3109 & 0.3106 & 0.3122 & 0.3116 \\
2 layers & 0.2965 & 0.2900 & 0.3078 & 0.3003 \\
3 layers & 0.2804 & 0.2460 & 0.3047 & 0.2744 \\
4 layers & 0.2669 & 0.2142 & 0.3005 & 0.2575 \\
5 layers & 0.2534 & 0.2013 & 0.2978 & 0.2506\\ \hline 
 & & & & \\ 
Model & \multicolumn{2}{c}{In-sample Accuracy}   & \multicolumn{2}{c}{Out-of-sample Accuracy}   \\
 & w/o Dropout & Dropout & w/o Dropout & Dropout \\ \hline
 & & & & \\ 
Logistic Regression & 0.8564 & 0.8564 & 0.8566 & 0.8566 \\
1 layer & 0.8687 & 0.8688 & 0.8681 & 0.8684 \\
2 layers & 0.8751 & 0.8787 & 0.8705 & 0.8736 \\
3 layers & 0.8829 & 0.9017 & 0.8729 & 0.8862 \\
4 layers & 0.8897 & 0.9163 & 0.8755 & 0.8943 \\
5 layers & 0.8968 & 0.9230 & 0.8785 & 0.8981 \\ \hline 
\end{tabular}
\end{center}

\msk

\footnotesize{In-sample and out-of-sample loss (categorical cross-entropy) and accuracy for neural networks of different depth and for logistic regression. Models are calibrated and evaluated on the pooled sample (2004Q1 - 2013Q4). Source: Authors' calculations based on Experian Data.}
\end{table}

\section{Cost Savings for Consumers}\label{app:cost_savings}

\begin{table}[htbp]
\caption{Distribution of Customers by Credit Score and Predicted Default}
\label{tab:value_consumer_shares}
\begin{center}
\begin{tabular}[c]{p{1.5in}p{0.75in}p{0.75in}p{0.75in}p{0.75in}p{0.85in}}
 & \multicolumn{4}{c}{Credit Score  } \\ \hline
   & & & & & \\ 
& & Subprime, Near Prime & Prime Low & Prime Mid &  Prime High, Superprime \\ \hline
  & & & & & \\ 
Predicted Default Bin & 1 & 36.20\% & 3.87\% & 1.10\% & 0.34\% \\
 & 2 & 3.64\% & 3.41\% & 2.51\% & 1.22\% \\
 & 3 & 1.25\% & 2.42\% & 4.02\% & 3.42\% \\
 & 4 & 0.41\% & 1.08\% & 3.48\% & 31.65\%\\ \hline 
\end{tabular}
\end{center}

\ssk
\footnotesize{
This table reports the share of customers in each predicted credit score categories and corresponding predicted default probability bins. Customers are classified based on credit scores and predicted default probabilities at account origination for each of their credit cards included in the balances. Source: Authors' calculations based on Experian data.
}
\end{table}

\begin{sidewaystable}[htbp]
\caption{Cost Savings for Consumers: Descriptive Statistics}
\label{tab:value_consumer_desc}
\begin{center}
\footnotesize
\begin{tabular}{llllllllllll}
 &  & \multicolumn{4}{c}{Credit Score} &  &  & \multicolumn{4}{c}{Credit Score} \\
 & & Subprime, Near Prime & Prime Low & Prime Mid &  Prime High, Superprime   & &  &  & \\  \hline 
 &  &  &  &  &  &  &  &  &  &  &  \\
 & \multicolumn{5}{l}{\textbf{Age}} &  & \multicolumn{5}{l}{\textbf{Credit History (Months)}} \\
\textbf{PP bin} & 1 & 40 & 44 & 47 & 52 & \textbf{PP bin} & 1 & 132 & 177 & 203 & 240 \\
 & 2 & 37 & 40 & 44 & 50 &  & 2 & 133 & 152 & 184 & 232 \\
 & 3 & 43 & 43 & 41 & 47 &  & 3 & 178 & 167 & 161 & 214 \\
 & 4 & 54 & 54 & 46 & 55 &  & 4 & 233 & 263 & 212 & 280 \\
 &  &  &  &  &  &  &  &  &  &  &  \\
 & \multicolumn{5}{l}{\textbf{Household Income (\$)}} &  & \multicolumn{5}{l}{\textbf{Credit Card Limit to Household Income}} \\
\textbf{PP bin} & 1 & 48655 & 63819 & 73405 & 93240 & \textbf{PP bin} & 1 & 0.062 & 0.144 & 0.195 & 0.195 \\
 & 2 & 57101 & 63826 & 72560 & 93274 &  & 2 & 0.226 & 0.214 & 0.238 & 0.238 \\
 & 3 & 76820 & 74341 & 73296 & 92944 &  & 3 & 0.321 & 0.273 & 0.273 & 0.273 \\
 & 4 & 87633 & 92734 & 94809 & 114604 &  & 4 & 0.281 & 0.286 & 0.346 & 0.346 \\
 &  &  &  &  &  &  &  &  &  &  &  \\
 & \multicolumn{5}{l}{\textbf{Total Debt Balances (\$)}} &  & \multicolumn{5}{l}{\textbf{Total 90+ days Debt Balances (\$)}} \\
\textbf{PP bin} & 1 & 44795 & 75847 & 80038 & 94085 & \textbf{PP bin} & 1 & 9627 & 1306 & 1506 & 2592 \\
 & 2 & 79933 & 86453 & 94346 & 119440 &  & 2 & 6 & 3 & 3 & 8 \\
 & 3 & 142295 & 108124 & 97620 & 120817 &  & 3 & 0 & 0 & 0 & 0 \\
 & 4 & 148323 & 143724 & 149229 & 92261 &  & 4 & 0 & 0 & 0 & 0 \\
 &  &  &  &  &  &  &  &  &  &  &  \\
 & \multicolumn{5}{l}{\textbf{Total bankcard balances (\$)}} &  & \multicolumn{5}{l}{\textbf{Total credit card balances (\$)}} \\
\textbf{PP bin} & 1 & 2678 & 3735 & 2870 & 2837 & \textbf{PP bin} & 1 & 3078 & 4345 & 3376 & 3234 \\
 & 2 & 10285 & 6938 & 4507 & 3320 &  & 2 & 10969 & 7518 & 5034 & 3738 \\
 & 3 & 16786 & 10408 & 6244 & 3870 &  & 3 & 17384 & 10795 & 6618 & 4222 \\
 & 4 & 12838 & 11169 & 9108 & 3036 &  & 4 & 13080 & 11364 & 9371 & 3205 \\
 &  &  &  &  &  &  &  &  &  &  &  \\
 & \multicolumn{5}{l}{\textbf{Credit Card Limit (\$)}} &  & \multicolumn{5}{l}{\textbf{Credit Card Utilization}} \\
\textbf{PP bin} & 1 & 4031 & 10066 & 14990 & 26119 & \textbf{PP bin} & 1 & 1.019 & 0.655 & 0.428 & 0.266 \\
 & 2 & 17009 & 15960 & 18183 & 28584 &  & 2 & 0.476 & 0.520 & 0.422 & 0.239 \\
 & 3 & 30403 & 24028 & 22483 & 32143 &  & 3 & 0.411 & 0.354 & 0.289 & 0.167 \\
 & 4 & 29367 & 30076 & 34851 & 39845 &  & 4 & 0.215 & 0.229 & 0.232 & 0.081 \\
 &  &  &  &  &  &  &  &  &  &  &  \\
 & \multicolumn{5}{l}{\textbf{90+ DPD on All Accounts, Next 24 Months}} &  & \multicolumn{5}{l}{\textbf{90+ DPD on Bankcards, Next 24 Months}} \\
\textbf{PP bin} & 1 & 0.77 & 0.42 & 0.38 & 0.42 & \textbf{PP bin} & 1.00 & 0.18 & 0.08 & 0.05 & 0.04 \\
 & 2 & 0.22 & 0.19 & 0.17 & 0.16 &  & 2.00 & 0.08 & 0.05 & 0.04 & 0.02 \\
 & 3 & 0.16 & 0.12 & 0.10 & 0.08 &  & 3.00 & 0.06 & 0.04 & 0.03 & 0.02 \\
 & 4 & 0.09 & 0.06 & 0.06 & 0.03 &  & 4.00 & 0.03 & 0.02 & 0.01 & 0.00\\
  &  &  &  &  &  &  &  &  &  &  &  \\
 & \multicolumn{5}{l}{\textbf{Predicted Probability, Current}} &  & \multicolumn{5}{l}{\textbf{Credit Score, Current}} \\
\textbf{PP bin}  & 1 & 0.762 & 0.448 & 0.416 & 0.445 & \textbf{PP bin}  & 1 & 554 & 677 & 717 & 763 \\
 & 2 & 0.191 & 0.186 & 0.179 & 0.175 &  & 2 & 625 & 680 & 718 & 763 \\
 & 3 & 0.106 & 0.101 & 0.096 & 0.091 &  & 3 & 640 & 681 & 721 & 767 \\
 & 4 & 0.044 & 0.044 & 0.044 & 0.025 &  & 4 & 642 & 680 & 725 & 798 \\ \hline 
\end{tabular}
\end{center}

\ssk
\footnotesize{
This table reports the descriptive statistics on customers based on their credit score and predicted default probability based on our hybrid model. Credit score categories and predicted default probability bins are computed at account origination for credit cards. Source: Authors' calculations based on Experian data.}
\end{sidewaystable}

\end{document}